\documentclass[amsmath,trackchanges,twocolumn]{aastex702}

\newcommand{\Msun}{ \mbox{M$_\odot$}}
\newcommand{\Mjup}{\mbox{M$_{\rm Jup}$}}
\newcommand{\Teff}{T$_{\rm{eff}}$}

\newcommand{\fsed}{$f_{\mathrm{sed}}$}

\usepackage{amssymb,gensymb,times,graphicx,morefloats,color}
\usepackage{amsmath}
\usepackage{graphics,graphicx,url,verbatim,xspace}

\usepackage{changepage}
\usepackage{mathtools} 
\usepackage{comment}
\usepackage{gensymb}
\usepackage{hyperref}
\usepackage{pgfpages}

\usepackage{tikz}   %TikZ is required for this to work.  Make sure this exists before the next line
\usepackage{tikz-3dplot} %requires 3dplot.sty to be in same directory, or in your LaTeX installation
\usepackage[dvipsnames]{xcolor}

\newcommand{\pmerr}[2]{^{+#1}_{-#2}}

\usepackage{listings}
\usepackage{xcolor}
\definecolor{codegreen}{rgb}{0,0.6,0}
\definecolor{codegray}{rgb}{0.5,0.5,0.5}
\definecolor{codepurple}{rgb}{0.58,0,0.82}
\definecolor{backcolour}{rgb}{0.95,0.95,0.92}
\lstdefinestyle{mystyle}{
  backgroundcolor=\color{backcolour},   commentstyle=\color{codegreen},
  keywordstyle=\color{magenta},
  numberstyle=\tiny\color{codegray},
  stringstyle=\color{codepurple},
  basicstyle=\ttfamily\footnotesize,
  breakatwhitespace=false,         
  breaklines=true,                 
  captionpos=b,                    
  keepspaces=true,                 
  numbers=left,                    
  numbersep=5pt,                  
  showspaces=false,                
  showstringspaces=false,
  showtabs=false,                  
  tabsize=2
}
\begin{document}
\defcitealias{macdonaldExploringH2OProminence2018}{M18}
\defcitealias{batalha_color_2018}{B18}
\defcitealias{Karkoschka1994JovianPlanets}{K94}
\defcitealias{Gupta2022JupitersPTVoyager}{G22}

\title{Proposed optimal targets for a Lazuli ExtraSolar Coronagraph reflected starlight exoplanet campaign}

\shorttitle{Lazuli ESC Targets}
\shortauthors{Pearce et al.}

\correspondingauthor{Logan A. Pearce}

\author[0000-0003-3904-7378]{Logan A. Pearce}
\affil{Department of Astronomy, University of Michigan, Ann Arbor, MI 48109, USA}
\email{lapearce@umich.edu}

\author[0000-0002-0813-4308]{Ewan Douglas}
\affil{Steward Observatory, University of Arizona, Tucson, AZ 85721, USA}
\email{douglase@arizona.edu}

\begin{abstract}

Direct detection and characterization of exoplanet atmospheres via reflected starlight enables probes of composition of mature exoplanet systems (since flux is decoupled from age) at closer separations than via thermal emission (due to observations in visible wavelengths), and even probes of biosignatures. New and upcoming ground and space-based platforms are enabling these observations for the first time. Recently the 3-m Lazuli Space Telescope, from the Eric and Wendy Schmidt Observatory System, was announced, which will be equipped with the ExtraSolar Coronagraph (ESC) designed for reflected light gas giant exoplanet observations in visible wavelengths. In this paper we present 27 known giant exoplanets that could make up an optimal first target list for these observations. This target list spans ages and temperatures of planets not probed before by direct imaging, but some of which will also be accessible to the \textsl{Nancy Grace Roman} Coronagraph instrument and the ELT-METIS thermal imager, providing comprehensive atmosphere characterization and synergies. We discuss the science enabled by ESC observations of this target list and enhancements provided by future astrometric and radial velocity measurements.

\end{abstract}


\section{Introduction} \label{sec:intro}

The latest direct imaging platforms are enabling direct detection and characterization of cold and mature exoplanets for the first time. Mid-infrared characterization of directly imaged giant planets is delivering gas giant planet bulk atmosphere abundances and starting to reveal a picture of giant planets dominated by clouds, disequilibrium chemistry, and a core accretion formation pathway \citep{Balmer2025LivingOnTheWedge, BardalezGagluiffi2025-14Herc, Balmer2026DirectImagesOf, Ferrer-Chavez2026Sub-JupitersSurvey, Hemmann2026GiantPlanetsNotFullyMixed, Xuan2026JWST-AFLepb, Xuan2026CompositionsOfHR8799}. JWST MIRI instrument sensitivity is permitting spatially resolved imaging of the coldest giant planets to date \citep{Matthews2024epsIndiAb, BardalezGagluiffi2025-14Herc} and promises sensitivity to planets even as cold as those in our own solar system \citep{Bowens-Rubin2025NIRCamYellsatCloud}.

To date all exoplanets have been imaged through their self-luminous thermal emission, but direct, spatially resolved imaging of planets via reflected starlight is on the near horizon. Reflected light imaging enables detection and characterization of planets decoupled from age (since reflected starlight luminosity does not decrease with age), and the smaller resolution limits provided by imaging shorter wavelengths ($\sim 0.5$--$2 \mu$m), where the star luminosity peaks, allows access to closer-in planets compared to thermal emission ($\sim 2$--$20 \mu$m). Characterization of the same planet in both reflected starlight and thermal emission captures the full energy budget of the planet atmosphere and offers an unprecedented probe of atmospheric processes and formation pathways. 
Exoplanet direct photometry and spectroscopy may also present the only method for detecting biosignatures \citep{hardegree-ullman_bioverse_2023, Hardegree-Ullman2025BioverseO2-ELTs}.

Direct imaging via reflected starlight is being enabled for the first time by the next generation of ground- and space-based visible wavelength coronagraph and extreme adaptive optics (ExAO) instruments. The \textsl{Nancy Grace Roman} Telescope just launched, equipped with the Coronagraph Instrument \citep[CGI,][]{Spergel2013WFIRST}, the first demonstration of visible light coronagraphy in space, planned to carry out the first space-based reflected light imaging campaigns in the next few years. Development of the ground-based visible light ExAO instruments GMagAO-X \citep{Males2024GMagAOX} for the 25.4~m Giant Magellan Telescope (GMT, \citealt{Johns2006GMT, Burgett2026GMTStatus}) and ELT-PCS \citep{Kasper2021-ELTPCS} for the 39~m Extremely Large Telescope (ELT, \citealt{Paolo2023ELT}) is well underway. The MagAO-X \citep{Males2024MagAOXPhaseIICommisioningResults} and SCExAO/VAMPIRES \citep{Lucas2024SCExAO-Vampires} instruments are serving as technology development platforms for this science case, and right now are moving towards the first ground-based reflected light imaging campaigns. Reflected light detection and characterization of potentially Earth-like planets is the one of the main driving goals of the flagship Habitable Worlds Observatory \citep[HWO][]{Astro2020Decadal}.

Recently the Lazuli Space Observatory was announced \citep{Roy2026LazuliArchitechture}, a 3-m space telescope that is part of the Eric and Wendy Schmidt Observatory System, designed for time-domain and multi-messenger astronomy, cosmology, and star and planet science in the 0.4--1.7$\mu$m range. Lazuli will be equipped with the ExtraSolar Coronagraph (ESC), designed with the goal of imaging giant planets in reflected light while building towards future flagship missions like HWO. The ESC is a two-channel vector-vortex coronagraph, leveraging the off-axis Lazuli telescope to push to short ($<0.5\mu$m) wavelengths and close to the host star ($\sim3\lambda/D$). The high-contrast field of view is set by the number of deformable mirror actuators across the telescope pupil. The baseline design calls for $10\lambda/D$, we have optimistically also studied 20$\lambda/D$ which may be an extended mode. For details of the coronagraph implementation, see \cite{Miller2026Lazuli-ESC} and Douglas et al. (in prep).

The high contrasts involved in reflected light detection ($\mathcal{O}\,10^{-7}-10^{-9}$) and subsequent long exposure times (10s--100s of hours) make uninformed searches challenging for a 3m class observatory. Campaign efficiency can be dramatically improved by targeting known planets and leveraging well constrained orbital priors in order to prioritize targets and time observations for the maximum likelihood of detection. This paper describes our first estimate of such ``best" planets for ESC reflected light imaging campaigns. This optimized target list will continue to evolve as the ESC design evolves, new planets are discovered, and orbital element estimates are refined.

In this paper we describe our approach to target selection (Section \ref{sec:target selection}), the properties of the selected top-priority planets and the science the list enables (Section \ref{sec:properties}), challenges to observing them and further observations to aid the campaign (Section \ref{sec:observation challenges}), and synergies with other platforms like CGI (Section \ref{sec:roman}), \textsl{Gaia}, extreme precision radial velocity instruments, and ELT thermal emission direct imaging (Section \ref{sec:other detection methods}).

\begin{deluxetable}{lcl}\label{tab:definitions}
\tablewidth{0pt}
\tablecaption{Definitions}
\tablehead{
\colhead{Parameter} & \colhead{Symbol} & \colhead{Description}
}
\startdata
Filter central wavelength & $\lambda$ & wavelength; central wavelength of the discovery filter $\lambda=0.63\mu$m \\
 & & except otherwise noted\\
Diameter & D & primary mirror diameter; D = 3~m\\
Inner Working Angle & IWA & smallest separation angle for coronagraph speckle suppression region;\\
& & IWA = 3$\lambda/D = 0.13$\arcsec at $\lambda=0.63\mu$m\\
Outer Working Angle & OWA & widest separation angle for coronagraph speckle suppression region \\
& & Design OWA = $10\lambda/D = 0.43$\arcsec at $\lambda=0.63\mu$m\\
& & Goal OWA = $20\lambda/D = 0.87$\arcsec at $\lambda=0.63\mu$m\\
Detection region & & IWA $<$ separation $<$ OWA, contrast $<$2$\times10^{-9}$\\
Semi-major axis & $a$ & orbital ellipse semi-major axis \\
Period & $P$ & orbital period \\
Eccentricity & $e$ & orbital ellipse eccentricity \\
Inclination & $i$ & orbit inclination with respect to sky plane; \\
& & face-on: $i=0^\circ \;\&\; i=180^\circ$, edge-on: $i = 90^\circ$, \\
& & prograde (ccw): $0^\circ<i<90^\circ$, retrograde (cw): $90^\circ<i<180 ^\circ$\\
Argument of periastron & $\omega$ & angle in orbit plane from ascending node to periastron \\
Longitude of ascending node & $\Omega$ & angle in sky plane from reference direction (+Decl) to ascending node \\
Time of periastron passage & T$_0$ & date at which planet is at periastron\\
Planet mass & M$_p$ & planet mass \\
\textsl{Gaia} G magnitude & G & host star magnitude in \textsl{Gaia} G band \\
& & Design limit: G $<$ 5\\
& & Goal limit: G $<$ 8 \\
Geometric albedo & A$_g$ & ratio of planet flux at full phase to an idealized flat, fully reflecting, \\
& & diffusive scattering disk of the same cross-section.\\
Viewing phase angle & $\alpha$ & angle formed between the incident light and light reflected to observer. \\
& & Gibbous phase: $0 ^\circ< \alpha < 90^\circ$, Crescent phase: $90 ^\circ< \alpha < 180^\circ$
\enddata
% \tablecomments{}
\end{deluxetable}

\section{Target selection}\label{sec:target selection}

We set out to find candidate target planets with the following characteristics:
\begin{enumerate}
    \item Host star brighter than the goal maximum magnitude for coronagraph operation, which we set at \textsl{Gaia} G mag $<$8; we also prioritized targets with G mag $<$5 for optimal coronagraph operation.
    \item Enters the detection region (within IWA/OWA and above goal contrast limit) for a significant fraction of the orbit
    \item Has well-characterized orbital elements. Planet must have semi-major axis ($a$), eccentricity ($e$), argument of periastron ($\omega$), and (minimum) mass constraints in literature.
    \item Planets with inclination ($i$) and longitude of nodes ($\Omega$) (requiring orbit constraints from multiple detection methods) will be prioritized as this constrains mass, separation, and sky-plane position angle. Planets with 3D velocity information (which removes node degeneracy) are of top priority as this also constrains viewing phase as a function of time.
\end{enumerate}

\begin{figure*}
\centering
\includegraphics[width=0.85\textwidth]{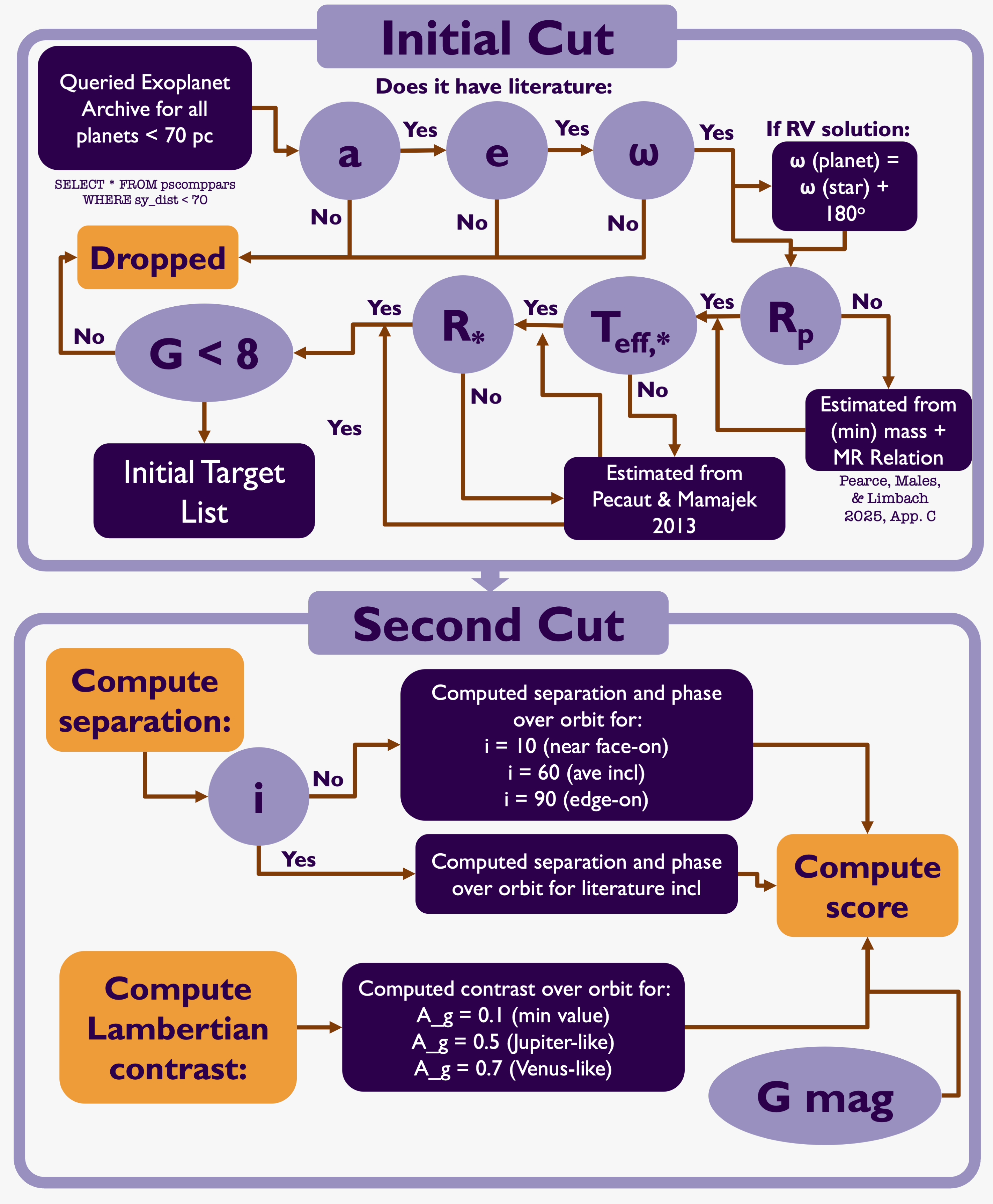}
\caption{\small{Illustration of the target list decision process, described in Section \ref{sec:target selection}.
}}
\label{fig:decision-tree}
\end{figure*}

\begin{figure}
\centering
\includegraphics[width=0.49\textwidth]{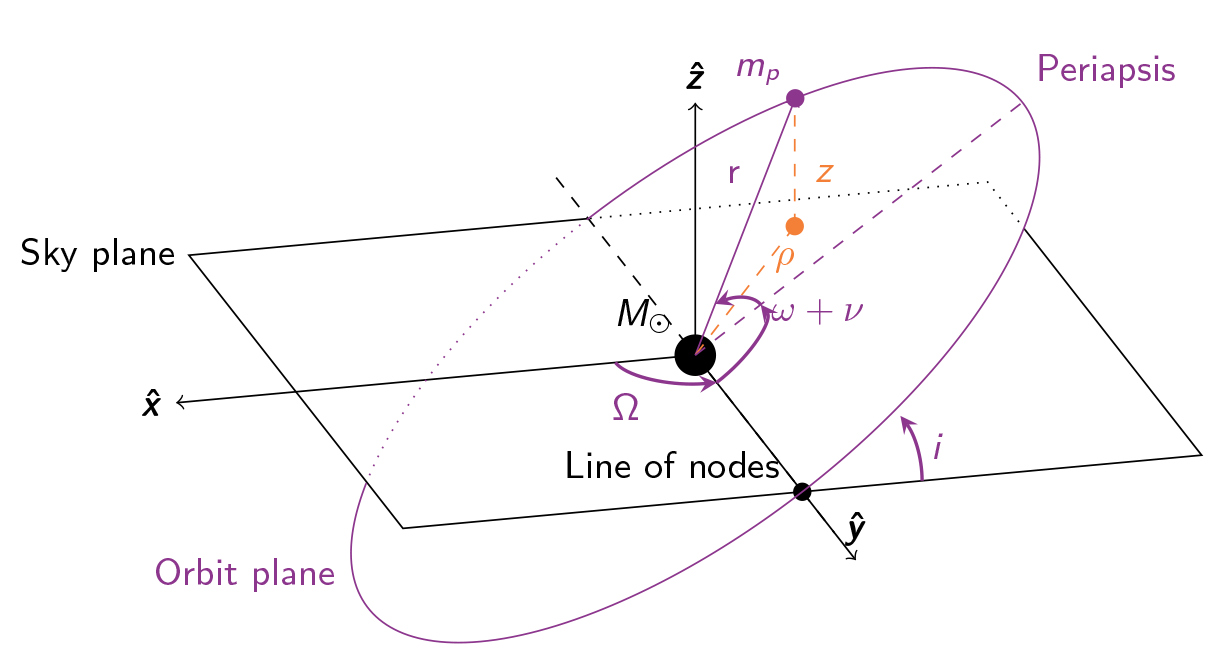}
\caption{\small{Diagram of the coordinate system and orbital elements used in this analysis. The $+\hat x$ and $+\hat y$ directions form the sky plane and correspond to the +Decl and +R.A directions respectively. The $+\hat z$ direction forms a right-handed coordinate system corresponding to towards the observer. (Note that this follows the direct imaging convention and opposite the typical RV convention of $+\hat z$ being away from the observer). M$_\odot$ and m$_p$ denote the host star and planet respectively, with $i, \omega$ and $\Omega$ describing the planet's angle with respect to the sky plane and reference direction of +Decl ($+\hat x$). $\nu$ is the true anomaly describing the angle in the orbit plane of the planet past periastron at a given time. The black dot notes the location of the ascending node, where the planet transitions from behind to in front of the sky plane. Orbital parameters are further described in Table \ref{tab:definitions}
}}
\label{fig:keplerianelements}
\end{figure}

Figure \ref{fig:decision-tree} illustrates our approach to identifying candidate targets, described in the following sections. In Table \ref{tab:definitions} we describe symbols and definitions used throughout this work. We used the coordinate system and calculations described in detail in \citet{Pearce2025projecc} and illustrated in Figure \ref{fig:keplerianelements}.

In this section we describe how we assessed candidate targets and arrived at the selected targets summarized in Table \ref{tab:inc-planets}.

\subsection{Initial cut}

To identify optimal known planets from which to select targets we queried the Exoplanet Archive\footnote{Accessed 2026 Sep 20} \citep{Christiansen2025ExoplanetArchive} for nearby exoplanets ($<$70 pc, arbitrarily selected) with literature constraints on semi-major axis (a), eccentricity (e), argument of periastron ($\omega$), period (P), and host star mass (M$_*$). We selected planets with host star \textsl{Gaia} G mag $<$5 (design) and G mag $<$8 (goal) for coronagraph operation. 

\begin{figure}
\centering
\includegraphics[width=0.49\textwidth]{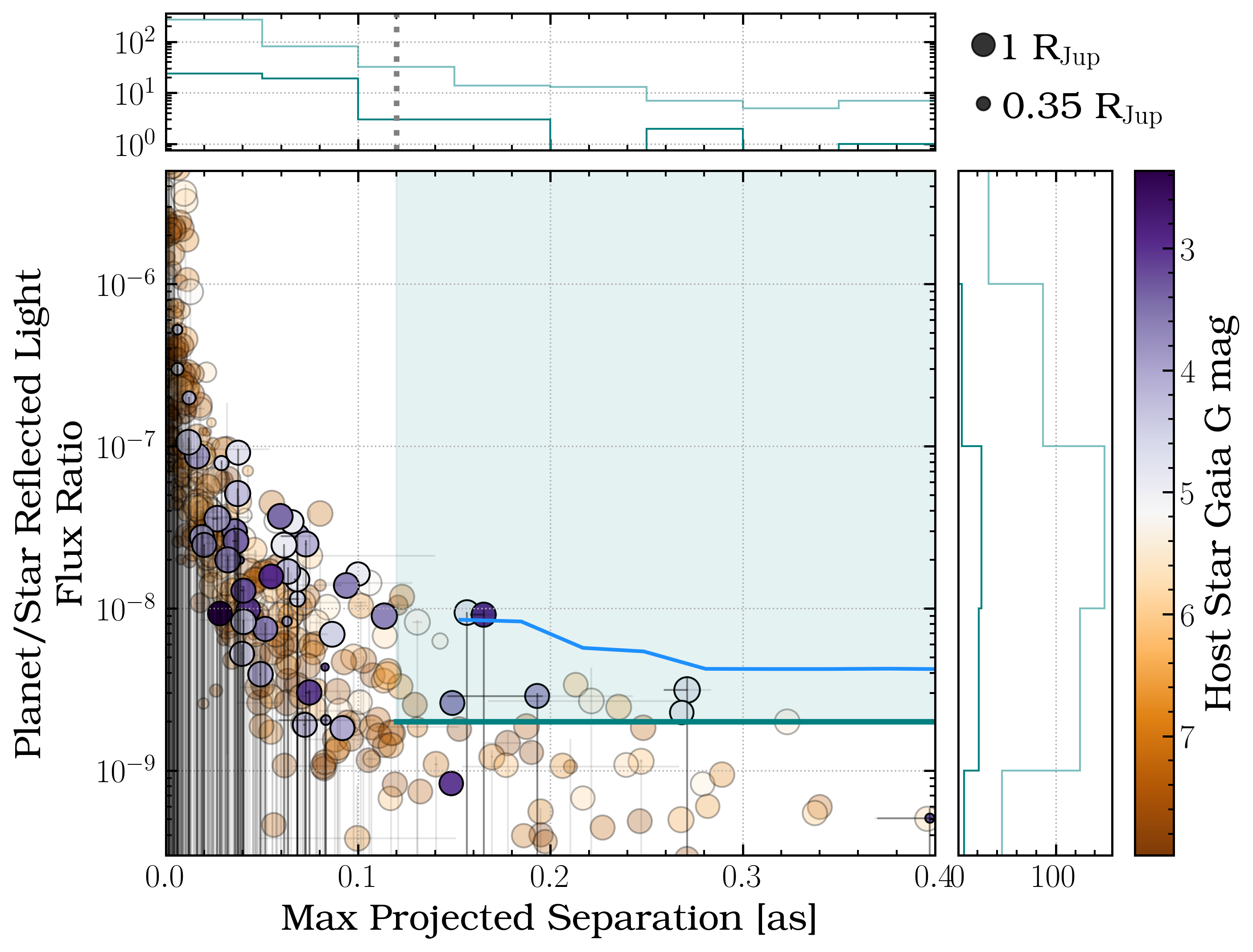}
\caption{\small{Candidate targets from Exoplanet Archive in maximum projected skyplane separation vs planet-star flux ratio in Lambertian contrast. Markers correspond to the maximum separation given the mean literature inclination constraint, or at $i=60^\circ$ if no inclination constraint, and at geometric albedo A$_g=0.5$, a Jupiter-like albedo. X-axis error bars correspond to max projected separation for $i=10^\circ$ and $i=90^\circ$ where no inclination constraint available. Y-axis error bars correspond to Lambertian contrast at A$_g=0.1$ and A$_g=0.7$ (Venus-like atmosphere). Markers are color coded by host star \textsl{Gaia} magnitude. Solid markers are for planets with host stars brighter than 5th magnitude; semi-transparent markers are for host stars brighter than 8th magnitude. Histograms show planets in bins of 0.05\arcsec\ and 1 dex of contrast. The teal shaded region shows the estimated inner and outer working angle (3$\lambda/D$ -- 20$\lambda/D$ at 0.63$\mu$m; 0.12 -- 0.86\arcsec) and the target contrast limit 2$\times10^{-9}$. The blue line marks the most optimistic \textsl{Roman} CGI contrast curve generated by \texttt{corgietc} \citep{corgietc} and retrieved from \citet{target_accessibility} at \url{https://github.com/roman-corgi/DI-flux-ratio-plot/}. \textbf{Takeaway:} Lazuli's detection region accesses planets slightly closer than \textsl{Roman} CGI given the larger primary mirror and potential to access shorter wavelength imaging; uncertainties on planet atmosphere and orbital elements move potential targets in and out of detection region.
}}
\label{fig:targets-plot}
\end{figure}

In Figure \ref{fig:targets-plot} we show our query results in separation vs. contrast ratio for host star G mag$<$5 (solid) and G mag$<$8 (semi-transparent). We computed planet-star contrast using the Lambertian phase function and
\begin{equation}\label{eqn: contrast}
    C(\alpha) = A_g \left(\frac{R_p}{r}\right)^2 \left[\frac{\sin\alpha + (\pi - \alpha)\cos\alpha}{\pi} \right]
\end{equation}
where $A_g$ is the geometric albedo, $R_p$ is the planet radius, $r$ is the planet-star separation in the orbit plane, and $\alpha$ is the viewing phase \citep{cahoyExoplanetAlbedoSpectra2010, cahoyErratumExoplanetAlbedo2017}.
We used the parametrization of \citet{Madhusudhan2012AnalyticModels} to compute phase angle:
\begin{equation}\label{eqn:alpha}
    \alpha = \cos^{-1} \left[ \sin(i) \sin(\nu + \omega_p)\right]
\end{equation}
where $i$ is inclination, $\nu$ is the true anomaly, and $\omega_p$ is the planet's argument of periastron\footnote{RV orbit solutions typically report the $\omega$ of the star, requiring a shift of 180$^\circ$ to convert to argument of periastron of the planet}. For planets without a radius estimate we used the (minimum) mass and the mass-radius relation of \citet{Pearce2025projecc} (Appendix C). Contrast is computed for the viewing phase at max projected separation.

In Figure \ref{fig:targets-plot} we show the planet's maximum projected separation in the sky plane. For planets without inclination constraints we computed separation for inclination values of $i=10^\circ$ (nearly face-on), $i=90^\circ$ (edge-on), and $i=60^\circ$ (average value for a uniform half-sphere). We plot the markers at the max projected separation for $i=60^\circ$ case, with error bars showing the $i=10^\circ$ and $i=90^\circ$ cases. On the y axis we plot contrast ratio assuming A$_g=0.5$, the geometric albedo for a Jupiter-like gas giant planet, with error bars for A$_g=0.7$ (a Venus-like atmosphere) and A$_g=0.1$ (most pessimistic case).

From this initial set of the candidates we selected the best candidate targets using the following method. Using the mean value of orbital parameters given in the Exoplanet Archive composite table (\texttt{pscompars}, \citealt{PSCompPars}, which may not be self-consistent as it may report parameters from different sources) we computed planet-star separation and contrast at 1000 points along a complete orbit (parameterized as mean anomaly from 0 to 2$\pi$). We computed contrast throughout an orbit for three geometric albedos: A$_g=0.1, 0.5, 0.7$, which spans a worst-case albedo, a Jupiter-like albedo, and a Venus-like albedo. For planets without a literature inclination estimate we computed separation for three inclination values: $i=10^\circ, 60^\circ, 90^\circ$, which spans nearly face-in, the average value for a uniform half-sphere, and edge-on. Given the added uncertainty if inclination is not constrained, for the remaining analysis we divided potential targets into two groups: those with and without inclination constraints. It should be noted here that many planets initially identified at having a literature inclination in \texttt{pscompars} table were found to be exactly 90$^\circ$ on scrutiny. In several RV orbit fits inclination was artificially fixed at 90$^\circ$ for the fit, and this was imported into the Exoplanet Archive as an inclination estimate. We removed all inclinations that were equal to exactly 90$^\circ$.

For each candidate target we computed the fraction of orbit within the detection region. For this initial pass we expanded the detection separation region to 0.1--0.95\arcsec\ in order to catch planets who's mean orbital parameters don't enter the IWA/OWA but for whom some fraction of the orbit might when orbital uncertainties are taken into account.

Appendix \ref{sec:appendix-scores} shows all planets in the Exoplanet Archive which spend any amount of their orbit in the detection region both for the design and goal OWAs.

\subsection{Second cut}

We applied a second cut to find the optimal targets to examine their suitability in more detail. We selected 27 top priority targets in four categories, tabulated in Table \ref{tab:inc-planets}, using the criteria described below.

\subsubsection{Planets with inclination constraint}\label{sec:second cut winc}

To quantify the suitability of a target we created a metric of the weighted average of the fraction of a planet's orbit that is above the goal contrast limit of 2$\times10^{-9}$ under various assumptions, within the separation range, and includes a penalty for guide star magnitude (to prioritize targets with brighter host stars). Our score is computed as:
\begin{equation*}\label{eqn:score}
    \rm{score} = \frac{w_1f_{\rho} + w_2 f_{C,A_g=0.1} + w_3 f_{C,A_g=0.5} + w_4\left(\frac{8.1}{G}\right)}{w_1 + w_2 + w_3 + w_4}
\end{equation*}
where f$_{\rho}$ is the fraction of orbit with projected separation within IWA/OWA (we assigned separate scores for the design and goal OWA), f$_{\rm{C},A_g=0.1}$ is the fraction of orbit above the contrast limit under the most pessimistic geometric albedo assumption of A$_g$ = 0.1, f$_{\rm{C},A_g=0.5}$ is the fraction of orbit above the contrast limit for a Jupiter-like geometric albedo A$_g$ = 0.5, G is the host star \textsl{Gaia} G magnitude, and w$_x$ is the weight assigned to each factor. We constructed the $\frac{8.1}{G}$ term to boost planets with bright host stars for best coronagraph performance, with G = 8.1 being the faintest host star in our initial selection. We chose w$_1$ = 1.5, w$_2$ = 1, w$_3$ = 1, and w$_4$ = 1, such that every metric is weighted equally with slight preference for planets that spend more time in the detection region.

Figure \ref{fig:inc-score-chart} shows the 49 planets with inclination constraints that fall within the detection region for a non-zero fraction of the mean orbit, sorted by score for the goal OWA of 20$\lambda/D$. We selected the nine highest scoring planets and two brightest host star planets with inclination constraints as optimal targets.

\subsubsection{Planets without inclination constraint}\label{sec:second cut woutinc}

While inclination and longitude of nodes estimates are strongly preferred due the mass and on-sky position angle constraints, expanding the target list to include planets with orbits characterized through RV measurements alone opens us to more planets with bright host stars and potential optimal targets if those parameters can be estimated. So we examine strong candidates here.

For planets without an inclination constraint we quantified suitability similar to Section \ref{sec:second cut winc} with additional terms:
\begin{equation*}\label{eqn:score2}
\begin{aligned}
    \rm{score} = {}&\left(\frac{1}{\sum w_x}\right) [w_1f_{\rho,i=10^\circ}+ w_2f_{\rho,i=60^\circ} + w_3f_{\rho,i=90^\circ} \\
    &+ w_4 f_{C,A_g=0.1,i=10^\circ} + w_5 f_{C,A_g=0.1,i=90^\circ} \\
    &+ w_6 f_{C,A_g=0.5,i=10^\circ}  + w_7 f_{C,A_g=0.5,i=90^\circ} \\
    & + w_8\left(\frac{8.1}{G}\right)]
\end{aligned}
\end{equation*}
where $\left(\frac{1}{\sum w_x}\right)$ is the normalization term of the sum of all the weights. We chose all weights = 1 except $w_1 = w_2 = w_3 = 1.5$ as before. Figure \ref{fig:woinc-score-chart} shows the planets with some fraction within detection region sorted by host star magnitude. Of these candidate targets, we identified $\gamma$~Cep~b and $\upsilon$~And~d as the two strongest given their combination of bright host star and time in the detection region. If inclination and longitude of nodes are constrained in the future, these will become the highest priority targets.

We selected five of the planets without inclination constraint with host star magnitude brighter than 5 for optimal coronagraph operation.

\subsubsection{Planets amenable to characterization in blue filters}\label{sec:second cut blue}

A strength of Lazuli ESC is its potential to characterize atmospheres in shorter (bluer) wavelengths ($0.3\mu\rm{m}<\lambda<0.5\mu\rm{m}$). This delivers some targets which score less favorably in the discovery filter $\lambda=0.63\mu$m but score higher at bluer wavelengths, so we elected to create a subset of targets specifically amenable to characterization in the blue. We repeated the above analysis for IWA/OWA using $\lambda=0.4\mu$m, making IWA~=~$3\lambda/D = 0.08$\arcsec\ and OWA~=~$20\lambda/D = 0.55$\arcsec, shown in Figure \ref{fig:blue-score-chart}

We selected eleven planets that score higher when the detection region is shifted to 3--20$\lambda/D$ when $\lambda = 0.4\mu$m.

\begin{deluxetable*}{lccccccc}\label{tab:inc-planets}
\tablewidth{10pt}
\tablecaption{Orbital Elements for prioritized targets}
\tablehead{
\colhead{Planet} & \colhead{$a$ [au]} & \colhead{$P$ [yr]} & \colhead{$e$} & \colhead{$i$ [deg]$^\dagger$} & \colhead{$\omega$ [deg]} & \colhead{$\Omega$ [deg]$^\dagger$} & \colhead{T$_0^*$ [JD]} 
}
\startdata
\multicolumn{8}{c}{Highest score}\\
\hline
14 Her b & 2.839$\pmerr{0.039}{0.041}$& 4.8362$\pmerr{0.0046}{0/0051}$ & 0.3683$\pm$0.0029 & 32.66$\pmerr{7.448}{4.011}$ & 202.68$\pmerr{0.45}{0.46}$ & 44.69$\pmerr{3.438}{4.011}$ & 2458853.207$\pmerr{0.012}{0.011}$  \\
GJ 832 b & 3.7$\pm$0.1 & 9.88$\pmerr{0.34}{0.33}$ & 0.05$\pm$0.03 & 51$\pm$3$^\dagger$ &  207$\pmerr{22}{31}$ & 61$\pmerr{16\dagger}{17}$ & 2456696$\pmerr{454}{338}$\\
HD 114783 c & 5.0$\pm$0.1 & 11.92$\pmerr{0.24}{0.21}$ & 0.05$\pmerr{0.04}{0.03}$ &21$\pmerr{7\dagger}{4}$ & 153$\pmerr{46}{48}$ & 307$\pmerr{14\dagger}{11}$ & 2455618$\pmerr{583}{555}$\\
$\beta$ Pic b & 10.07$\pm$0.03 & 23.770$\pmerr{0.151}{0.158}$ & 0.105$\pmerr{0.004}{0.003}$ & 88.993$\pmerr{0.010}{0.011}$ & 23.7$\pmerr{3.1}{2.6}$ & 211.787$\pmerr{0.010}{0.008}$ & 2447991$\pm$14 \\
HD 190360 b & 3.965$\pm{0.047}$ & 7.9055$\pmerr{0.0096}{0.0099}$& 0.3342$\pmerr{0.0061}{0.0062}$ & 71$\pmerr{42}{17}$ & 15.0$\pm1.2$ & 108$\pmerr{60}{22}$ & 2456423.6$\pmerr{8.6}{8.9}$ \\
$\pi$ Men b & 3.24$\pm$0.03 & 5.7194$\pm{0.0011}$ & 0.637$\pm{0.002}$ & 51.1$\pmerr{3.0^\dagger}{2.8}$ &330.1$\pm$0.3 & 282.5$\pmerr{4.3^\dagger}{4.2}$ &  2458389$\pm$3 \\
HD 154345 b & 4.158$\pmerr{0.066}{0.067}$ & 8.980$\pmerr{0.079}{0.077}$ & 0.058$\pm$0.019 & 88$\pm$20 & 309$\pmerr{18}{21}$ & 68$\pmerr{26}{31}$ & 2458230$\pmerr{147}{263}$ \\
HD 87883 b & 4.055$\pmerr{0.147}{0.161}$ & 9.060$\pmerr{0.075}{0.083}$ & 0.707$\pm$0.007 & 25.446$\pmerr{1.612}{1.053}$ & 289.351$\pm$1.286 & 99.951$\pmerr{4.233}{3.299}$ & 2450345$\pmerr{62}{56}$ \\
HD 100508 b & 6.11$\pmerr{0.25}{0.28}$ & 5681$\pm$42 & 0.420$\pm$0.02 & 62$\pmerr{19\dagger}{16}$ & 99.1$\pmerr{4.1}{4.2}$ & 82$\pmerr{254}{58}$ & 2447556$\pmerr{87}{84}$  \\
\hline
\multicolumn{8}{c}{Guide star $<$ 5}\\
\hline
$\epsilon$ Eri b & 3.52$\pm0.03$ & 7.34$\pm$0.08 & 0.04$\pmerr{0.04}{0.03}$ & 43$\pm4$$^\dagger$ & 202$\pmerr{122}{114}$ & -165$\pmerr{4}{5}$ & 2458017$\pmerr{1740}{447}$ \\
$\beta$ Pic c & 2.70$\pm$0.03 & 3.29$\pmerr{0.07}{0.03}$& 0.307$\pmerr{0.019}{0.025}$ & 88.93$\pmerr{0.11}{0.09}$ & -114.9$\pmerr{2.6}{2.4}$ & 210.08$\pmerr{0.05}{0.04}$ & 2458663.4$\pmerr{9.9}{6.6}$ \\
\hline
\multicolumn{8}{c}{No incl constraint (M$_p$ = M$_p \sin i$)}\\
\hline
$\gamma$ Cep b & 2.05$\pm$0.06 & 2.473$\pm$0.004 & 0.049$\pm$0.034 & -- & 94.6$\pm$34.6 & -- & 2453227$\pm$87 \\
$\upsilon$ And d & 2.585$\pmerr{0.048}{0.05}$& 3.5882$\pm$0.0019 & 0.3139$\pmerr{0.0095}{0.0096}$ & -- & 72.0$\pm{1.8}$ & -- & 2455905.2$\pm$7.9$^{**}$ \\
47 UMa b & 2.104$\pmerr{0.039}{0.040}$& 2.9460$\pmerr{0.0041}{0.0025}$ & 0.030$\pm$0.008 & -- & 142$\pmerr{16}{15}$ &  -- & 2452409$\pm$47$^{**}$ \\
47 UMa c & 3.453$\pmerr{0.065}{0.067}$ & 6.20$\pm$0.04 & 0.247$\pmerr{0.086}{0.150}$ & -- & 236$\pmerr{22}{21}$ & -- & 2456240$\pm$470$^{**}$ \\
HD 160691 c & 5.52$\pmerr{0.17}{0.15}$ & 11.61$\pmerr{0.44}{0.30}$ & 0.039$\pm$0.011 &  -- & 292$\pmerr{21}{22}$ & -- & 2452070$\pm$1380$^{**}$ \\
\hline
\multicolumn{8}{c}{Blue subset}\\
\hline
HD 219134 h & 2.828$\pmerr{0.068}{0.071}$ & 5.722$\pm$0.038& 0.05$\pm$0.03 & -- & 330$\pmerr{13}{35}$ & -- & 2456222$\pm$376$^{**}$ \\
HD 160691 b & 1.570$\pmerr{0.029}{0.030}$ & 1.7657$\pm$00007 & 0.0505$\pmerr{0.0094}{0.0093}$ & -- & 203$\pmerr{9.7}{8.6}$ & -- & 2456995$\pm$28$^{**}$  \\
7 CMa c & 2.153$\pmerr{0.003}{0.08}$ & 2.727$\pmerr{0.004}{0.143}$ & 0.08$\pmerr{0.05}{0.04}$ & -- & 46.6$\pmerr{7.7}{40.2}$ & -- & 24526473$\pm$73  \\
7 CMa b & 1.758$\pmerr{0.024}{0.001}$ & 2.013$\pmerr{0.041}{0.003}$ & 0.06$\pm$0.03 & -- & 345.3$\pmerr{5.1}{70.8}$ & -- & 2452249$\pm$174  \\
HD 10647 b & 2.073$\pmerr{0.051}{0.053}$ & 2.713$\pmerr{0.010}{0.012}$ & 0.126$\pmerr{0.069}{0.071}$ & -- & 49$\pmerr{23}{28}$ & -- & 2456718$\pm$66$^{**}$  \\
HD 192310 c & 1.183$\pm$0.022 & 1.488$\pm$0.009 & 0.0530$\pmerr{0.0610}{0.0380}$ & -- & 10$\pmerr{100}{120}$ & -- & 2455224$\pm$146  \\
HD 216437 b & 2.57$\pm$0.04 & 3.721$\pm$0.003 & 0.31$\pm$0.01 & 36$\pmerr{8\dagger}{6}$ & 65$\pm$2 & 129$\pm$8$^\dagger$ & 2454667$\pm$8 \\
HD 164922 b & 2.16$\pm$0.03 & 3.305$\pmerr{0.011}{0.014}$ & 0.08$\pmerr{0.06}{0.05}$ & -- & 296$\pmerr{48}{274}$ & -- & 2457978$\pmerr{20}{26}$ \\
HD 70642 b & 3.295$\pm$0.021 & 5.751$\pmerr{0.038}{0.037}$ & 0.04$\pmerr{0.034}{0.027}$ & 29.9$\pmerr{2.6^\dagger}{2.4}$ & 258.0$\pmerr{51.0}{66.0}$ & 60.7$\pmerr{8.8\dagger}{8.6}$ & 2455918.0$\pmerr{358.0}{337.0}$ \\
$\psi^1$ Dra B b & 4.43$\pm$0.04 & 8.53$\pm$0.11 & 0.40$\pm$0.05 & -- & 244$\pm$9 & -- & 2449344$\pm$76 \\
HD 128311 c & 1.742$\pm$0.020 & 2.511$\pm$0.005 & 0.196$\pmerr{0.047}{0.057}$ & -- & 160$\pm$20 & -- & 2455350$\pm$50 \\
\enddata
\end{deluxetable*}

\begin{deluxetable*}{lcccccccc}\label{tab:inc-planets}
\addtocounter{table}{-1} 
\tablewidth{10pt}
\tablecaption{Orbital Elements for prioritized targets (cont.)}
\tablehead{
\colhead{Planet} & \colhead{M$_p$ [\Mjup]} & \colhead{M$_*$ [\Msun]} & \colhead{G mag} & \colhead{Ref.} & \colhead{Age [Gyr]} & \colhead{Age Ref.} & \colhead{Score} & \colhead{Score} \\
\colhead{} & \colhead{} & \colhead{} & \colhead{} & \colhead{} & \colhead{} & \colhead{} & \colhead{[Design OWA]} & \colhead{[Goal OWA]}
}
\startdata
\multicolumn{9}{c}{Highest score}\\
\hline
14 Her b  & 8.9$\pmerr{1.3}{1.7}$ & 0.97$\pm{0.04}$ & 6.4 & 1 & 4.6$\pmerr{3.8}{1.3}$ & 17 & 0.61 & 0.61 \\
GJ 832 b  & 0.99$\pmerr{0.09}{0.08}$ & 0.45$\pm$0.01 & 7.7 & 2 & 5.27$\pm$2.14 & 18 & 0.29 & 0.61\\
HD 114783 c & 1.9$\pmerr{0.5}{0.4}$ & 0.90$\pm$0.04 &  7.3 & 2 & 6.2$\pm$4.9 & 19 & 0.56 & 0.56 \\
$\beta$ Pic b & 8.7$\pmerr{0.8}{0.7}$ &  1.789$\pmerr{0.024}{0.027}$ & 3.8 & 5 &  0.593$\pm$0.332 & 20 & 0.43 & 0.55\\
HD 190360 b &  1.69$\pmerr{0.25}{0.15}$ & 0.996$\pmerr{0.036}{0.036}$ & 5.6 & 3 & 8.85$\pmerr{6.57}{4.05}$ & 21 &0.55 &  0.55\\
$\pi$ Men b &  12.4$\pm$0.3$^\dagger$ & 1.02$\pm$0.03 &  5.5 & 7 & 5.10$\pmerr{4.75}{1.51}$ & 21 &0.49 &  0.49\\
HD 154345 b &  1.186$\pmerr{0.095}{0.059}$ & 0.890$\pm$0.040 &  6.6 & 4 & 3.1$\pm$0.4 & 22 & 0.48 & 0.48 \\
HD 87883 b & 5.409$\pmerr{0.469}{0.635}$ & 0.810$\pm$0.091 & 7.3 & 6 & 9.1$\pm$3.5 & 22 & 0.48 & 0.48\\
HD 100508 b &  1.20$\pmerr{0.30}{-0.18}$ & 0.94$\pm$0.12 & 7.52 & 8 & 3.902$\pm$3.334 & 23 &0.47 & 0.47 \\
\hline
\multicolumn{9}{c}{Guide star $<$ 5}\\
\hline
$\epsilon$ Eri b  & 0.91$\pm$0.04 & 0.819$\pm0.02$  & 3.5 & 9 & 1.1$\pm$0.1 & 24 & 0.00 & 0.47\\
$\beta$ Pic c &  8.7$\pmerr{0.8}{0.7}$ & 1.789$\pmerr{0.024}{0.027}$ &  3.8 & 5 & 0.593$\pm$0.332 & 19 & 0.46 & 0.46 \\
\hline
\multicolumn{9}{c}{No incl constraint (M$_p$ = M$_p \sin i$)}\\
\hline
$\gamma$ Cep b & 1.85$\pm$0.16 & 1.4$\pm$0.12 &  2.9 & 11 & 6.6$\pmerr{2.6}{0.7}$ & 25 & 0.46 & 0.46\\
$\upsilon$ And d & 4.336$\pmerr{0.173}{0.170}$ & 1.29$\pm$0.04 &  4.0 & 12 & 3.80$\pm$1.00 & 25 & 0.44 & 0.44 \\
47 UMa c &  0.5160$\pm$0.0378 & 1.063$\pm$0.029 &  4.9 & 12 & 7.30$\pm$1.90 & 26 &  0.47 & 0.47\\
47 UMa b &  2.492$\pmerr{0.107}{0.104}$ & 1.063$\pm$0.029 &  4.9 & 12 & 7.30$\pm$1.90 & 26 & 0.45 & 0.45\\
HD 160691 c & 2.262$\pmerr{0.198}{0.151}$ &1.13$\pm$0.02 & 4.9 & 12 & 5.666$\pmerr{4.401}{6.963}$ & 27 & 0.44 & 0.44\\
\hline
\multicolumn{9}{c}{Blue subset}\\
\hline
HD 219134 h & 0.277$\pm{0.017}$ & 0.81$\pm$0.03 & 5.2 & 12 & 11.00$\pm$2.20 & 26 & 0.50 & 0.50 \\
HD 160691 b &  1.749$\pm$0.066 & 1.1$\pm$0.1 & 4.9 & 12 & 5.666$\pmerr{4.401}{6.963}$ & 27 & 0.49 & 0.49 \\
7 CMa c & 0.87$\pm$0.06 & 1.34$\pmerr{0.11}{0.12}$ & 3.7 & 13 & 4.90$\pm$1.00 & 28 & 0.49 & 0.49 \\
7 CMa b & 1.85$\pmerr{0.06}{0.04}$ & 1.34$\pmerr{0.11}{0.12}$ & 3.7 & 13 & 4.90$\pm$1.00 & 28 & 0.47 & 0.47 \\
HD 10647 b & 0.9880$\pmerr{0.0818}{0.0787}$ & 1.11$\pm$0.02 & 5.4 & 12 & 1.40$\pm$0.90 & 26 & 0.48 & 0.48\\
HD 192310 c & 0.045$\pm$0.0038 & 0.85$\pm$0.02 & 5.5 & 12 & 7.81$\pm$0.90 & 26 & 0.47 & 0.47\\
HD 216437 b & 4.2$\pm$0.7 & 1.22$\pm$0.06 & 5.89 & 2 & 4.75$\pm$1.06 & 26 & 0.44 & 0.44\\
HD 164922 b & 0.041$\pmerr{0.009}{0.006}$ & 0.926$\pmerr{0.029}{0.035}$ & 6.8 & 14 & 9.580$\pmerr{1.990}{1.550}$ & 14 & 0.46 & 0.46\\
HD 70642 b &3.9$\pmerr{0.29}{0.27}$ & 1.078$\pm$0.015 & 6.99 & 10 & 0.92$\pm$0.73 & 26 & 0.49 & 0.49\\
$\psi^1$ Dra B b & 1.53$\pm$0.10 & 1.19$\pm$0.07 & 5.6 & 15 & 3.30$\pm$1.00 & 26 & 0.43 & 0.43 \\
HD 128311 c & 3.25$\pm$0.12 & 0.83$\pm$0.03 & 7.2 & 16 & 1.0$\pm$0.5 & 29 & 0.43 & 0.43\\
\enddata
\tablecomments{$^\dagger$Where plane-of-sky direction is not constrained we selected the prograde solution; $^{*}$Orbit timing parameters are reported in a variety of ways. For each orbit solution we have computed the epoch of periastron passage from what was reported in the literature to standardize our report here. $^\ddagger$For RV-only constrained orbits, unless otherwise stated, we assume the reported $\omega$ is for the star and apply a 180$^\circ$ offset to convert to argument of periastron of the planet, which we report here and use in the analysis. $^{**}$Time of inferior conjunction (T$_c$) was reported, which we have converted to T$_0$ including propagation of uncertainty from T$_c$, $e$, $\omega$, and $P$. References: 1: \citet{BardalezGagluiffi2025-14Herc} 2: \citet{Philipot2023Multitechnique}; 
3: \citet{Giovinazzi2026NEIDEarthTwinIV}; 4: \citet{Giovinazzi2025NEIDEarthTwinII}; 5: \citet{Sutlieff2026betaPicd}; 6: \citet{Feng2022-3dSelection}; 7: \citet{Piccinini2026piMen}; 8: \citet{Wu2026DetectionOfFourColdJupiters}; 9:\citet{Llop-Sayson2026EpsEri} using the solution including JWST imaging; 10: \citet{Xioa2023MassesOfASample}; 11: \citet{Endl2011NewsFromGammaCep}; 12: \citet{Harada2025SPORES-HWOII}; 13: \citet{Luque2019PreciseRadialVelocities}; 14: \citet{Benatti2020HD164922}; 15: \citet{Endl2016TwoNewLongPeriod}; 16: \citet{Rosenthal2021CalLegSurvey}; 17: \citet{BardalezGagliuffi2021-14Her}; 18: \citet{Gaido2023TimeTable}; 19: \citet{Adibekyan2021StellarClustering}; 20: \citet{Maldonado2025IntermeidateMassStars}; 21: \citet{Harada2024SettingTheStage}; 22: \citet{Bonfanti2015RevisingAges}; 23: \citet{CostaSilvia2020ChemicalAbuncances}; 24: \citet{Sanghi2026epsEri}; 25: \citet{Schlegenhauf2024SearchForStellar}; 26: \citet{Winter2020StarClustering}; 27: \citet{Lehmann2025ChemodynamicEvolution}; 28: \citet{Bonfanti2016AgeConsistency}; 29: \citet{delaReza2023Searching}
}
\end{deluxetable*}

\subsection{Notes on specific systems}

55 Cancri d: \texttt{pscompars} lists $i = 89.73\pmerr{24.49}{24.54}$ for d from \citet{Nelson201455Cancri} however this is the inclination constraint for planet b (see Table 3); no inclination constraint exists for d in that work.

HD 181433 d: This planet scored highly from the \texttt{pscompars} summary table, however no single orbit solution exists with all required parameters. \citet{Standing2026DMPP} constrains M$\sin{i}$, $\omega$, T$_0$ but not $i$; \citet{An2025OrbitsMassesFor156Companions} provides dynamical mass and inclination constraint but does not give $\omega$ and T$_0$. The two precise and recent solutions (\citealt{Standing2026DMPP} from ESO/HARPS, RV only; \citealt{An2025OrbitsMassesFor156Companions} from RV with Keck/HIRES and ESO/HARPS and proper motion anomaly, PMa, from the \textsl{Hipparcos-Gaia} Catalog of Accelerations, HGCA, \citealt{Brandt2021_HGCA_ERD3}) disagree on $a$, $e$, and period by more than 1$\sigma$. Given the significant discrepancies on important parameters we chose to exclude this planet from the current sample. Hopefully \textsl{Gaia} DR4 will refine and clarify the solution and we will add it onto the list if appropriate. It is one of the faintest host stars in our list so it is not as strong a candidate as others in the end.

$\beta$ Pic b: Using photometry and model fitting of effective temperature with AMES-DUSTY models \citep{Allard2001-AMES-Dusty}, \citet{Currie2013betaPicbAtm} estimated a radius of 1.4 -- 1.8~R$_\textrm{Jup}$, and \citet{Bonnefoy2014betaPicb} estimated a radius of 1.1 -- 1.9~R$_\textrm{Jup}$ with \texttt{Phoenix} \citep[see][]{Witte2011DustInBDs} and LESIA models \citep{Baudino2014-betaPicb-models, Baudino2015-BetaPicb-models}, both of which are significantly larger than a radius derived from our mass-radius relation and the mass constraint of \citet{Sutlieff2026betaPicd}, which gives 1.05$\pm$0.02~M$_\textrm{Jup}$. For our analysis presented here we use the Exoplanet Archive tabulated radius estimate of \citet{Currie2013betaPicbAtm} of 1.65$\pm$0.06~R$_\textrm{Jup}$; a smaller radius may make the planet too faint for Lazuli detection. The radius should be reexamined before targeting this planet with Lazuli.

$\epsilon$ Eri b: Our analysis assumed a geometric albedo from 0.1--0.7, with a Jupiter-like 0.5 assumed to be the most likely, however \citet{Llop-Sayson2026EpsEri} recently found that models of cloudy atmospheres with disequilibrium chemistry had higher geometric albedoes closer to 0.6--0.7. Given that recent JWST results \citep[e.g.][]{Balmer2025LivingOnTheWedge, Balmer2026DirectImagesOf, BardalezGagluiffi2025-14Herc, Balmer202514Herc, Xuan2026JWST-AFLepb, Xuan2024Gl229, Fu2026JWSTColorSequence, Hemmann2026GiantPlanetsNotFullyMixed} have found cold directly imaged planets and brown dwarfs show clouds and stratification, the true geometric albedo of $\epsilon$ Eri b is likely higher than we have assumed, making our contrast estimate pessimistic. This is not likely to affect our conclusion in the end as the majority of orbit realizations are below the contrast limit even at higher albedoes.

$\gamma$ Cep b: \citet{Benedict2018MassGammaCepb} constrained an inclination, longitude of nodes, and mass for $\gamma$ Cep b from relative astrometry using the \textsl{Hubble} Space Telescope (HST) Fine Guidance Sensor (FGS), however they do not quote the full orbit solution with all parameters and uncertainties from their fit. We elected to evaluate the orbit on the posteriors from a full RV orbit fit of \citet{Endl2011NewsFromGammaCep}. Given the brightness of this host star this is a top candidate for ESC observations and a full orbit fit to all available data (RV+astrometry) should be a top priority.

HD 160691 b \& c: As with $\gamma$ Cep b, \citet{Benedict2022muArae} measured an inclination with HST FGS but did not report complete orbital elements and uncertainties.

$\upsilon$ And d \& HD 128311 c: \citet{McArthur2010UpsAndHSTFGS} ($\upsilon$ And d) and \citet{McArthur2014} (HD 128311 c) report complete orbits fit from RV + HST FGS, however the fitting method employed may introduce bias in inclination \citep{Pourbaix2001Hip-Planet-Masses, Benedict2022muArae, Benedict2023-14Her}, so we elected to wait on using these inclination estimates until they can be independently verified through another method.

HD~147513~b: Discovered via RV in \citet{Mayor2004} and originally identified as a promising target (score = 0.44, host star 5.2 mag), it is listed as controversial in the Exoplanet Archive, so we removed it from our analysis.

HD 216437 b: \citet{Philipot2023Multitechnique} derives an orbit from 100 HARPS RV, 31 CORALIE RV, 39 UCLES RV, and a custom code \citep{Philipot2023UpdatedCharacterization} based on \texttt{htof} \citep{Brandt2021HOTF} to fit Hipparcos-Gaia accelerations; \citet{Xioa2023MassesOfASample} uses orvara \citep{Brandt2021orvara} and \texttt{htof} to fit 33 HARPS RV, 21 AAT RV and Hipparcos IAD and synthetic Gaia data. Both solutions agree within uncertainties on all parameters. We arbitrarily chose the \citet{Philipot2023Multitechnique} solution for our analysis.

7 CMa b \& c: \citet{Luque2019PreciseRadialVelocities} reports the periastron passage as the mean anomaly since the reference date of JD 2451808.021, which we converted to T$_0$. However their posterior summary statistics are highly non-Gaussian, and without the full posterior it is difficult to propagate uncertainties accurately. In the table we report the posteriors of \citet{Luque2019PreciseRadialVelocities}, and to compute T$_0$ we elected to use a Gaussian centered their ``best fit" value with standard deviation equal to the largest of the two uncertainties for period and mean anomaly. Our tabulated T$_0$ value therefore has overly inflated uncertainties and we direct readers to their article for more accurate timing parameters. 

\begin{figure*}[t!]
\centering
\includegraphics[width=0.9\textwidth]{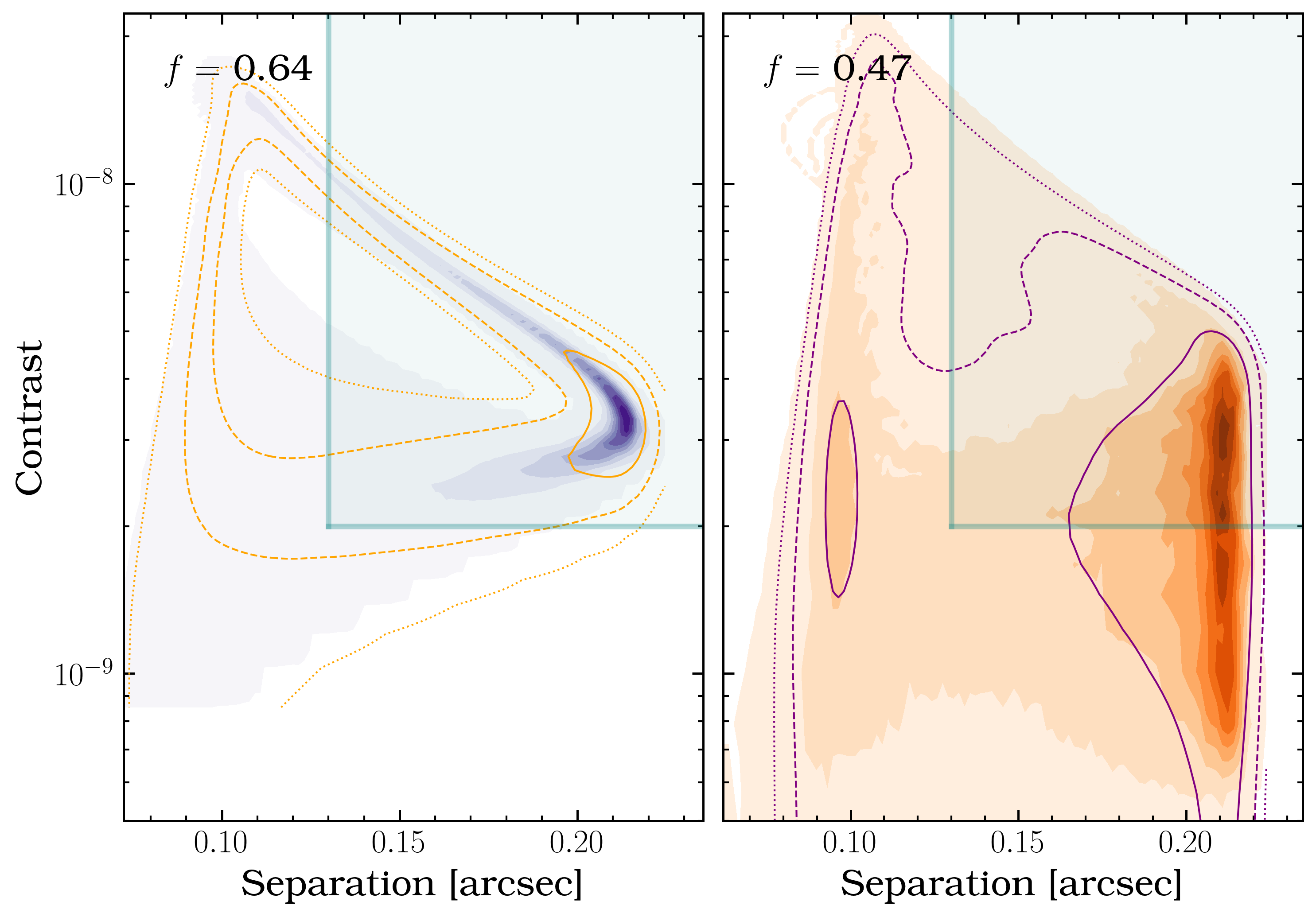}\\
\includegraphics[width=0.9\textwidth]{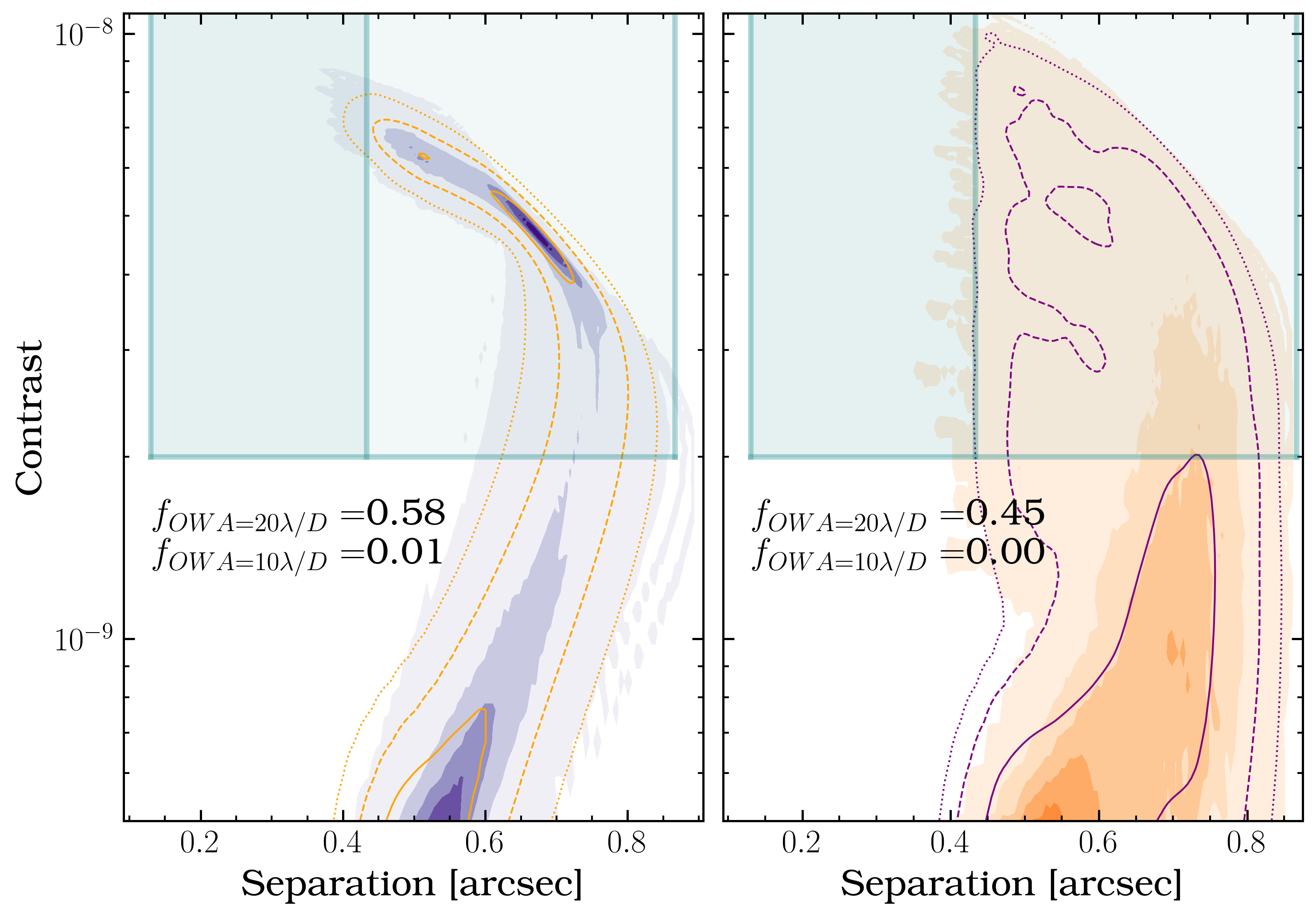}\\
\caption{Separation vs planet-star Lambertian contrast for the top two candidate target planets, 14 Her b and GJ 832 b, for A$_g = 0.5$ (left, purple) and A$_g = 0.1 -0.7$ (right, orange) computed from orbital elements tabulated in Table \ref{tab:inc-planets}. Colormap traces density of orbit realizations described in Section \ref{sec:orbit-sim}. Orange (left) and purple (right) contours trace regions containing 68\% (solid), 95\% (dashed), and 99.7\% (dotted) of orbit realizations. The top of each axis shows the fraction $f$ of points within the detection region, marked by the teal box. For 14 Her b the fraction is the same for the design (10$\lambda/D$) and goal (20$\lambda/D$) OWAs, but GJ~832~b is undetectable at the design OWA of 10$\lambda/D$ }
\label{fig:inc-sep-cont}
\end{figure*}

\begin{figure*}[t!]
\centering
\includegraphics[width=0.9\textwidth]{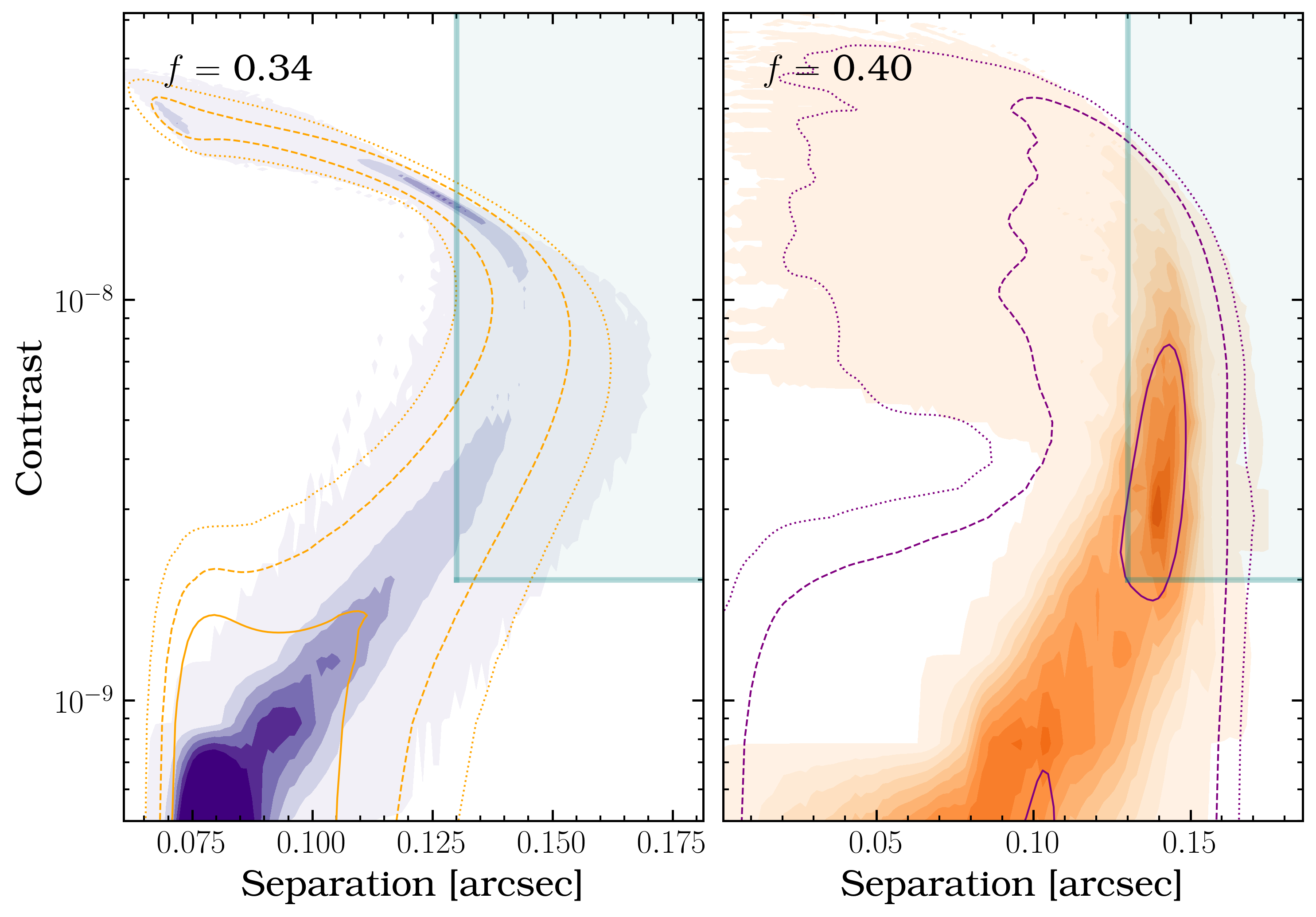}\\
\includegraphics[width=0.9\textwidth]{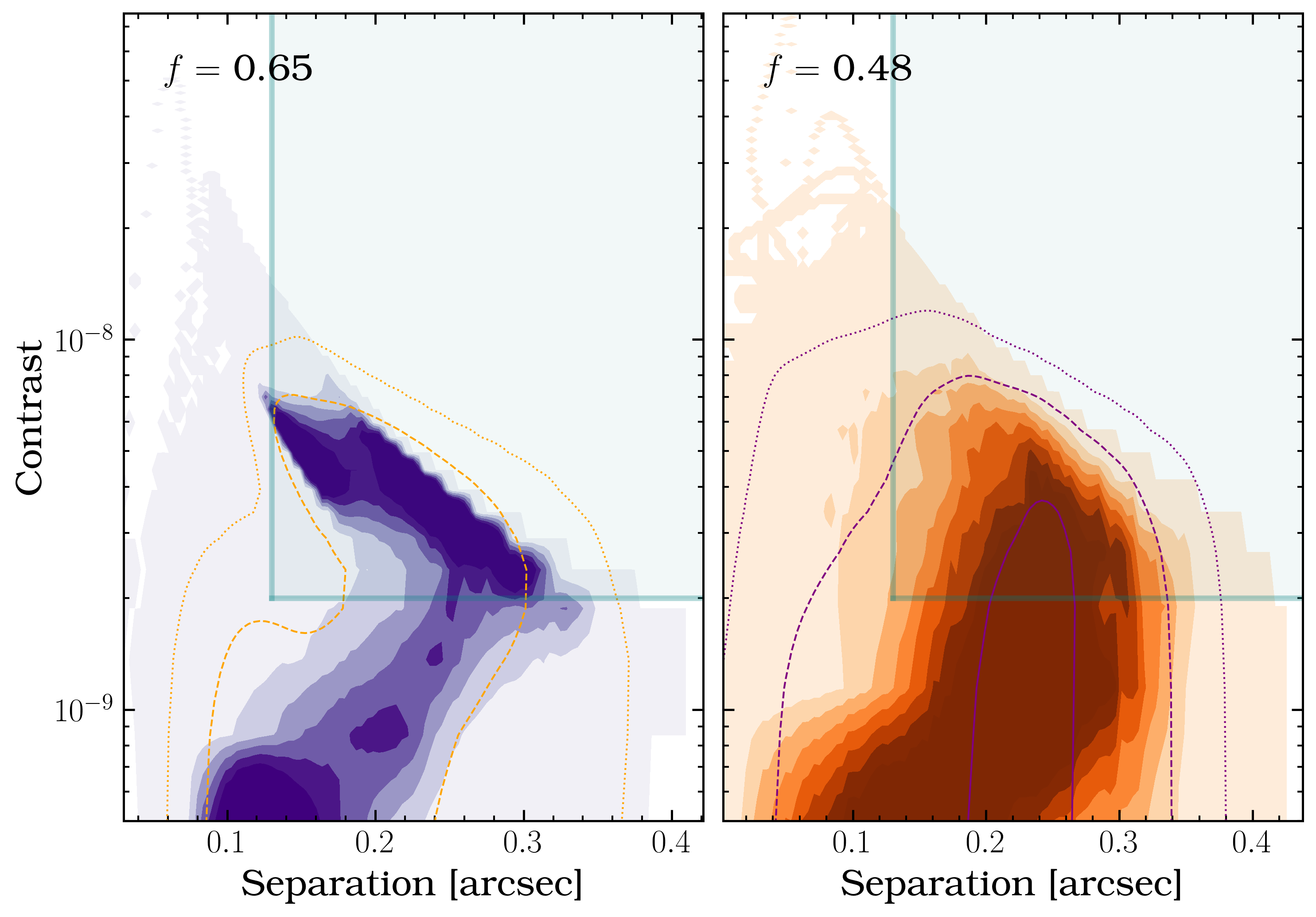}\\
\caption{Same as Figure \ref{fig:inc-sep-cont} for the two top priority candidate targets without inclination constraints: $\gamma$ Cep b (top) and 47 UMa c (bottom). Left: Orbit simulation for A$_g=0.5$ and $i = 60^\circ$. Orbit simulation for A$_g=0.1 - 0.7$ and inclination drawn from $\cos i$ on uniform distribution on [-1.0,1.0]. \textbf{Takeaway:} Both targets have a high percentage of orbit within the detection region, even with all unknowns folded in. However the lack of $i$ and $\Omega$ would complicate observations and interpretation so these are not the top priority targets without those constraints.}
\label{fig:noinc-sep-cont}
\end{figure*}

\subsection{Orbit simulations}\label{sec:orbit-sim}

We further scrutinized the published orbits of the selected targets to estimate the fraction of an orbit within the detection region given orbital parameter and albedo uncertainties. Table \ref{tab:inc-planets} tabulates the most recent self-consistent complete orbit solutions for each selected planet. From these parameters we performed a Monte Carlo simulation to compute separation and contrast through one orbital period. We generated an array of 10,000 trial orbits drawn from a Gaussian distribution described by the mean and (largest) uncertainty of each orbital parameter and computed the sky plane planet-star separation at 500 points along the orbit period. For planets without inclination constraints we performed this simulation for two inclination conditions: 1) $i=60^\circ$ (the average inclination for a uniform half-sphere) and 2) for 10,000 values drawn from a uniform distribution of $\cos i$ on [0, 0.98] ($i = 10^\circ - 90^\circ$). For all planets we computed contrast at each point in the orbit for two geometric albedo assumptions: 1) A$_g = 0.5$ (Jupiter-like) and 2) A$_g$ drawn from a uniform distribution on [0.1, 0.7] spanning a worst case albedo scenario to a best case, which is Venus-like.

Figure \ref{fig:inc-sep-cont} shows the results for the top two candidates: 14~Her~c and GJ~832~b. The left (purple) column shows the orbit for A$_g = 0.5$ and the right (orange) for A$_g = 0.1-0.7$, with the detection region marked in teal and 1-, 2-, and 3$\sigma$ contours in solid, dashed, and dotted lines respectively. In each axes we show the fraction of simulated points within the detection region. We see that both targets have around 50\% of their simulated points in the detection region. However GJ~832~b will not be detectable under the design OWA of 10$\lambda/D$. Figure \ref{fig:inc-planets-orbits} shows all planets in Table \ref{tab:inc-planets}.

Figure \ref{fig:noinc-sep-cont} shows the same for the planets without inclination estimates with the two brightest host stars: $\gamma$ Cep b and $\upsilon$ And d, with left (purple) showing results for A$_g = 0.5, i = 60^\circ$ and left (orange) showing A$_g = 0.1-0.7, i = 10^\circ - 90^\circ$. Both have high fractions of the orbit in the detection region making them top targets if inclination can be estimated in the future.

Figure \ref{fig:noinc-planets-orbits} shows the orbit simulation results for all planets without inclination constraints in Table \ref{tab:inc-planets}, and Figure \ref{fig:blue-planets-orbits} shows the same for planets in the blue wavelength subset.

\section{Discussion}\label{sec:discussion}

Here we examine the properties of the stars and planets that makeup the target list and what a survey that observed all of these targets could provide.

\subsection{Characteristics of the targets}\label{sec:properties}

In Figures \ref{fig:star-cmd-radec} and \ref{fig:system-architecture} we explored trends among the selected target planets and here we discuss science enabled by our selected targets.

\subsubsection{System architectures}
 Most host stars are AFGK type, with only one (GJ~832) being an M dwarf, likely due to the bright host star requirement for coronagraph operation. The targets span all regions of the night sky. Almost all have (minimum) mass larger than Neptune and are (most likely) less massive than 13~\Mjup\ (one often-used boundary separating giant planets from brown dwarfs), making them all gas giant planets. HD~192310~c has a minimum mass $\sim$0.8 Neptune masses, making it a possible sub-Neptune.

\begin{figure}[t!]
\centering
\includegraphics[width=0.48\textwidth]{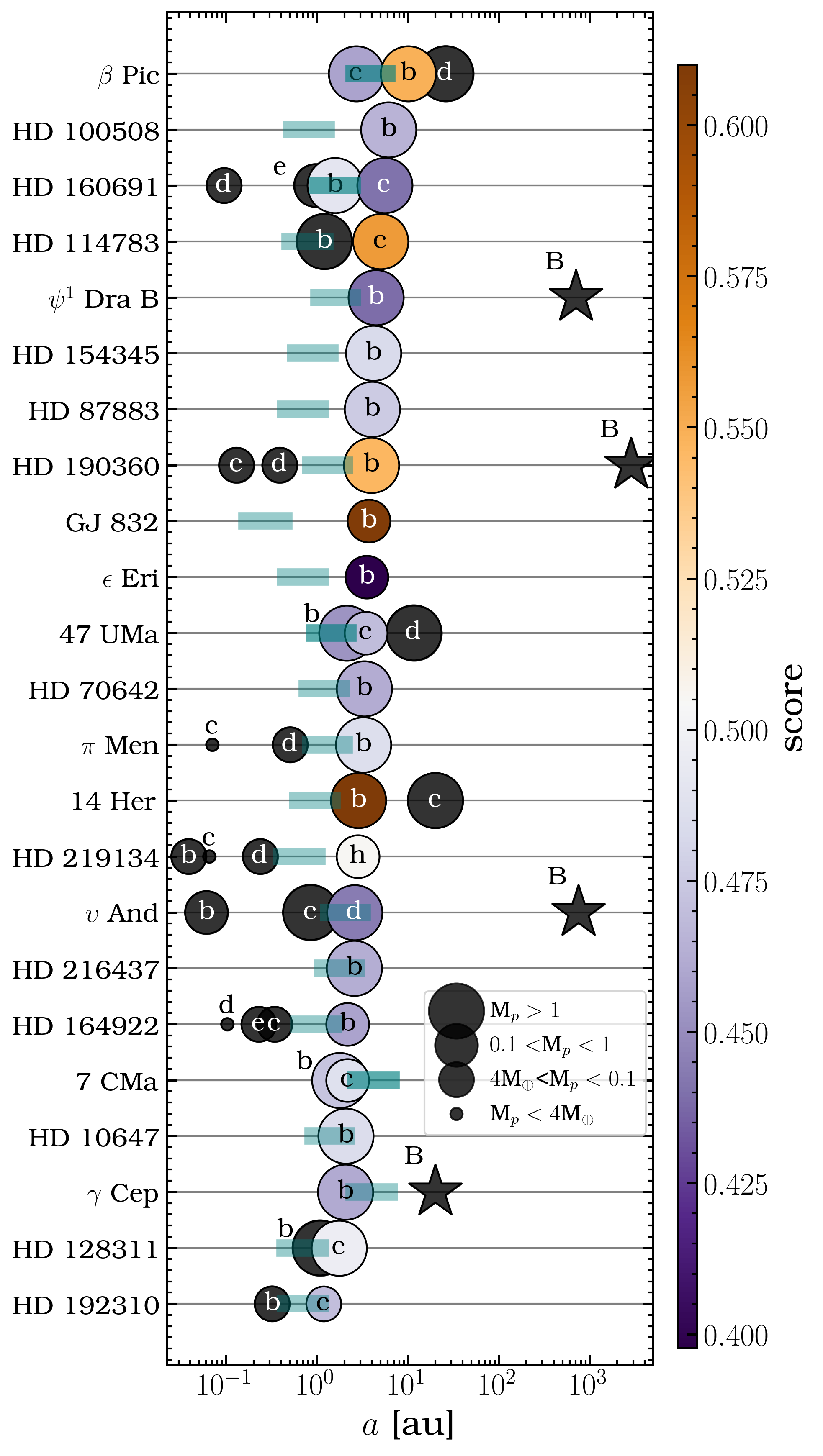}
\caption{Architectures of target systems. Selected targets are color coded by score, additional known bodies in the system are colored black, with semi-major axis on the x-axis. Planet-mass objects (M$_p<13$\Mjup) are shown by circles, stellar companions are marked with stars. Planet marker size corresponds to (minimum) mass. The optimistic habitable zone \citep[][Eqn 4 \& 5]{Kopparapu2014-HZ} for each host star is shown in teal. $\beta$ Pic b \& c, HD~160691~b, 47~UMa~b, $\pi$~Men~b, $\upsilon$~And~d, HD~216437~b, 7~CMa~b \& c, HD~10647~b $\gamma$~Cep b, and HD~192310~c are in their star's HZ, while HD~114783~c, HD~164922~b, HD~128311~c, and HD~192310~c have interior smaller planets in the HZ. }
\label{fig:system-architecture}
\end{figure}

Figure \ref{fig:system-architecture} shows the system architecture for our selected targets, with all known companions in each system and the habitable zone (HZ) for each star (from the definition of \citealt{Kopparapu2014-HZ}). Most targets have additional known companions in the system. Several targets are in their star's HZ, and while they are gas giants and unlikely astrobiology candidates, this is a unique population not seen in the solar system, and may host moons more favorable to astrobiology. Many also have interior smaller planets in the HZ. Outer giant planets play a key role in shaping the architecture of our Solar System \citep{KaibChambers2016Fragility}, and may be the key to delivering the right amount of water to the inner solar system for an Earth-like water content \citep{Bitsch2021DryOrWaterWorld}, making these interesting systems to understand from a habitability perspective. 

\subsubsection{Cold and evolved systems}

\begin{figure*}[t!]
\centering
\includegraphics[width=0.3\textwidth]{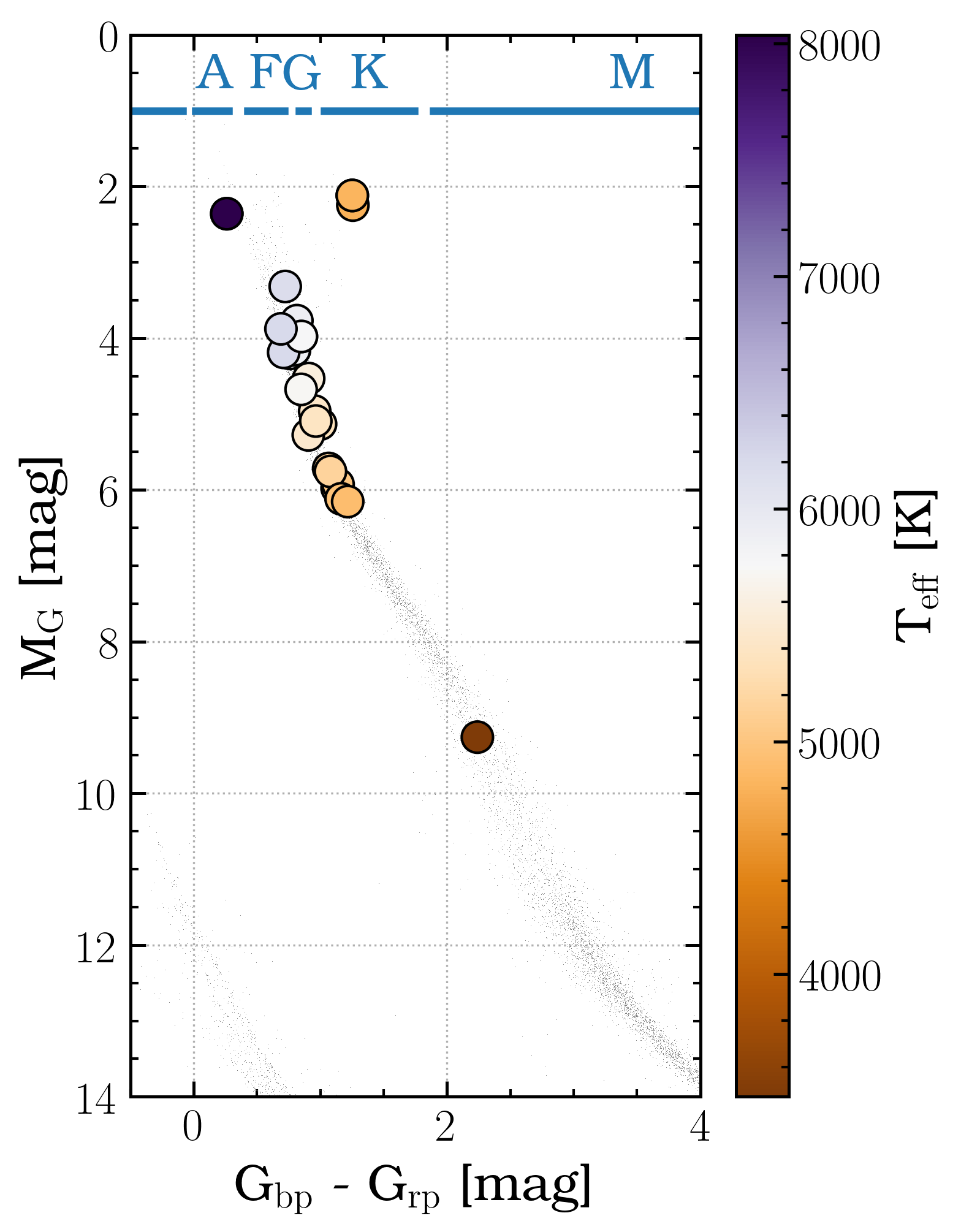}
\includegraphics[width=0.6\textwidth]{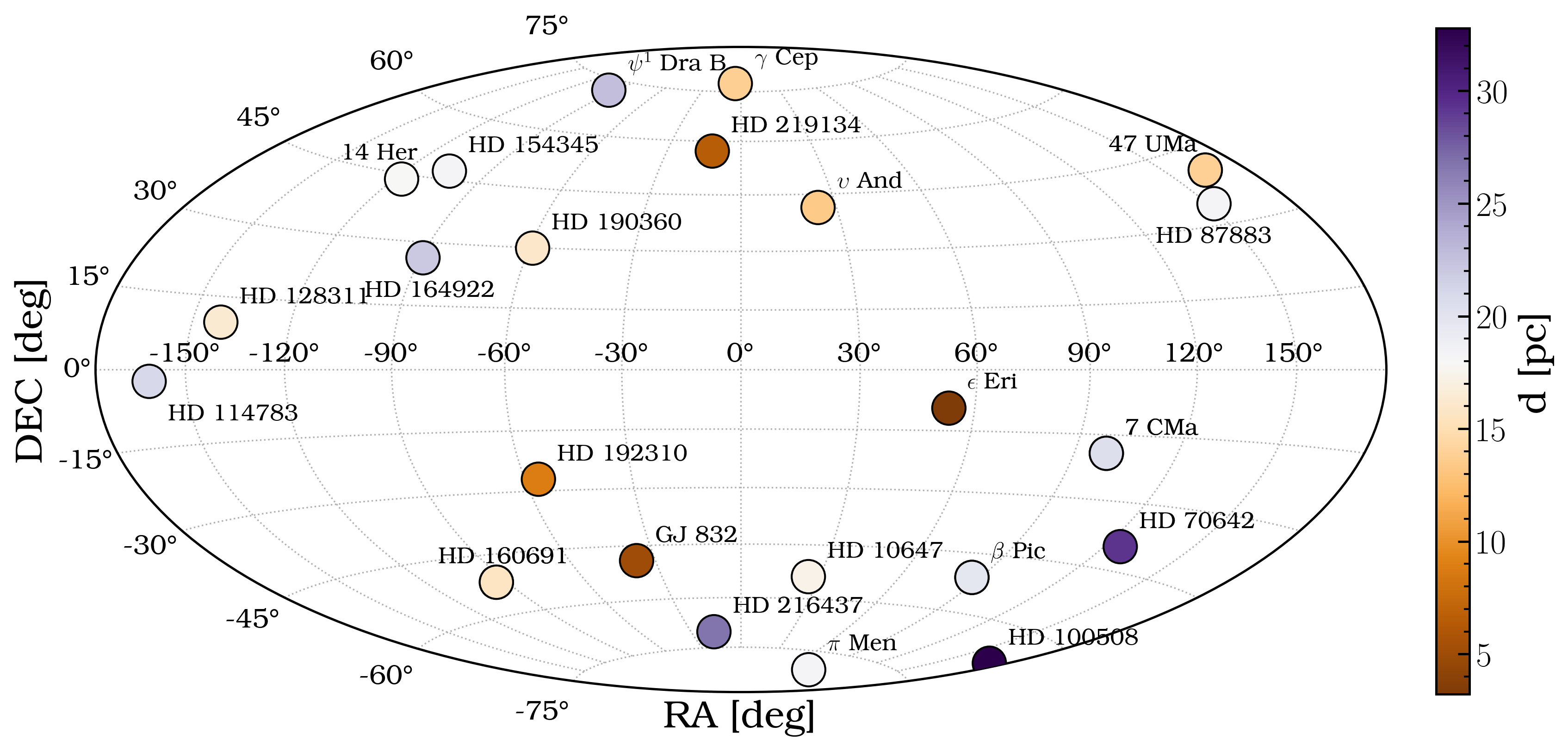}\\
\includegraphics[width=0.9\textwidth]{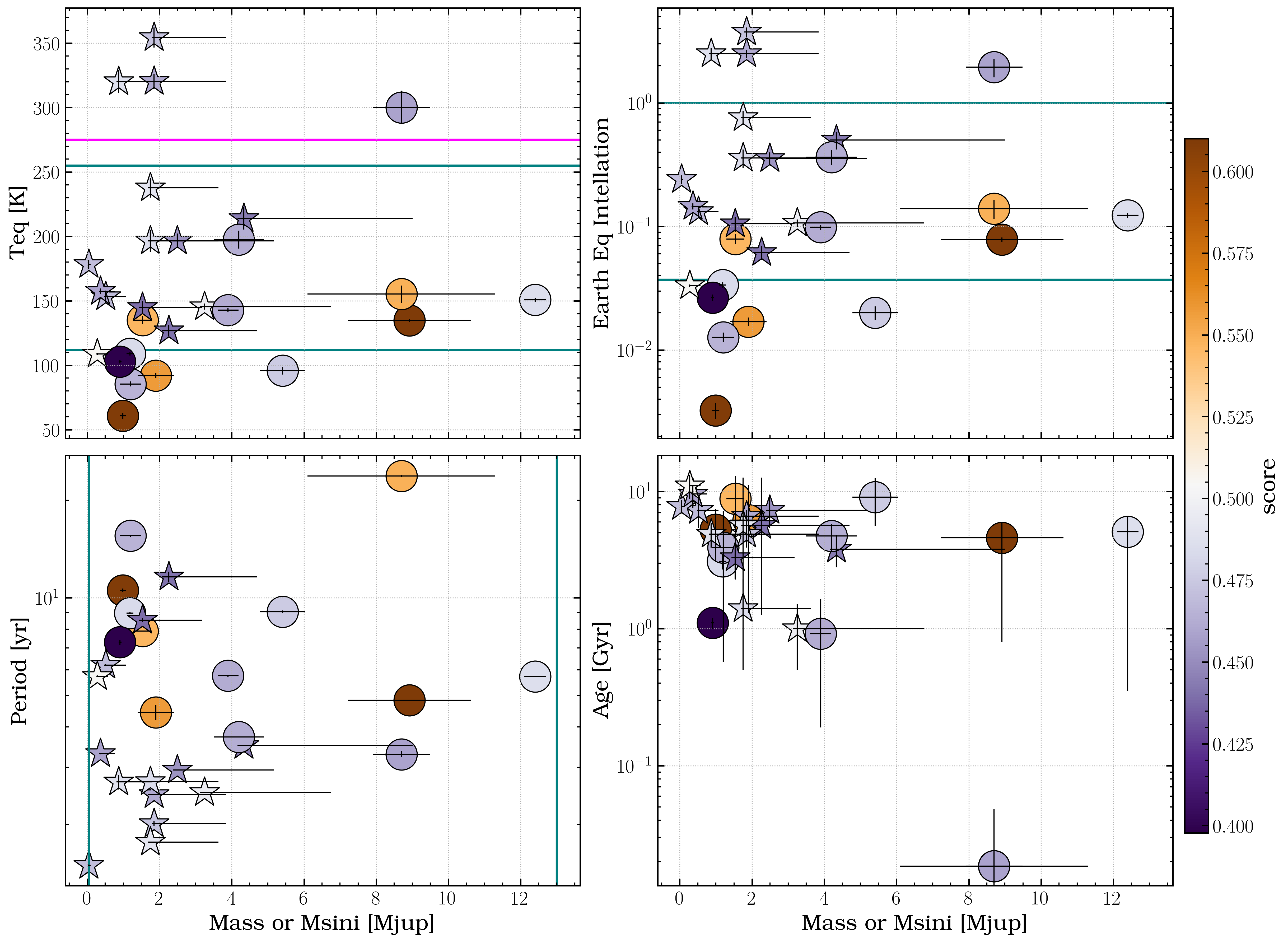}
\caption{Properties of the selected targets and their host stars. \textit{Top left}: \textsl{Gaia} color-magnitude diagram showing the effective temperature (colormap) and spectral type of host stars compared to a sample of \textsl{Gaia} sources \citep{Pearce2025SLS}. Most targets orbit AFGK stars, with one (GJ~832~b) orbiting an M dwarf star. \textit{Top right}: Host star R.A. and Decl. on an Aitoff projection color coded by distance. $\epsilon$ Eri is the closest system (3.2 pc) and HD~75302 is the furthest (30.3 pc). \textsl{Middle}: planet equilibrium temperature (T$_{\rm{eq}}$, left) and Earth-equivalent instellation flux (right) as a function of mass (circle markers) or M$\sin i$ (star markers), color coded by score. For planets without inclination estimates mass uncertainties span literature minimum mass uncertainty to maximum mass (M$\sin i / \sin(10^\circ)$). Earth and Jupiter values are marked with teal lines; the magenta line marks the coldest effective temperatures of planets directly imaged to date. Most, but not all, gas giant planet targets are cooler and receive less stellar flux than Earth but more than Jupiter. \textit{Bottom}: Orbital period (left) and star age (right) as a function of mass as above, color coded by score. On bottom left we show a vertical teal line at the mass of Neptune ($\sim$0.05\Mjup) and at 13~\Mjup to illustrate that our targets are all likely gas giants. Many systems have periods shorter than a few years, meaning the planet will likely move significantly during a long observation. Reflected light direct imaging is decoupled from age, and our target list spans systems at all stages of evolution, with $\beta$~Pic the youngest at $\sim$0.6~Gyr and HD~219134 the oldest at $\sim$11~Gyr.
}
\label{fig:star-cmd-radec}
\end{figure*}

Most selected targets have a cooler equilibrium temperature (where T$_{\rm{eq}} = $\Teff$\sqrt{R_*/a} \,[\frac{1}{4}(1-A_B)]^{1/4}$ with $A_B$ = Bond albedo = 0.3, Jupiter-like) and less instellation flux (where Earth equivalent instellation flux $= \left(\frac{R_*}{a}\right)^2 \left(\frac{\rm{T}_{\rm{eff}}}{5780 \rm{K}}\right)$) than Earth, but seven are also cooler with less instellation flux than Jupiter. The magenta line in Figure \ref{fig:star-cmd-radec} (middle, left) marks 275~K, approximate (effective) temperature of the coldest directly imaged planets to date: 14~Her~c (\Teff = 275--300~K, \citealt{BardalezGagluiffi2025-14Herc}), $\epsilon$~Indi~Ab (275--300~K, \citealt{Matthews2024epsIndiAb}). Four overlap that region ($\gamma$~Cep~b, 7 CMa b \& c, and $\beta$~Pic~c) while most are colder than what direct imaging via self-luminous thermal emission has reached to date (although the incredible power of JWST/MIRI and NIRCam for this science case promises colder planets in the near future, \citealt{Bowens-Rubin2025NIRCamYellsatCloud}). Figure \ref{fig:star-cmd-radec} (bottom, right) shows the targets as a function of literature age estimates of their host stars. Since reflected light flux is decoupled from age, our target list spans systems from $\sim$18~Gyr ($\beta$ Pic b \& c) to $\sim$11~Gyr (HD~219134~h), allowing direct imaging of systems much more evolved than previously available. Atmospheric characterization of all the targets on this list would provide a comprehensive survey of gas giant planet atmospheres in visible wavelengths spanning temperatures and ages above and below those in our solar system, and characterize planets cooler and older than what has been directly imaged via thermal emission to date.

\subsection{Difficulties in observing these targets}\label{sec:observation challenges}

We discuss a few challenges involved in carrying out these observations here.

\subsubsection{Inclination and Longitude of Nodes are essential for observation}

As discussed above, a lack of inclination constraint provides only a minimum mass estimate and adds to uncertainty in separation over time. Inclination constraint requires an astrometric measurement, either wobble in the plane of the sky, such as proper motion anomaly or imaging time series such as has been done with HST FGS \citep[e.g.][]{Benedict2022muArae}, or a relative position via direct imaging epoch \citep[e.g.][]{Matthews2024epsIndiAb}. 

This also provides a constraint on the longitude of nodes, which is necessary to predict the sky plane azimuthal position (position angle East of North) as a function of time. ESC's dark hole will not be azimuthally symmetric, requiring knowing the planet's position angle to plan efficient observing times and avoiding trying to observe it when it is not in the dark hole \citep{Pearce2025projecc}.

However even with this constraint it will not necessarily distinguish which node is which. Without knowing where the planet goes from in front of to behind the sky plane ($-\hat{z}$) and vice versa ($+\hat{z}$) we cannot know which is the ascending and which is descending node. This is usually not an important distinction for self-luminous direct imaging, however if you care about the viewing phase of the planet it becomes critical. Swapping which node is the ascending and descending node swaps which part of the orbit the planet is in a gibbous versus crescent phase, affecting the predicted contrast at observing time. In Table \ref{tab:inc-planets} we tabulated the prograde orbit where the node is unconstrained, but the retrograde orbit is also a valid solution. To break this degeneracy requires multiple direct imaging epochs, or measurement of velocity in three dimensions at a single epoch.

\subsubsection{Planets will move during long observations}

The reflected light detection regime opens direct imaging to planets on shorter periods than previous surveys. The shortest period in our target list is HD~192310~c at 1.5~yrs, and as ELTs and HWO begin to target small potentially habitable planets periods can drop to as short as 11.2 days in the case of Proxima Centauri b. Shorter periods coupled with longer observation times due to high contrasts (10's to 100's of hours) means direct imagers will have to worry more about planet motion than we are used to. In this analysis we have considered only the fraction of an orbit within the detection region and did not consider the amount of time that fraction encompasses (and also ignored sky-plane orientation). A more detailed analysis needs to consider the fraction of orbits that will leave the detection region during long integration times, either passing out of IWA/OWA or passing into phases with contrasts below the detection limit, termed Integration Time Adjusted Completeness in \citet{Keithly2021IntegrationTimeAdjustedCompleteness}. They found that traditional completeness metrics can overestimate yield by as much as 10\% when integration time is not considered.

Additionally long integration times and shorter periods gives the need to ``deorbit" images to build up enough signal-to-noise for a detection. \citet{malesDirectImagingHabitable2013} showed that planets can move significant fractions of a resolution element in a single observation, smearing out the planet's signal, which cannot be overcome by longer integration times. Shorter exposures (where the planet moves less than a resolution element) that are rotated to counter the orbital motion, dubbed Orbital Differential Imaging (ODI), can be used to obtain sufficient S/N for detection, at the cost of having to run a large number of trial orbits \citep[$\mathcal{O}\,10^8$][]{LeCoroller2020KStacker} and generating false alarms. Well constrained orbital parameters can alleviate some of both, as well as disentangle planet signal from speckle noise in a way similar to Angular Differential Imaging \citep{Males2015ODI}. Several orbit fitting and image reduction algorithms currently implement ODI: K-Stacker \citep{LeCoroller2015KStacker,Nowak2018KStacker, LeCoroller2020KStacker}, \texttt{octofitter} \citep{thompsonOctofitterFastFlexible2023}, and PACOME \citep{Dallant2023PACOME} use Keplerian orbit-informed image reduction and extract orbit parameter constraints. With a well-constrained orbit prior to observations, applying the principles of ODI to signal recovery, as well as further refinement of orbital parameters, will make Keplerian orbit-informed image stacking more efficient.

\subsubsection{Rings and moons confuse interpretation}

All giant planets in our Solar System have rings and moons, and in some cases those rings and moons contribute significantly to the flux when not resolved from the planet (S. Hopper et al., in press). It is reasonable to anticipate many (most?) giant exoplanets will also have rings and moons; the fact that none have been confirmed to date likely reflects observational challenges rather than an astrophysical absence. Some fraction will likely be bright enough to complicate interpretation of a reflected light detection once made. 

There already exists a phase-radius degeneracy in that a small planet at full phase can have the same reflected light flux as a large planet at a crescent phase \cite{nayak_atmospheric_2017}; knowing the radius or orbital parameters a priori can break this degeneracy, but rings add another complicating factor in that a large ringless planet and a small ringed planet may have the same observed flux at the same phase. A ringed and ringless planet can be distinguished through observation of the phase curve, because a ring produces two local maxima while a ringless planet produces one \citep{ArnoldSchneider2004, Lane2026RomanRings, Hasler2026RingDirectImaging} but this requires observing multiple epochs along the orbit. Rings also impart polarization \citep{LietzowWol2023RingPolarization, Veenstra2025RingPolarizationModel} and could be distinguished via polarimetery.

Additionally bright rings and moons complicate interpretation of photometry and spectroscopy. As the sources will be unresolved it is unclear if features originate from the moon/ring or planet (S. Hopper et al., in press) and eclipse and shadowing events can mimic weather \citep{Limbach2024ExomoonsRingHWO-I}. Spectral retrieval codes able to fit two bodies and identify where a two-body fit is preferred would be valuable. Moons could even be detected from shadowing and transit events \citep{Limbach2024ExomoonsRingHWO-I}. The presence of moons around a gas giant is a benefit to reflected light science return in that they are interesting in their own right and may be a promising astrobiology target.

\subsection{Synergies with CGI}\label{sec:roman}

The \textsl{Nancy Grace Roman} Space Telescope Coronagraph Instrument (CGI) will demonstrate space-based visible light coronagraphy for the first time \citep{Bailey2023RomanOverview} and is expected to deliver planet-star contrast ratios as high as 10$^{-9}$, enabling spatially-resolved reflected light exoplanet direct imaging for the first time. Wide field of view (0.3\arcsec--0.95\arcsec) imaging mode with the Shaped Pupil Coronagraph is planned to target $\epsilon$ Eri b in Band 1 (0.575$\mu$m), and the narrow field of view imaging mode (0.14\arcsec--0.45\arcsec) with the Hybrid Lyot Coronagraph are targeting $\upsilon$~And~d and 14~Her~b \citep{Wolff2026RomanCGI-CPP} for reflected light detection and orbit constraints. 

\begin{figure}[t!]
\centering
\includegraphics[width=0.48\textwidth]{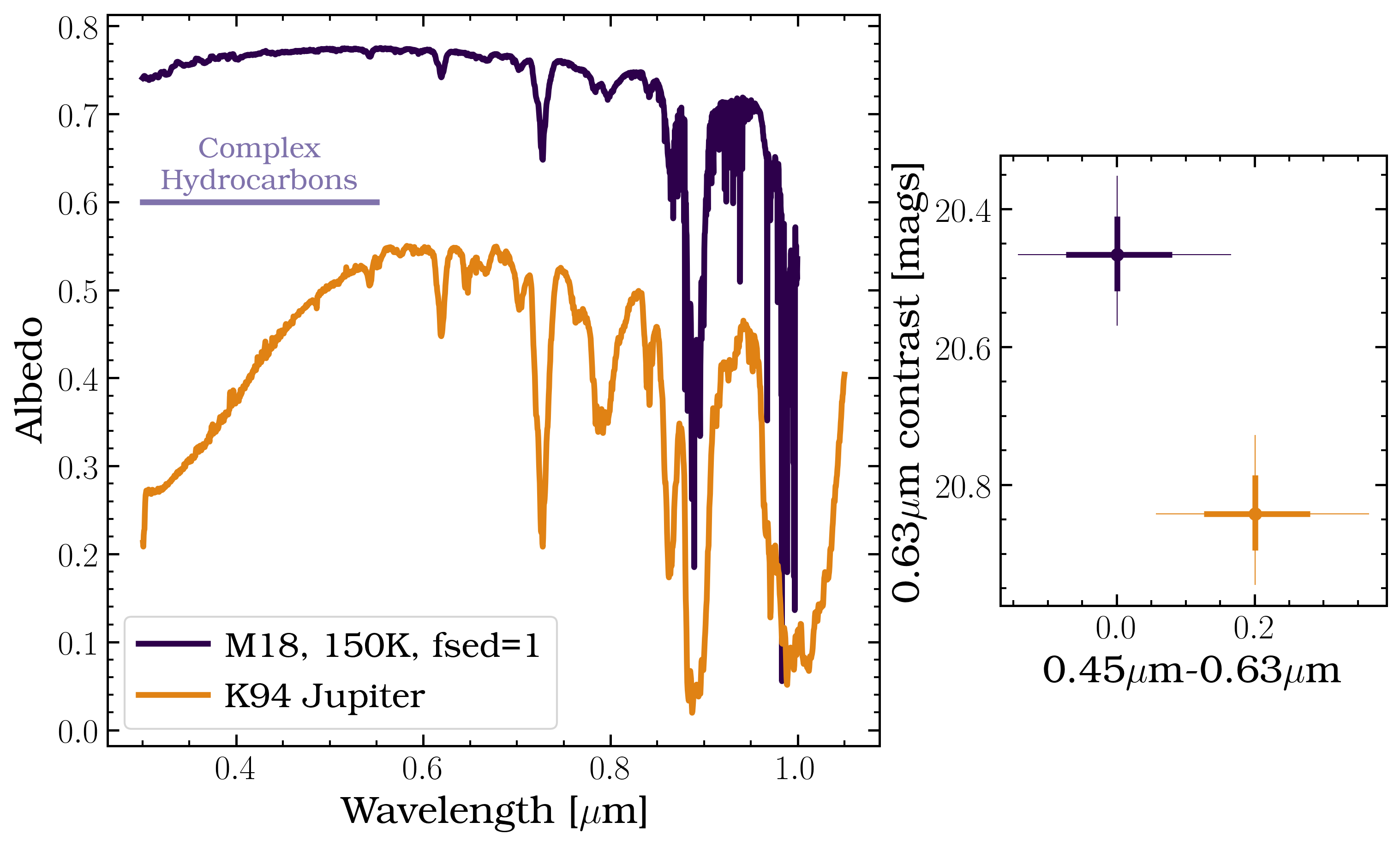}\\
\caption{\textit{Left}: Full-disk reflected light albedo spectrum of Jupiter (blue; \citealt{Karkoschka1994JovianPlanets}) compared to a model Jupiter-like spectrum at 150~K from \citet{macdonaldExploringH2OProminence2018}. Complex hydrocarbon hazes in Jupiter's atmosphere scatter blue light, suppressing Jupiter's flux in a way not easily reflected in models. \textit{Right}: broadband color-magnitude at 0.45 and 0.63~$\mu$m, with error bars for S/N~=~10 (thin) and 20 (thick). We computed the photometric uncertainty as $\sigma = \rm{S/N} \times f$, where f is the planet-star flux contrast at the filter central wavelength. Hazes make the 0.45--0.63~$\mu$m color redder than without, distinguishable at $>1\sigma$ at S/N~=~20. \textbf{Takeaway}: observing targets at wavelengths $<$0.5$\mu$m can probe the presence of hydrocarbon hazes outside the solar system, and is potentially distinguishable at moderate S/N.}
\label{fig:jup-vs-m18}
\end{figure}

All three targeted planets are also on our priority target list (although $\epsilon$ Eri b is only detectable at the goal OWA of 20$\lambda/D$). ESC will be able to complement CGI observations at 0.575, 0.730 (Band 3), and 0.825$\mu$m (Band 4), and expand the wavelength range shorter than 0.5$\mu$m. Interesting science can be done in the regime of 0.3--0.5$\mu$m. Figure \ref{fig:jup-vs-m18} shows the reflectance spectrum of Jupiter \citep{Karkoschka1994JovianPlanets} compared to a Jupiter-like planet model from \citet{macdonaldExploringH2OProminence2018}. Jupiter shows a characteristic drop in reflected flux from 0.3--0.5$\mu$m due to complex hydrocarbon hazes that are difficult to model and are typically absent from gas giant models. It is unknown if gas giants outside the solar system host similar complex hazes, but broadband photometry at $<$0.5$\mu$m can distinguish their presence even at relatively low S/N (Figure \ref{fig:jup-vs-m18}, right). If observed by both CGI and ESC, this will be a powerful constraint on atmosphere composition.

\begin{figure}[t!]
\centering
\includegraphics[width=0.48\textwidth]{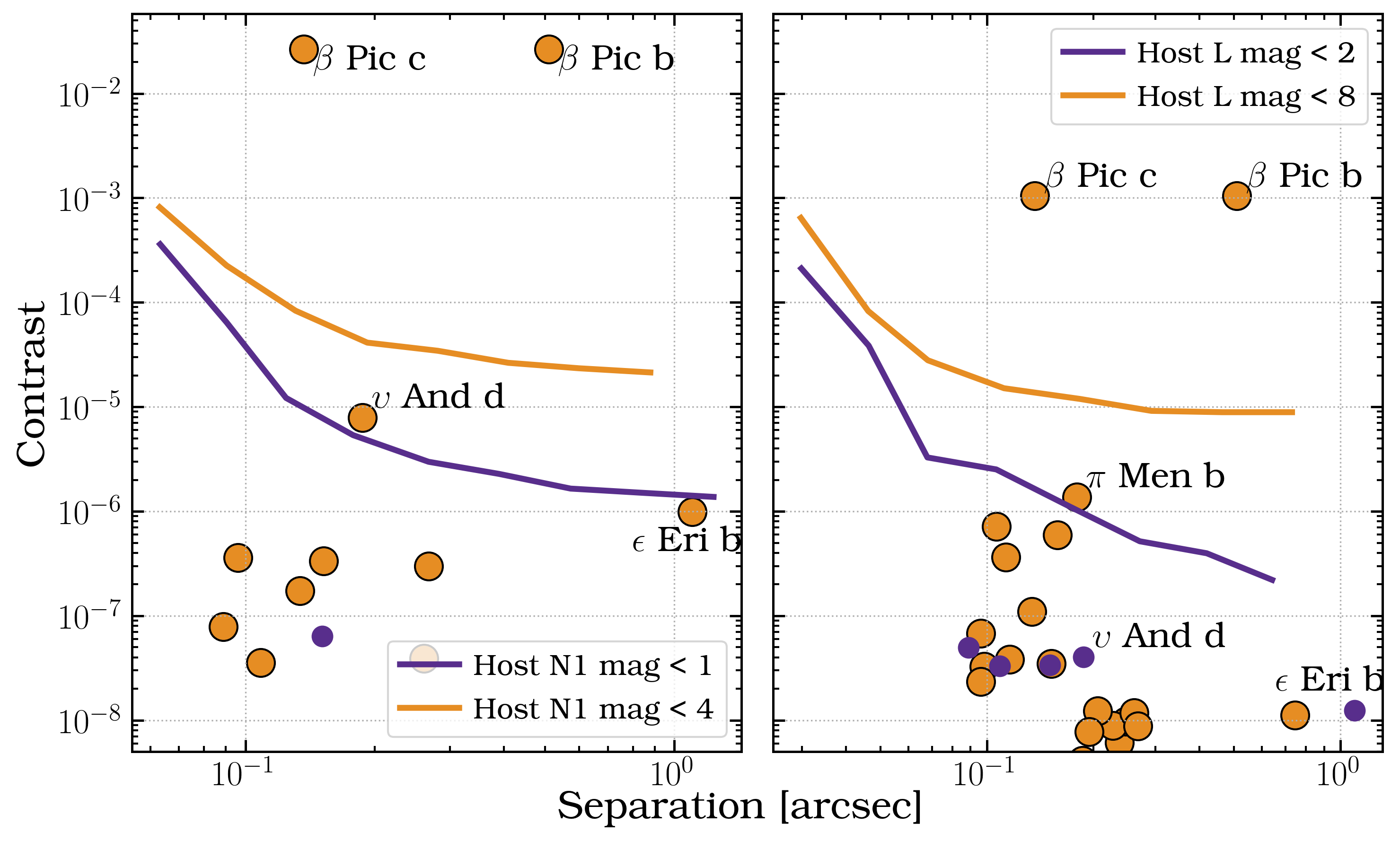}
\caption{Target list planets in thermal emission compared to ELT-METIS contrast curves from \citet{deXivry2026METIS-simultations}. Left: ADI-post processed contrast curves from \citet{deXivry2026METIS-simultations} Fig. 12 for the classical vortex coronagraph (CVC) for the N1 band ($\lambda_c = 8.7\mu$m, bandwidth~=~13\%) for guide star with N1 mag brighter than 2 (purple) and N1 mag brighter than 8 (orange). Right: Same as left for L band ($\lambda_c = 3.8\mu$m, bandwidth~=~6\%). Circles indicate the predicted contrast at corresponding bands for our selected targets with circle color indicating corresponding host star magnitude. \textbf{Takeaway:} Only the $\beta$ Pic planets will be accessible to METIS from this target list.}
\label{fig:metis}
\end{figure}

\subsection{Synergies with IR direct imaging platforms}

As JWST and future platforms (e.g. ELT-METIS, \citealt{Brandl2021METIS}) continue to image cold planets through thermal emission, we can begin to characterize the same planet atmosphere in both reflected light (visible) and thermal emission (IR) and fully capture the energy budget of the planet. Multicolor photometry and spectroscopy of the same planet in both reflected light and thermal emission will provide 1) a full understanding of the energy budget through probing the interior and irradiative heat input to the atmosphere, 2) testing for greenhouse processes, photochemical hazes, and bulk gas abundances across wavelength regimes, and 3) tests of mass/separation/composition predictions from formation theory. Thermal and reflective direct imaging of the same planet represents a significant milestone in exoplanet science enabled for the first time by the next generation of ground- and space-based platforms.

We examined which of our targets might be accessible to IR direct imaging characterization by comparing a model of IR contrast to simulated contrast curves for ELT-METIS from \citet{deXivry2026METIS-simultations}. In Figure \ref{fig:metis} we show estimated METIS contrast curves for the N1 band ($\lambda_c = 8.7\mu$m) and L band ($\lambda_c = 3.8\mu$m) for the classical vortex coronagraph (CVC) as representative of METIS sensitivity.

We estimated planet flux as the sum of flux contribution from gravitational contraction (i.e. self-luminous thermal emission), flux due to heating from the host star at a given wavelength, and reflected starlight: $\rm{F}_{\rm{IR}}(\lambda) = \rm{F_{stellar\,heating}}(\lambda) + \rm{F_{self-luminous}}(\lambda) + \rm{F_{refl}}(\lambda) $. For stellar heating, $\rm{F_{stellar\,heating}}(\lambda) = \rm{F_{star}}(\lambda) \frac{1-A_B}{4}\left(\frac{R_p}{a}\right)$, where A$_B$ is the Bond albedo and $\rm{F_{star}}(\lambda)$ is the star's flux at a given wavelength. We used \texttt{Phoenix} models \citep{Allard2003Phoenix1, Allard2007Phoenix2} to estimate the star flux as a function ot wavelength. 
% \citet{Li2018InternalHeatJupiter} indicates a Bond albedo near zero for the 3--4$\mu$m range for Jupiter, so 
We adopted a Jupiter-like A$_B = 0.3$. For self-luminous thermal emission, we used the spectra and evolutionary tables of Sonora Flame Skimmer \citep{Mang2026FlameSkimmer} cold (50--500~K) cloud-free model suite. First we used the Flame Skimmer evolutionary tables, combined with planet mass or M$_p(\sin i)/(\sin{60^\circ})$ and star ages to estimate the planet's effective temperature. We then used the corresponding Flame Skimmer disequilibrium model (with [M/H] = 3$\times$ solar, C/O = 1, K$_{\rm{zz}} = 10^7$, and corresponding gravity) to estimate planet flux at the METIS L and N1 bands. $\beta$ Pic~b and c are the youngest targets and are the only ones likely to be hotter than Flame Skimmer models, so we used Sonora Diamondback \citep{morleySonoraSubstellarAtmosphere2024} models for 900--2400~K cloudy chemical equilibrium substellar objects. We used parameters derived from \citet{Kammerer2024JWST/NIRCam-Beta-Pic-b} model fits to JWST/NIRCam photometry to select the closest Diamondback model (\Teff~=~1700~K, logg~=~4.0 [cgs], [M/H]~=~0.0, C/O~=~1.0, \fsed~=~1) to predict 3.8 and 87$\mu$m contrast for $\beta$~Pic~b. \citet{Kammerer2025ExoGRAVITYSurvey} found a similar temperature and spectral type for $\beta$~Pic~c so we used the same model for that planet. We then took the ratio $\rm{F}_{\rm{th}}(\lambda)/\rm{F_{star}}(\lambda)$ as the planet-star contrast in each band due to thermal flux. 
For reflected flux contrast we used Eqn \ref{eqn: contrast}. 
We estimated the geometric albedo at 3.8 and 8.7$\mu$m for a Jupiter-like planet using \texttt{PICASO} \citep{Mang2026PICASO40} radiative transfer code and \texttt{Virga} cloud modeling code \citep{Batalha2025Virga1, Moran2025Virga2}, using the supplied Jupiter-like default climate model and cloud species. We assumed a most optimistic scenario with cloud sedimentation efficiency \fsed~=~8 and full phase. Under these assumptions \texttt{PICASO} estimates A$_g \approx 0.12$ at 8.7$\mu$m and A$_g \approx 0.17$ and 3.8$\mu$m. We applied Eqn \ref{eqn: contrast} using these values to estimate reflected flux contrast at those wavelengths, and added it to the thermal flux contrast to estimate the total flux contrast in each filter.

We show the results in Figure \ref{fig:metis}. We predict that of our selected targets only $\beta$~Pic~b and c are accessible to METIS. We anticipate that as target list and instrument design is refined more dual-detection targets may present themselves.

\subsection{\textsl{Gaia} DR4 will improve target selection and observation planning}

\begin{figure}[t!]
\centering
\includegraphics[width=0.48\textwidth]{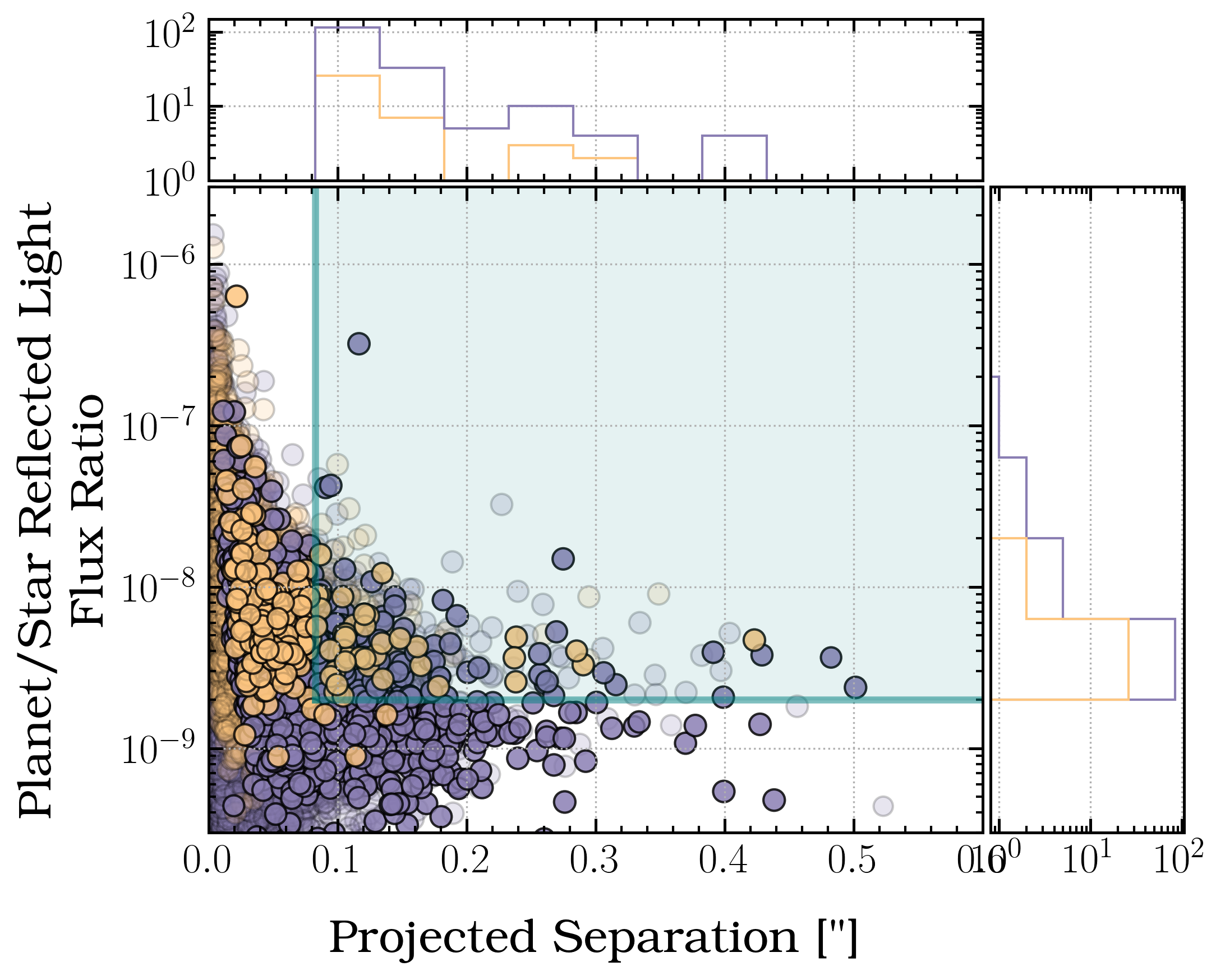}\\
\caption{The mock \textsl{Gaia} DR4 (orange) and DR5 (purple) planet catalog of \citet{LammersWinn2026GaiaDR4} converted into projected separation vs. planet-star flux ratio. We used the catalog true planet mass and period to compute $a$, and assumed $a =$ orbital separation to estimate the projected separation. We used the true mass, the mass-radius relation of \citet{Pearce2025projecc}, Appendix C, and an albedo of A$_g = 0.5$ for the contrast. Solid markers denote mock planets with G$<8$, semi-transparent markers show mock planets with fainter host stars. The teal overlay shows the detection region, using the IWA = $3\times \lambda$/D at $\lambda = 0.4\mu$m, and OWA = $20\times \lambda$/D at $\lambda = 0.63\mu$m, to capture the full detection region explored in this work. The histograms show number of mock planets in bins of contrast and separation within the detection region. 39 DR4 mock planets and 135 of DR5 mock planets fall within the detection region. These are planets with full orbit solutions, we expect a greater number of potential targets with partial (acceleration) solutions. 
\textbf{Takeaway}: \textsl{Gaia} exoplanet detection pipeline will deliver new potential Lazuli ESC reflected light targets.}
\label{fig:gaia-pl}
\end{figure}

The upcoming (2026 Dec) fourth data release of \textsl{Gaia} astrometry will be a revolution in all fields, and a boon to the reflected light imaging project. \textsl{Gaia} DR4 is expected to deliver thousands of potential new planet detections through precision astrometry, and is particularly sensitive to giant planets (3--13~\Mjup) around GKM stars with $a\approx$2--5~au \citep{LammersWinn2026GaiaDR4}, exactly the range where many of our targets lie. In Figure \ref{fig:gaia-pl} we used the mock DR4 and DR5 planet catalog of \citet{LammersWinn2026GaiaDR4} and converted the simulated planets into projected separation vs. planet-star flux ratio for a Jupiter-like albedo (A$_g=0.5$) and the the same mass-radius relation as before, with orange for the DR4 planets, purple for DR5. Solid markers show candidates with G$<$8. We overlayed the detection region for 0.4 -- 0.63$\mu$m filters in teal. Of the 7545 planets in the mock DR4 catalog, 39 fall in the detection region, and 135 of 117305 for DR5. We note that these are candidate planets with full orbit solutions; we expect a large number of candidate signals with partial orbit coverage (i.e. acceleration terms) and wider separations will also fall in the detection region. While as much as 50\% of candidate DR4 planets can be expected to be stellar binary interlopers \citep{LammersWinn2026GaiaDR4}, we can reasonably expect a large number of potential ESC targets with orbits in the DR4 catalog.

Another improvement promised by DR4 is well-constrained orbital parameters for numerous planets from fits to astrometry and RV (where available), as well as the published epoch measurements (intermediate astrometric data, IAD) allowing independent verification of orbits. Velocity measurements in three dimensions mitigates the missing inclination and longitude of ascending node issues discussed above and tightens precision of orbital parameters. Recently \citet{Llop-Sayson2026EpsEri} used simulations of DR4 IAD to show an improvement of a factor of $\sim$2 on orbital parameter uncertainties for $\epsilon$ Eri b.

However these data must be used with caution. Fitting orbits using Thiele-Innes (TI) parameters, as the \textsl{Gaia} pipeline does, linearizes the fit parameters but introduces a bias away from face-on inclinations on translating TI into Campbell elements \citep{Makarov2025GaiaInclBias}. Additionally, planet multiplicity biases retrieved orbits towards high eccentricities, particularly when the planets are close in mass and near a 2:1 resonance, with the higher S/N planet being retrieved in a 1-planet fit but with an unphysically high eccentricity posterior \citep{Greenfield2026GaiaOrbitMultiplicityBias}. Several systems in our target list have multiple giant planets, so we should expect to use orbit solutions, especially of new planets discovered in \textsl{Gaia}, with caution and scrutinize potential targets in detail. Where possible orbits should be independently verified and combined with other measurements to increase accuracy.

\subsection{Synergies with other detection methods are vital for carrying out observations}\label{sec:other detection methods}

In addition to \textsl{Gaia} astrometry and RV, other platforms will be vital partners in preparing for observations. Extreme precision radial velocity (EPRV) instruments such as NEID \citep{Schwab2016NEID,Gupta2021TargetPrioritizationNEID} and PFS \citep{Crane2010PFS} provide vital synergy. Orbital parameter uncertainties compound over time \citep{Pearce2025projecc}, so a check of orbital parameters close to when an observation is conducted is essential for efficiency.

The mock planet catalog of \citet{LammersWinn2026GaiaDR4} represents planets discovered in \textsl{Gaia} through full orbit fits, however many sources will have accelerations only, pointing to wider companions which may also fall in our detection region. Additionally, proper motion anomaly (PMa) between \textsl{Hipparcos} and \textsl{Gaia} has been used to great effect in pointing direct imaging discoveries through thermal emission \citep[e.g.][]{Currie2021SurveyofAcceleratingStars, Currie2023-HIP99770b, Franson2023-AFLepb} and estimating evolutionary model-independent dynamical masses when coupled with RV \citep[e.g.][]{Franson2022DynamicalMass, Piccinini2026piMen, Pinamonti2026HD128717}. Accelerations and PMa for reflected light targets can help resolve inclination and longitude of nodes for RV planets and refine position angle predictions in the absence of full orbit solutions. Several targets on our list have significant acceleration in the HGCA \citep{Brandt2021_HGCA_ERD3}. The \textsl{Gaia} DR4 acceleration catalog will also yield potential targets.

Finally, many targets on our list have been observed through relative astrometry using the HST FGS. Updated RV and joint RV+astrometry orbit fitting with packages like \texttt{orbitize!} \citep{blunt_orbitize_2020} and \texttt{octofitter} \citep{thompsonOctofitterFastFlexible2023} that do not impose external criteria \citep{Benedict2022muArae} can help refine observation planning.

\section{Conclusion}

We identified 27 target planets for a potential Lazuli ESC reflected light imaging campaign based on the current best literature estimates for their orbits, host star brightness, and the anticipated detection region. We determined that 14~Her~b, GJ~832~b, HD~114783~c, HD~1090360~c, and $\beta$ Pic~b are the top five suggested targets for Lazuli ESC. $\gamma$~Cep~b, $\upsilon$~And~b, 47~Uma~b and c, and HD~160691~c will become the top priority targets if inclination and longitude of nodes can be constrained, requiring astrometric measurement and joint astrometric + RV orbit fitting. Several promising targets, including overlap with \textsl{Roman} CGI, will be inaccessible at the designed outer working angle, so best efforts should be made to reach the stretch goal ESC speckle suppression region. \textsl{Gaia} DR4 could provide astrometric constraints these and other potential targets, as well as reveal new candidate targets. Our target list expands direct imaging of gas giant planets into regimes colder than previously imaged, probing mature exoplanet systems, with some overlap with \textsl{Roman} CGI and ELT-METIS imaging, providing more comprehensive atmosphere characterization through synergies with those platform capabilities. Updated RV and astrometric measurements and joint orbit fit of all target planets close to ESC observation campaigns should be a community priority to maximize observational efficiency.

%% Please use the acknowledgment and contribution environments. This will 
%% be anonomyized when the "anonymous" style option is used. 
\begin{acknowledgments}
L.A.P.~acknowledges research support from the University of Michigan through the ELT Fellowship Program. This work began as part of an internship through the NSF INTERN program.

This research has made use of the NASA Exoplanet Archive, which is operated by the California Institute of Technology, under contract with the National Aeronautics and Space Administration under the Exoplanet Exploration Program.

This work has made use of data from the European Space Agency (ESA) mission
{\it Gaia} (\url{https://www.cosmos.esa.int/gaia}), processed by the {\it Gaia}
Data Processing and Analysis Consortium (DPAC,
\url{https://www.cosmos.esa.int/web/gaia/dpac/consortium}). Funding for the DPAC
has been provided by national institutions, in particular the institutions
participating in the {\it Gaia} Multilateral Agreement.
\end{acknowledgments}

\begin{contribution}
L.P. compiled the target list, performed the analysis, produced all the figures, and wrote the manuscript. E.D. edited the manuscript and wrote the specifics of the ESC design.
\end{contribution}

\facilities{Exoplanet Archive \citep{PSCompPars}}

\software{Numpy \citep{harris_array_2020}, Astropy \citep{astropy_collaboration_astropy_2018}, Matplotlib \citep{hunter_matplotlib_2007}, pysynphot \citep{pysynphot-citation}, projecc \citep{Pearce2025projecc}}

\appendix
\restartappendixnumbering
\section{Summary of scores for all considered planets}\label{sec:appendix-scores}

Here we show figures for all planets in the Exoplanet Archive which spend some fraction of an orbit (described by the mean parameters) in the detection region. We drew our selected targets from these summary figures.

\begin{figure*}
\centering
\includegraphics[width=0.9\textwidth]{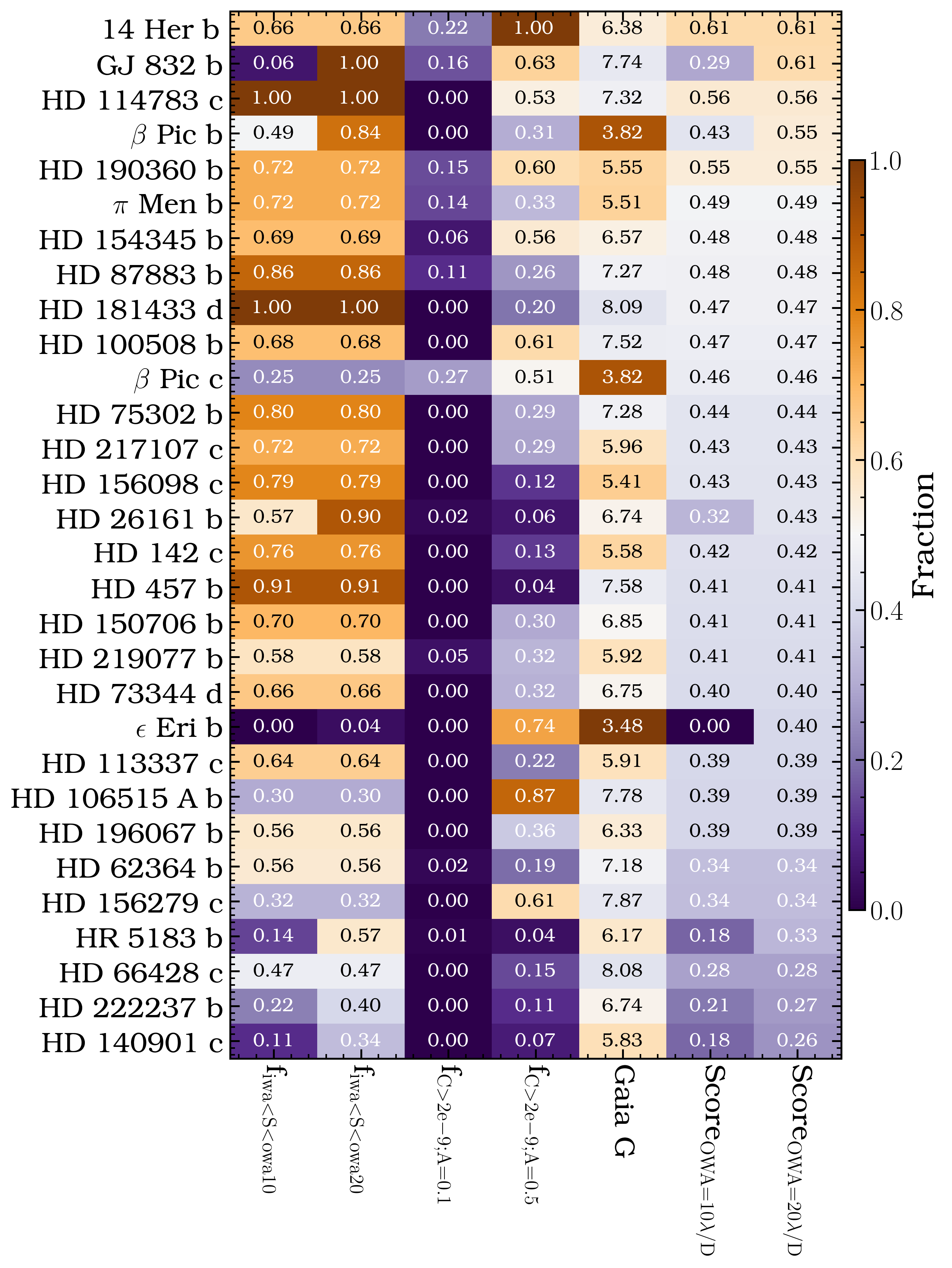}
\caption{\small{Potential ESC targets with a literature inclination constraint sorted by score. Color and numbers correspond to fraction terms in Eqn \ref{eqn:score} We tabulated scores for each planet based on the designed OWA = 10$\lambda/D$ and the goal OWA = 20$\lambda/D$ separately.
}}
\label{fig:inc-score-chart}
\end{figure*}

\begin{figure*}
\centering
\includegraphics[width=0.95\textwidth]{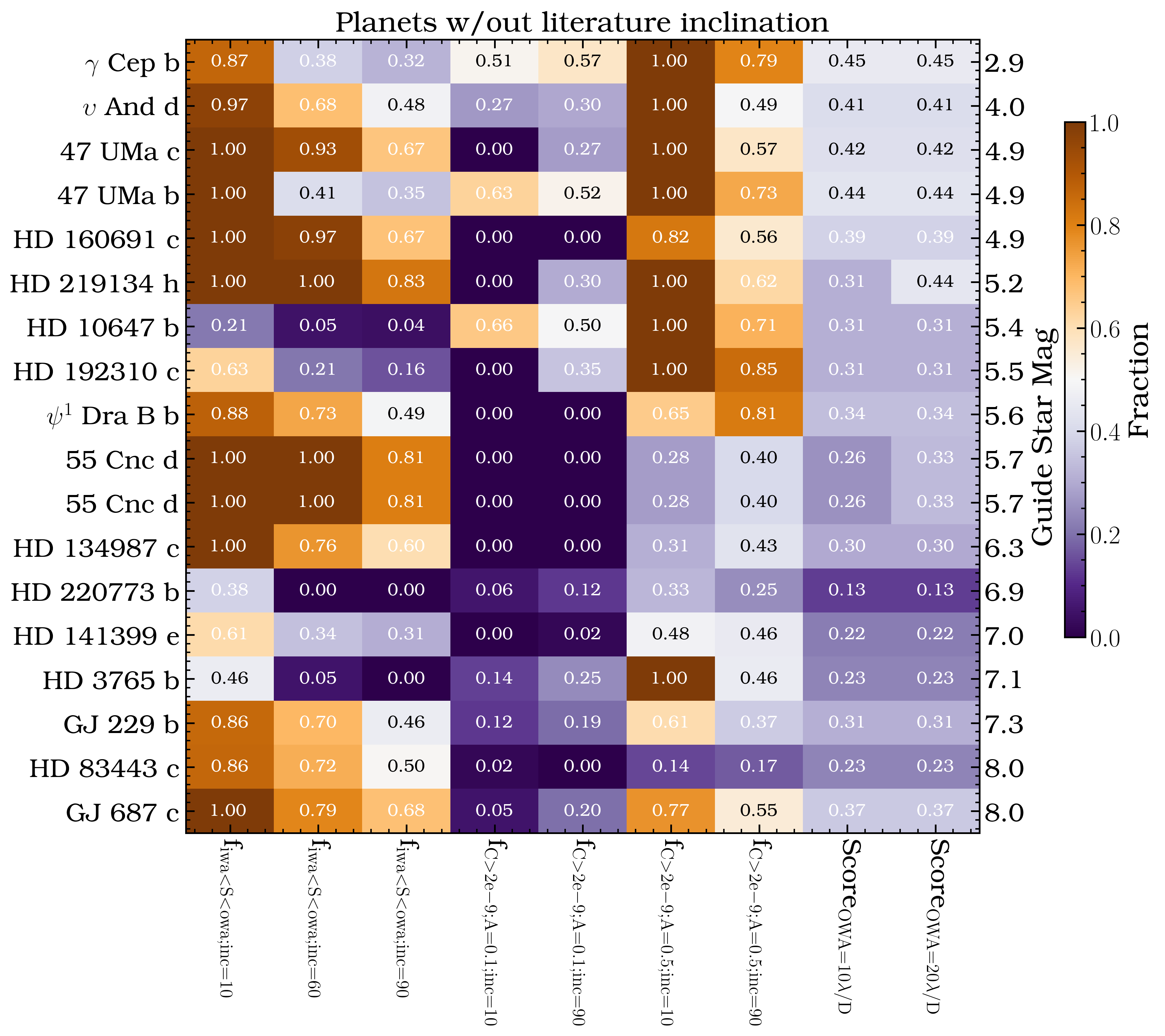}
\caption{\small{Potential ESC targets without a literature inclination constraint sorted by \textsl{Gaia} G magnitude. Color and numbers correspond to fraction terms in Eqn \ref{eqn:score2}.
}}
\label{fig:woinc-score-chart}
\end{figure*}

\begin{figure*}
\centering
\includegraphics[width=0.9\textwidth]{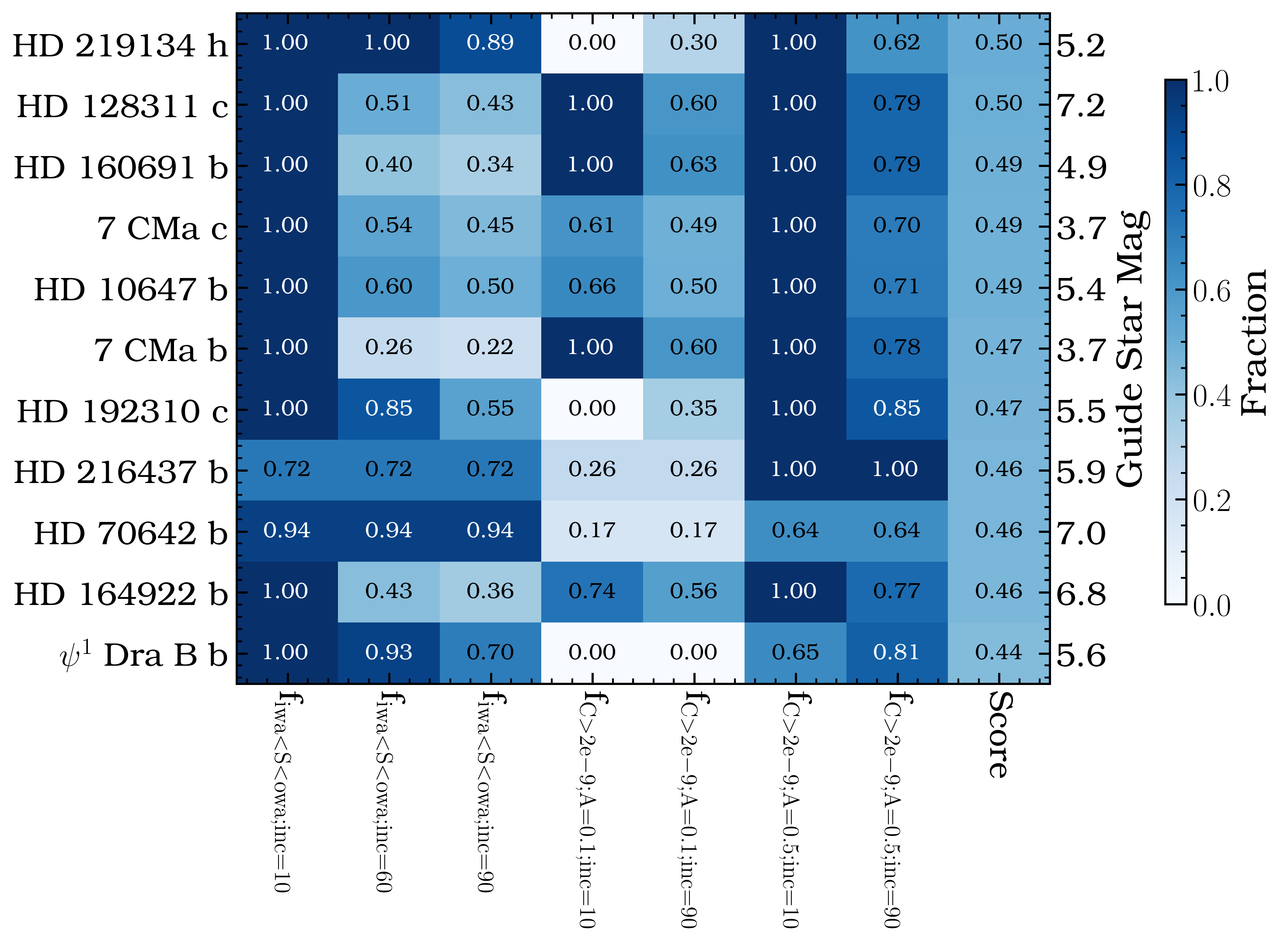}
\caption{\small{Potential ESC targets which score higher at 0.4$\mu$m than in the discovery filter, making them amenable to characterization in short wavelengths..Color and numbers correspond to fraction terms in Eqn \ref{eqn:score2}. We did not differentiate the score based on OWA angle here since these are not significantly impacted by the OWA angle options.
}}
\label{fig:blue-score-chart}
\end{figure*}

\clearpage
\restartappendixnumbering
\section{Orbit simulation results for all selected targets}\label{sec:appendix-orbit-sim}

Here we show orbit simulation results for all planets in Table \ref{tab:inc-planets}

\begin{figure*}[t!]
\centering
\includegraphics[width=\textwidth]{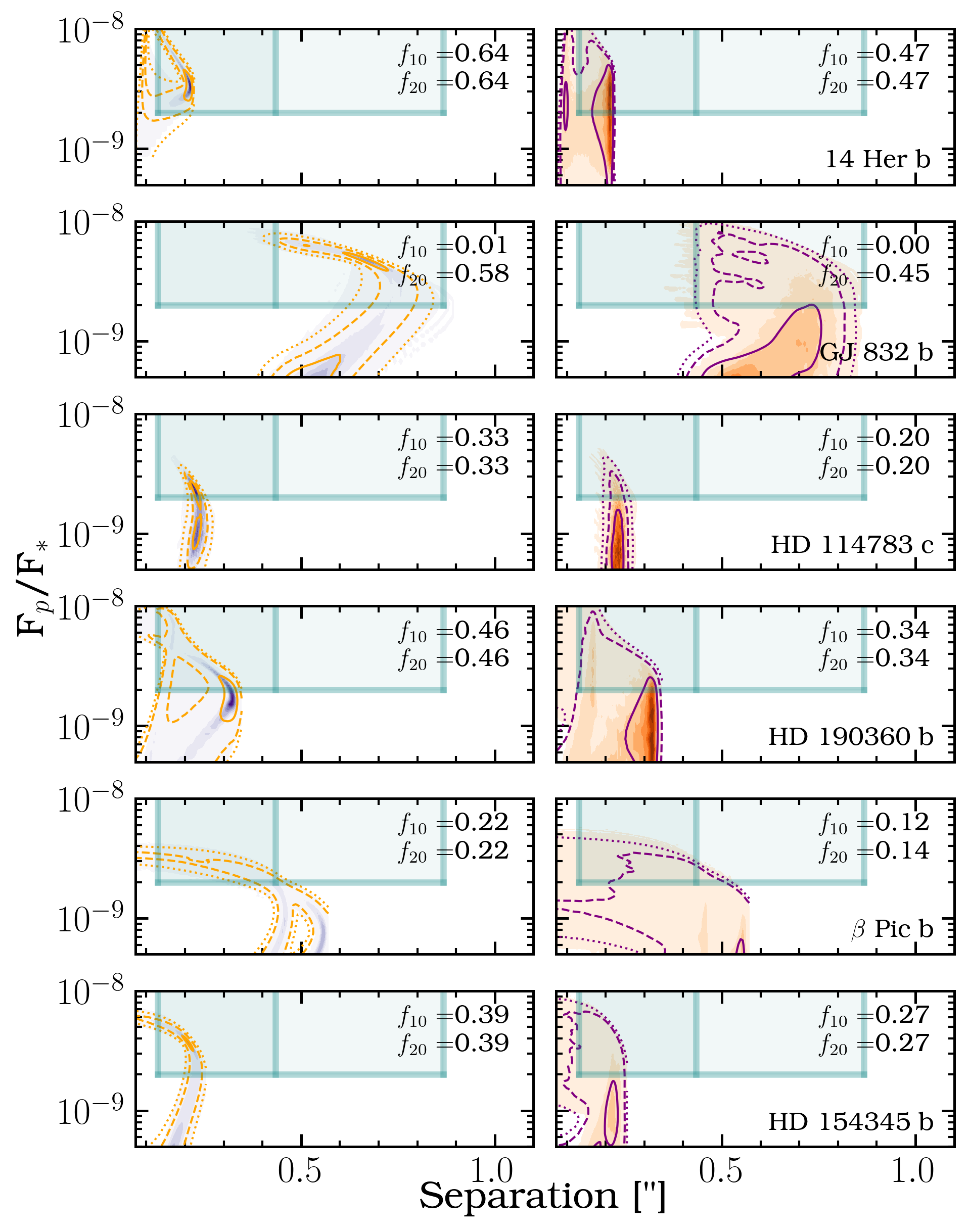}
\caption{Cont. next page}
\end{figure*}

\begin{figure*}[t!]
\addtocounter{figure}{-1} 
\centering
\includegraphics[width=\textwidth]{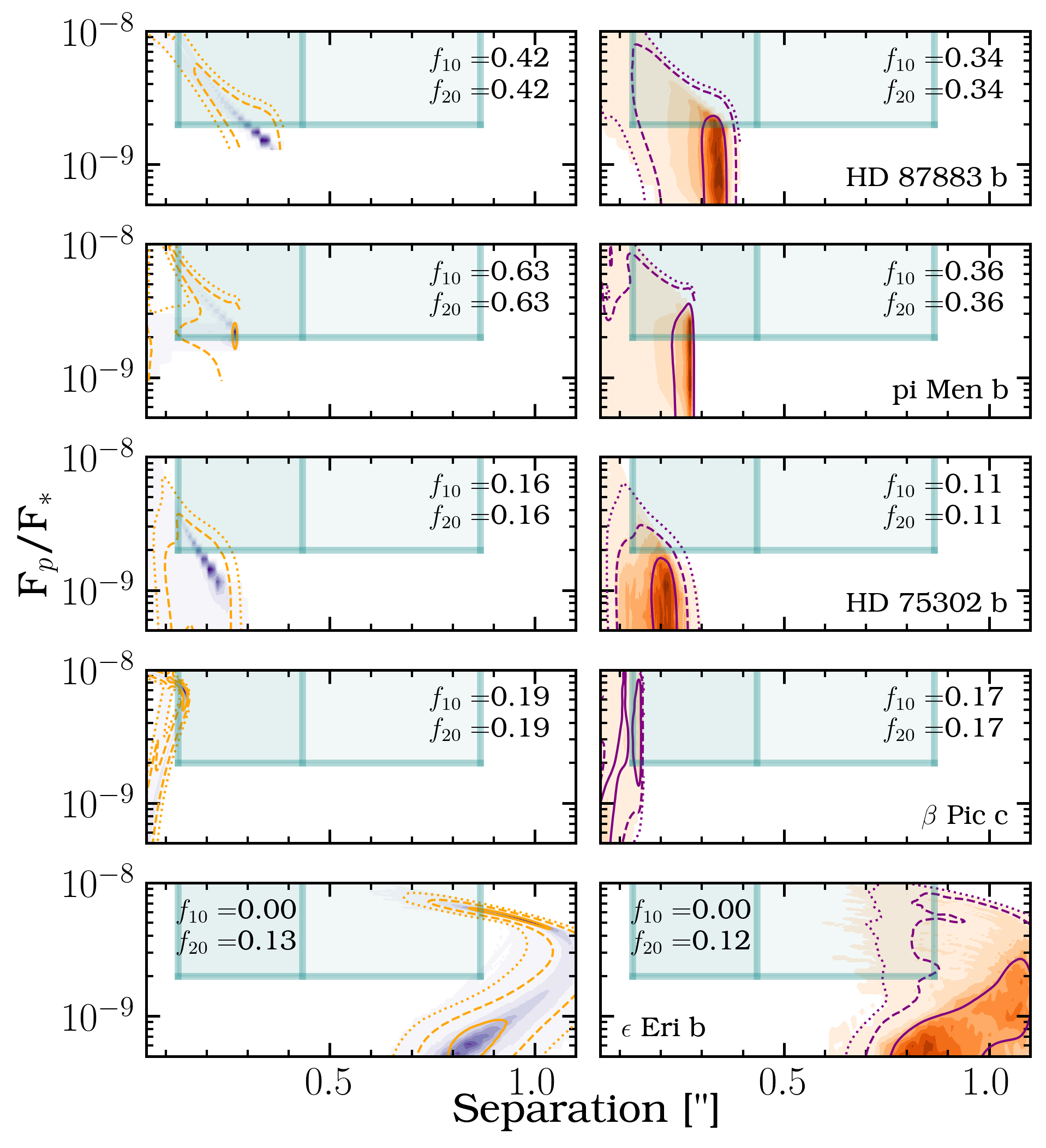}
\caption{Same as Figure \ref{fig:inc-sep-cont} for the 11 planets with inclination in Table \ref{tab:inc-planets}. The teal area shows the detection region for OWA = 10$\lambda/D$ (darker teal) and 20$\lambda/D$ (lighter teal).
}
\label{fig:inc-planets-orbits}
\end{figure*}

\begin{figure*}[t!]
\centering
\includegraphics[width=\textwidth]{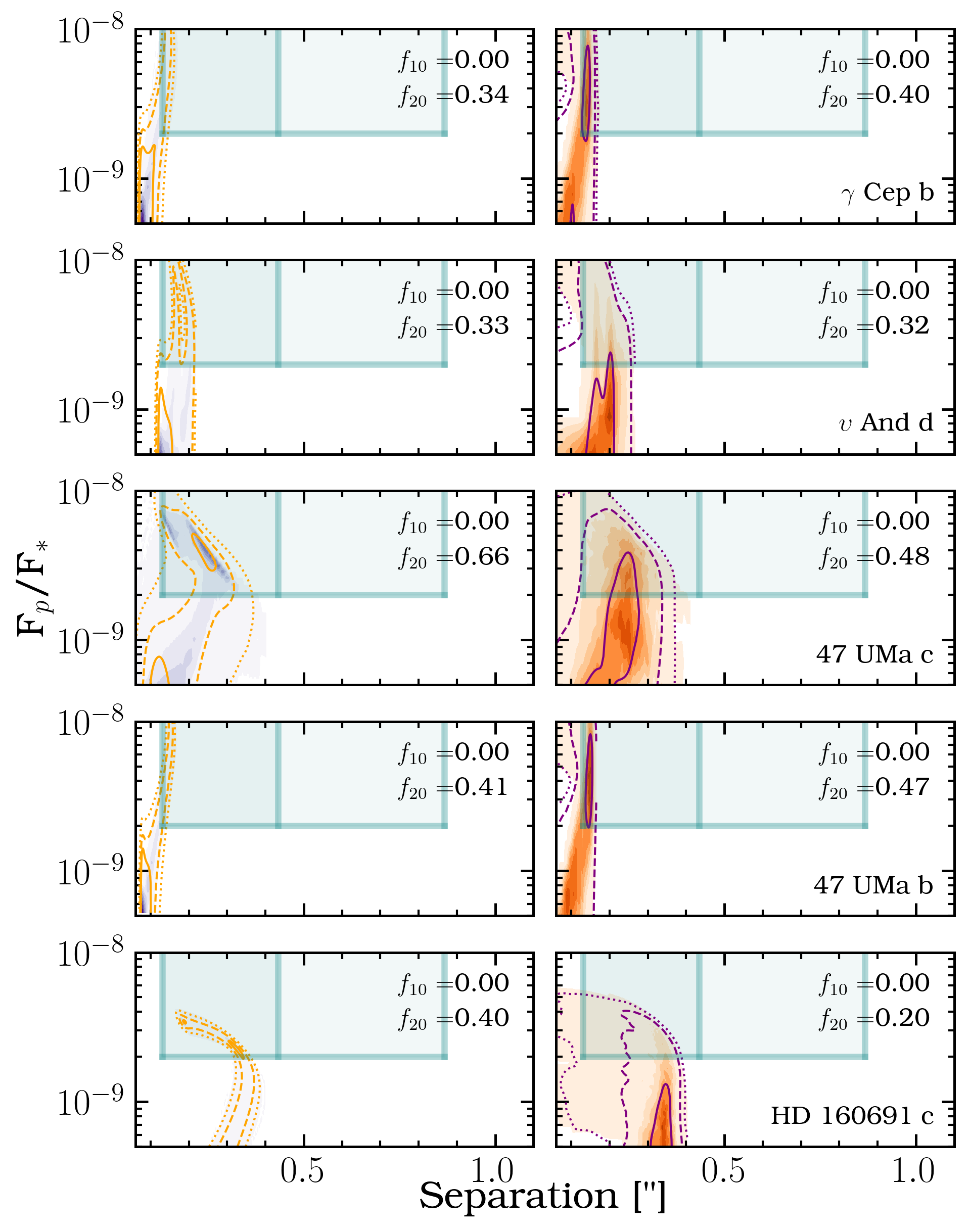}
\caption{Same as Figure \ref{fig:inc-planets-orbits} for the five planets without inclination constraint in Table \ref{tab:inc-planets}
}
\label{fig:noinc-planets-orbits}
\end{figure*}

\begin{figure*}[t!]
\centering
\includegraphics[width=\textwidth]{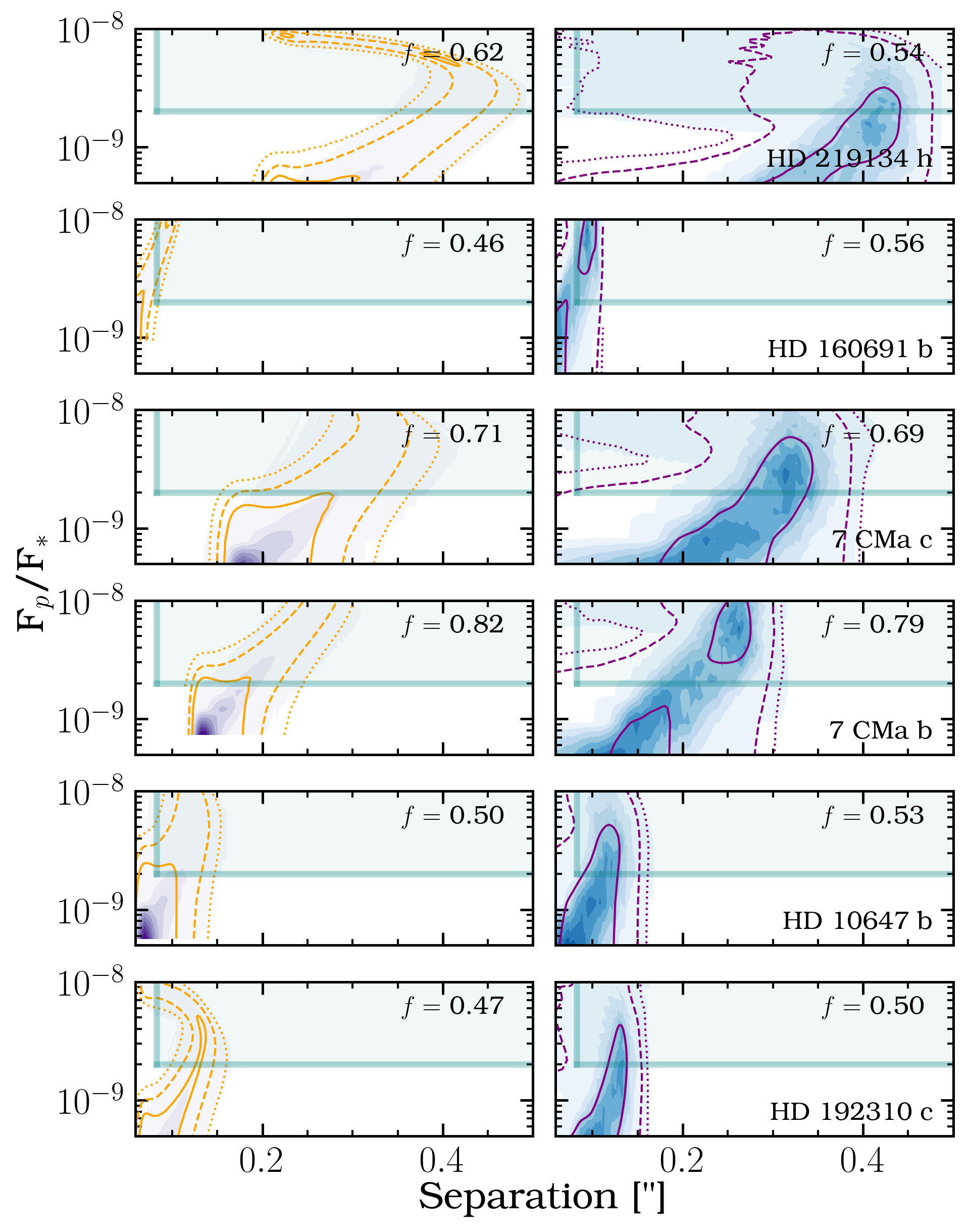}
\caption{Cont. next page}
\end{figure*}

\begin{figure*}[t!]
\addtocounter{figure}{-1} 
\centering
\includegraphics[width=\textwidth]{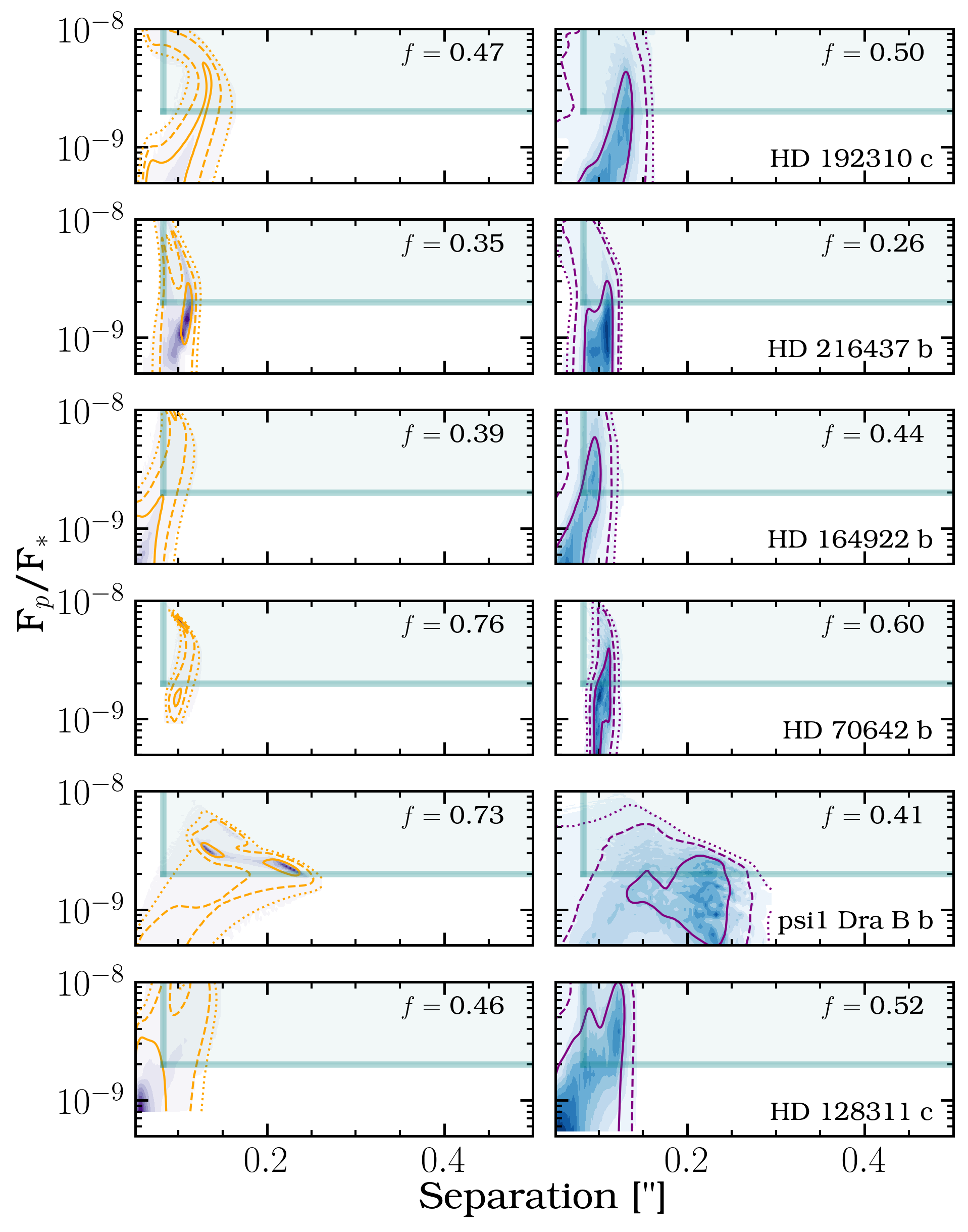}
\caption{Same as Figure \ref{fig:inc-planets-orbits} for the blue subset planets in Table \ref{tab:inc-planets}
}
\label{fig:blue-planets-orbits}
\end{figure*}

\clearpage

\bibliography{references}{}

\begin{thebibliography}{}
\expandafter\ifx\csname natexlab\endcsname\relax\def\natexlab#1{#1}\fi
\providecommand{\url}[1]{\href{#1}{#1}}
\providecommand{\dodoi}[1]{doi:~\href{http://doi.org/#1}{\nolinkurl{#1}}}
\providecommand{\doeprint}[1]{\href{http://ascl.net/#1}{\nolinkurl{http://ascl.net/#1}}}
\providecommand{\doarXiv}[1]{\href{https://arxiv.org/abs/#1}{\nolinkurl{https://arxiv.org/abs/#1}}}

% type= article
\bibitem[{V. {Adibekyan} {et~al.}(2021){Adibekyan}, {Santos}, {Demangeon},
  {Faria}, {Barros}, {Oshagh}, {Figueira}, {Delgado Mena}, {Sousa},
  {Israelian}, {Campante}, \& {Hakobyan}}]{Adibekyan2021StellarClustering}
{Adibekyan}, V., {Santos}, N.~C., {Demangeon}, O.~D.~S., {et~al.} 2021,
  \bibinfo{title}{{Stellar clustering and orbital architecture of planetary
  systems},} \aap, 649, A111, \dodoi{10.1051/0004-6361/202040201}

% type= article
\bibitem[{F. {Allard} {et~al.}(2007){Allard}, {Allard}, {Homeier}, {Kielkopf},
  {McCaughrean}, \& {Spiegelman}}]{Allard2007Phoenix2}
{Allard}, F., {Allard}, N.~F., {Homeier}, D., {et~al.} 2007,
  \bibinfo{title}{{K-H$_{2}$ quasi-molecular absorption detected in the T-dwarf
  \textbackslashvarepsilon Indi Ba},} A\&A, 474, L21,
  \dodoi{10.1051/0004-6361:20078362}

% type= inproceedings
\bibitem[{F. {Allard} {et~al.}(2003){Allard}, {Guillot}, {Ludwig},
  {Hauschildt}, {Schweitzer}, {Alexander}, \& {Ferguson}}]{Allard2003Phoenix1}
{Allard}, F., {Guillot}, T., {Ludwig}, H.-G., {et~al.} 2003,
  \bibinfo{title}{{Model Atmospheres and Spectra: The Role of Dust},} in Brown
  Dwarfs, ed. E.~{Mart{\'\i}n}, Vol. 211, 325

% type= article
\bibitem[{F. {Allard} {et~al.}(2001){Allard}, {Hauschildt}, {Alexander},
  {Tamanai}, \& {Schweitzer}}]{Allard2001-AMES-Dusty}
{Allard}, F., {Hauschildt}, P.~H., {Alexander}, D.~R., {Tamanai}, A., \&
  {Schweitzer}, A. 2001, \bibinfo{title}{{The Limiting Effects of Dust in Brown
  Dwarf Model Atmospheres},} \apj, 556, 357, \dodoi{10.1086/321547}

% type= article
\bibitem[{Q. {An} {et~al.}(2025){An}, {Brandt}, {Brandt}, \&
  {Venner}}]{An2025OrbitsMassesFor156Companions}
{An}, Q., {Brandt}, T.~D., {Brandt}, G.~M., \& {Venner}, A. 2025,
  \bibinfo{title}{{Orbits and Masses for 156 Companions from Combined
  Astrometry and Radial Velocities, and a Validation of Gaia Non-single-star
  Solutions},} \apjs, 280, 61, \dodoi{10.3847/1538-4365/adfa99}

% type= article
\bibitem[{L. {Arnold} {\&} J. {Schneider}(2004){Arnold} \&
  {Schneider}}]{ArnoldSchneider2004}
{Arnold}, L., \& {Schneider}, J. 2004, \bibinfo{title}{{The detectability of
  extrasolar planet surroundings. I. Reflected-light photometry of unresolved
  rings},} \aap, 420, 1153, \dodoi{10.1051/0004-6361:20035720}

% type= article
\bibitem[{{}{Astropy Collaboration} {et~al.}(2018){Astropy Collaboration},
  Price-Whelan, Sipőcz, Günther, Lim, Crawford, Conseil, Shupe, Craig,
  Dencheva, Ginsburg, VanderPlas, Bradley, Pérez-Suárez, de~Val-Borro,
  Aldcroft, Cruz, Robitaille, Tollerud, Ardelean, Babej, Bach, Bachetti,
  Bakanov, Bamford, Barentsen, Barmby, Baumbach, Berry, Biscani, Boquien,
  Bostroem, Bouma, Brammer, Bray, Breytenbach, Buddelmeijer, Burke, Calderone,
  Cano~Rodríguez, Cara, Cardoso, Cheedella, Copin, Corrales, Crichton,
  D'Avella, Deil, Depagne, Dietrich, Donath, Droettboom, Earl, Erben, Fabbro,
  Ferreira, Finethy, Fox, Garrison, Gibbons, Goldstein, Gommers, Greco,
  Greenfield, Groener, Grollier, Hagen, Hirst, Homeier, Horton, Hosseinzadeh,
  Hu, Hunkeler, Ivezić, Jain, Jenness, Kanarek, Kendrew, Kern, Kerzendorf,
  Khvalko, King, Kirkby, Kulkarni, Kumar, Lee, Lenz, Littlefair, Ma, Macleod,
  Mastropietro, McCully, Montagnac, Morris, Mueller, Mumford, Muna, Murphy,
  Nelson, Nguyen, Ninan, Nöthe, Ogaz, Oh, Parejko, Parley, Pascual, Patil,
  Patil, Plunkett, Prochaska, Rastogi, Reddy~Janga, Sabater, Sakurikar,
  Seifert, Sherbert, Sherwood-Taylor, Shih, Sick, Silbiger, Singanamalla,
  Singer, Sladen, Sooley, Sornarajah, Streicher, Teuben, Thomas, Tremblay,
  Turner, Terr\'on, van Kerkwijk, de~la Vega, Watkins, Weaver, Whitmore,
  Woillez, Zabalza, \& {Astropy
  Contributors}}]{astropy_collaboration_astropy_2018}
{Astropy Collaboration}, Price-Whelan, A.~M., Sipőcz, B.~M., {et~al.} 2018,
  \bibinfo{title}{The {Astropy} {Project}: {Building} an {Open}-science
  {Project} and {Status} of the v2.0 {Core} {Package},} The Astronomical
  Journal, 156, 123, \dodoi{10.3847/1538-3881/aabc4f}

% type=
\bibitem[{V. Bailey(2021)Bailey}]{target_accessibility}
Bailey, V. 2021, {CGI\_target\_accessibility}, Github.
\newblock \url{https://github.com/nasavbailey/CGI_target_accessibility}

% type= inproceedings
\bibitem[{V.~P. {Bailey} {et~al.}(2023){Bailey}, {Bendek}, {Monacelli},
  {Baker}, {Bedrosian}, {Cady}, {Douglas}, {Groff}, {Hildebrandt}, {Kasdin},
  {Krist}, {Macintosh}, {Mennesson}, {Morrissey}, {Poberezhskiy}, {Subedi},
  {Rhodes}, {Roberge}, {Ygouf}, {Zellem}, {Zhao}, \&
  {Zimmerman}}]{Bailey2023RomanOverview}
{Bailey}, V.~P., {Bendek}, E., {Monacelli}, B., {et~al.} 2023,
  \bibinfo{title}{{Nancy Grace Roman Space Telescope coronagraph instrument
  overview and status},} in Society of Photo-Optical Instrumentation Engineers
  (SPIE) Conference Series, Vol. 12680, Society of Photo-Optical
  Instrumentation Engineers (SPIE) Conference Series, 126800T,
  \dodoi{10.1117/12.2679036}

% type= inproceedings
\bibitem[{W. {Balmer} {et~al.}(2025){Balmer}, {Bardalez Gagliuffi}, {Pueyo},
  {Brandt}, {Giovinazzi}, {Millholland}, {Black}, {Lu}, {Rice}, {Mang},
  {Morley}, {Lacy}, {Girard}, {Matthews}, {Carter}, {Bowler}, {Faherty},
  {Fontanive}, \& {Rickman}}]{Balmer202514Herc}
{Balmer}, W., {Bardalez Gagliuffi}, D., {Pueyo}, L., {et~al.} 2025,
  \bibinfo{title}{{JWST Coronagraphic Images of 14 Her c: a Cold Giant Planet
  in a Dynamically Hot, Multi-planet System},} in American Astronomical Society
  Meeting Abstracts, Vol. 246, 246th Meeting of the American Astronomical
  Society, 116.02

% type= article
\bibitem[{W.~O. {Balmer} {et~al.}(2025){Balmer}, {Kammerer}, {Pueyo}, {Perrin},
  {Girard}, {Leisenring}, {Lawson}, {Dennen}, {van der Marel}, {Beichman},
  {Bryden}, {Llop-Sayson}, {Valenti}, {Lothringer}, {Lewis}, {M{\^a}lin},
  {Rebollido}, {Rickman}, {Hoch}, {Soummer}, {Clampin}, \&
  {Mountain}}]{Balmer2025LivingOnTheWedge}
{Balmer}, W.~O., {Kammerer}, J., {Pueyo}, L., {et~al.} 2025,
  \bibinfo{title}{{JWST-TST High Contrast: Living on the Wedge, or, NIRCam Bar
  Coronagraphy Reveals CO$_{2}$ in the HR 8799 and 51 Eri Exoplanets'
  Atmospheres},} \aj, 169, 209, \dodoi{10.3847/1538-3881/adb1c6}

% type= article
\bibitem[{W.~O. {Balmer} {et~al.}(2026){Balmer}, {Pueyo}, {Messier},
  {Bruinsma}, {Jones}, {Matuszewska}, {Perrin}, {Girard}, {Leisenring},
  {Lawson}, {van der Marel}, {Kammerer}, {Carter}, {M{\^a}lin}, {Ward-Duong},
  {Hoch}, {Rickman}, \& {Seager}}]{Balmer2026DirectImagesOf}
{Balmer}, W.~O., {Pueyo}, L., {Messier}, A., {et~al.} 2026,
  \bibinfo{title}{{Direct Images of CO$_{2}$ Absorption in the Atmosphere of a
  Super-Jupiter: Enhanced Metallicity Suggestive of Formation in a Disk},}
  \apjl, 1001, L26, \dodoi{10.3847/2041-8213/ae374a}

% type= article
\bibitem[{D.~C. {Bardalez Gagliuffi} {et~al.}(2021){Bardalez Gagliuffi},
  {Faherty}, {Li}, {Brandt}, {Williams}, {Brandt}, \&
  {Gelino}}]{BardalezGagliuffi2021-14Her}
{Bardalez Gagliuffi}, D.~C., {Faherty}, J.~K., {Li}, Y., {et~al.} 2021,
  \bibinfo{title}{{14 Her: A Likely Case of Planet-Planet Scattering},} \apjl,
  922, L43, \dodoi{10.3847/2041-8213/ac382c}

% type= article
\bibitem[{D.~C. {Bardalez Gagliuffi} {et~al.}(2025){Bardalez Gagliuffi},
  {Balmer}, {Pueyo}, {Brandt}, {Giovinazzi}, {Millholland}, {Black}, {Lu},
  {Rice}, {Mang}, {Morley}, {Lacy}, {Girard}, {Matthews}, {Carter}, {Bowler},
  {Faherty}, {Fontanive}, \& {Rickman}}]{BardalezGagluiffi2025-14Herc}
{Bardalez Gagliuffi}, D.~C., {Balmer}, W.~O., {Pueyo}, L., {et~al.} 2025,
  \bibinfo{title}{{JWST Coronagraphic Images of 14 Her c: A Cold Giant Planet
  in a Dynamically Hot Multiplanet System},} \apjl, 988, L18,
  \dodoi{10.3847/2041-8213/ade30f}

% type= article
\bibitem[{N.~E. {Batalha} {et~al.}(2025){Batalha}, {Rooney}, {Visscher},
  {Moran}, {Marley}, {Sengupta}, {Kiefer}, {Lodge}, {Mang}, {Morley},
  {Mukherjee}, {Fortney}, {Gao}, {Lewis}, {Mayorga}, {Pearce}, \&
  {Wakeford}}]{Batalha2025Virga1}
{Batalha}, N.~E., {Rooney}, C.~M., {Visscher}, C., {et~al.} 2025,
  \bibinfo{title}{{Condensation Clouds in Substellar Atmospheres with Virga},}
  arXiv e-prints, arXiv:2508.15102, \dodoi{10.48550/arXiv.2508.15102}

% type= inproceedings
\bibitem[{J.-L. {Baudino} {et~al.}(2014){Baudino}, {B{\'e}zard}, {Boccaletti},
  {Bonnefoy}, \& {Lagrange}}]{Baudino2014-betaPicb-models}
{Baudino}, J.-L., {B{\'e}zard}, B., {Boccaletti}, A., {Bonnefoy}, M., \&
  {Lagrange}, A.-M. 2014, \bibinfo{title}{{A radiative-convective equilibrium
  model for young giant exoplanets: Application to {\ensuremath{\beta}}
  Pictoris b},} in IAU Symposium, Vol. 299, Exploring the Formation and
  Evolution of Planetary Systems, ed. M.~{Booth}, B.~C. {Matthews}, \& J.~R.
  {Graham}, 277--278, \dodoi{10.1017/S1743921313008570}

% type= article
\bibitem[{J.-L. {Baudino} {et~al.}(2015){Baudino}, {B{\'e}zard}, {Boccaletti},
  {Bonnefoy}, {Lagrange}, \& {Galicher}}]{Baudino2015-BetaPicb-models}
{Baudino}, J.-L., {B{\'e}zard}, B., {Boccaletti}, A., {et~al.} 2015,
  \bibinfo{title}{{Interpreting the photometry and spectroscopy of directly
  imaged planets: a new atmospheric model applied to {\ensuremath{\beta}}
  Pictoris b and SPHERE observations},} \aap, 582, A83,
  \dodoi{10.1051/0004-6361/201526332}

% type= article
\bibitem[{S. {Benatti} {et~al.}(2020){Benatti}, {Damasso}, {Desidera},
  {Marzari}, {Biazzo}, {Claudi}, {Di Mauro}, {Lanza}, {Pinamonti}, {Barbato},
  {Malavolta}, {Poretti}, {Sozzetti}, {Affer}, {Bignamini}, {Bonomo}, {Borsa},
  {Brogi}, {Bruno}, {Carleo}, {Cosentino}, {Covino}, {Frustagli}, {Giacobbe},
  {Gonzalez}, {Gratton}, {Harutyunyan}, {Knapic}, {Leto}, {Lodi}, {Maggio},
  {Maldonado}, {Mancini}, {Martinez Fiorenzano}, {Micela}, {Molinari},
  {Molinaro}, {Nardiello}, {Nascimbeni}, {Pagano}, {Pedani}, {Piotto},
  {Rainer}, \& {Scandariato}}]{Benatti2020HD164922}
{Benatti}, S., {Damasso}, M., {Desidera}, S., {et~al.} 2020,
  \bibinfo{title}{{The GAPS programme at TNG. XXIII. HD 164922 d: close-in
  super-Earth discovered with HARPS-N in a system with a long-period Saturn
  mass companion},} \aap, 639, A50, \dodoi{10.1051/0004-6361/202037939}

% type= article
\bibitem[{G.~F. {Benedict} {et~al.}(2018){Benedict}, {Harrison}, {Endl}, \&
  {Torres}}]{Benedict2018MassGammaCepb}
{Benedict}, G.~F., {Harrison}, T.~E., {Endl}, M., \& {Torres}, G. 2018,
  \bibinfo{title}{{A Mass for {\ensuremath{\gamma}} Cep Ab},} Research Notes of
  the American Astronomical Society, 2, 7, \dodoi{10.3847/2515-5172/aabe7e}

% type= article
\bibitem[{G.~F. {Benedict} {et~al.}(2023){Benedict}, {McArthur}, {Nelan}, \&
  {Bean}}]{Benedict2023-14Her}
{Benedict}, G.~F., {McArthur}, B.~E., {Nelan}, E.~P., \& {Bean}, J.~L. 2023,
  \bibinfo{title}{{The 14 Her Planetary System: Companion Masses and
  Architecture from Radial Velocities and Astrometry},} \aj, 166, 27,
  \dodoi{10.3847/1538-3881/acd93a}

% type= article
\bibitem[{G.~F. {Benedict} {et~al.}(2022){Benedict}, {McArthur}, {Nelan},
  {Wittenmyer}, {Barnes}, {Smotherman}, \& {Horner}}]{Benedict2022muArae}
{Benedict}, G.~F., {McArthur}, B.~E., {Nelan}, E.~P., {et~al.} 2022,
  \bibinfo{title}{{The {\ensuremath{\mu}} Arae Planetary System: Radial
  Velocities and Astrometry},} \aj, 163, 295, \dodoi{10.3847/1538-3881/ac6ac8}

% type= article
\bibitem[{B. {Bitsch} {et~al.}(2021){Bitsch}, {Raymond}, {Buchhave},
  {Bello-Arufe}, {Rathcke}, \& {Schneider}}]{Bitsch2021DryOrWaterWorld}
{Bitsch}, B., {Raymond}, S.~N., {Buchhave}, L.~A., {et~al.} 2021,
  \bibinfo{title}{{Dry or water world? How the water contents of inner
  sub-Neptunes constrain giant planet formation and the location of the water
  ice line},} \aap, 649, L5, \dodoi{10.1051/0004-6361/202140793}

% type= article
\bibitem[{S. Blunt {et~al.}(2020)Blunt, Wang, Angelo, Ngo, Cody, De~Rosa,
  Graham, Hirsch, Nagpal, Nielsen, Pearce, Rice, \&
  Tejada}]{blunt_orbitize_2020}
Blunt, S., Wang, J.~J., Angelo, I., {et~al.} 2020, \bibinfo{title}{orbitize!:
  {A} {Comprehensive} {Orbit}-fitting {Software} {Package} for the
  {High}-contrast {Imaging} {Community},} The Astronomical Journal, 159, 89,
  \dodoi{10.3847/1538-3881/ab6663}

% type= article
\bibitem[{A. {Bonfanti} {et~al.}(2016){Bonfanti}, {Ortolani}, \&
  {Nascimbeni}}]{Bonfanti2016AgeConsistency}
{Bonfanti}, A., {Ortolani}, S., \& {Nascimbeni}, V. 2016, \bibinfo{title}{{Age
  consistency between exoplanet hosts and field stars},} \aap, 585, A5,
  \dodoi{10.1051/0004-6361/201527297}

% type= article
\bibitem[{A. {Bonfanti} {et~al.}(2015){Bonfanti}, {Ortolani}, {Piotto}, \&
  {Nascimbeni}}]{Bonfanti2015RevisingAges}
{Bonfanti}, A., {Ortolani}, S., {Piotto}, G., \& {Nascimbeni}, V. 2015,
  \bibinfo{title}{{Revising the ages of planet-hosting stars},} \aap, 575, A18,
  \dodoi{10.1051/0004-6361/201424951}

% type= article
\bibitem[{M. {Bonnefoy} {et~al.}(2014){Bonnefoy}, {Marleau}, {Galicher},
  {Beust}, {Lagrange}, {Baudino}, {Chauvin}, {Borgniet}, {Meunier}, {Rameau},
  {Boccaletti}, {Cumming}, {Helling}, {Homeier}, {Allard}, \&
  {Delorme}}]{Bonnefoy2014betaPicb}
{Bonnefoy}, M., {Marleau}, G.-D., {Galicher}, R., {et~al.} 2014,
  \bibinfo{title}{{Physical and orbital properties of {\ensuremath{\beta}}
  Pictoris b},} \aap, 567, L9, \dodoi{10.1051/0004-6361/201424041}

% type= article
\bibitem[{R. {Bowens-Rubin} {et~al.}(2025){Bowens-Rubin}, {Mang}, {Limbach},
  {Carter}, {Stevenson}, {Wagner}, {Strampelli}, {Morley}, {Kennedy},
  {Matthews}, {Vanderburg}, \& {Salama}}]{Bowens-Rubin2025NIRCamYellsatCloud}
{Bowens-Rubin}, R., {Mang}, J., {Limbach}, M.~A., {et~al.} 2025,
  \bibinfo{title}{{NIRCam Yells at Cloud: JWST MIRI Imaging Can Directly Detect
  Exoplanets of the Same Temperature, Mass, Age, and Orbital Separation as
  Saturn and Jupiter},} \apjl, 986, L26, \dodoi{10.3847/2041-8213/addbde}

% type= article
\bibitem[{B. {Brandl} {et~al.}(2021){Brandl}, {Bettonvil}, {van Boekel},
  {Glauser}, {Quanz}, {Absil}, {Amorim}, {Feldt}, {Glasse}, {G{\"u}del}, {Ho},
  {Labadie}, {Meyer}, {Pantin}, {van Winckel}, \& {METIS
  Consortium}}]{Brandl2021METIS}
{Brandl}, B., {Bettonvil}, F., {van Boekel}, R., {et~al.} 2021,
  \bibinfo{title}{{METIS: The Mid-infrared ELT Imager and Spectrograph},} The
  Messenger, 182, 22, \dodoi{10.18727/0722-6691/5218}

% type= article
\bibitem[{G.~M. {Brandt} {et~al.}(2021){Brandt}, {Michalik}, {Brandt}, {Li},
  {Dupuy}, \& {Zeng}}]{Brandt2021HOTF}
{Brandt}, G.~M., {Michalik}, D., {Brandt}, T.~D., {et~al.} 2021,
  \bibinfo{title}{{htof: A New Open-source Tool for Analyzing Hipparcos, Gaia,
  and Future Astrometric Missions},} \aj, 162, 230,
  \dodoi{10.3847/1538-3881/ac12d0}

% type= article
\bibitem[{T.~D. {Brandt}(2021){Brandt}}]{Brandt2021_HGCA_ERD3}
{Brandt}, T.~D. 2021, \bibinfo{title}{{The Hipparcos-Gaia Catalog of
  Accelerations: Gaia EDR3 Edition},} ApJs, 254, 42,
  \dodoi{10.3847/1538-4365/abf93c}

% type= article
\bibitem[{T.~D. {Brandt} {et~al.}(2021){Brandt}, {Dupuy}, {Li}, {Brandt},
  {Zeng}, {Michalik}, {Bardalez Gagliuffi}, \&
  {Raposo-Pulido}}]{Brandt2021orvara}
{Brandt}, T.~D., {Dupuy}, T.~J., {Li}, Y., {et~al.} 2021,
  \bibinfo{title}{{orvara: An Efficient Code to Fit Orbits Using Radial
  Velocity, Absolute, and/or Relative Astrometry},} AJ, 162, 186,
  \dodoi{10.3847/1538-3881/ac042e}

% type= inproceedings
\bibitem[{W.~S. {Burgett} {et~al.}(2026){Burgett}, {Bernstein}, {Bigelow},
  {Brossus}, {Cox}, {Demers}, {Figueroa}, {Fischer}, {Groark}, {Lam}, {Laskin},
  {Millan-Gabet}, {Park}, {Ruggerio}, {Turner}, \&
  {Walls}}]{Burgett2026GMTStatus}
{Burgett}, W.~S., {Bernstein}, R., {Bigelow}, B., {et~al.} 2026,
  \bibinfo{title}{{The Giant Magellan Telescope project in 2026: status and
  plan forward},} in Society of Photo-Optical Instrumentation Engineers (SPIE)
  Conference Series, Vol. 14147, Ground-based and Airborne Telescopes XI, ed.
  H.~K. {Marshall}, T.~{Usuda}, B.~{Onodera}, \& J.~{Spyromilio} (SPIE),
  141471C, \dodoi{10.1117/12.3105768}

% type= article
\bibitem[{K.~L. Cahoy {et~al.}(2017)Cahoy, Marley, \&
  Fortney}]{cahoyErratumExoplanetAlbedo2017}
Cahoy, K.~L., Marley, M.~S., \& Fortney, J. 2017, \bibinfo{title}{Erratum:
  “{Exoplanet} {Albedo} {Spectra} and {Colors} as a {Function} of {Planet}
  {Phase}, {Separation}, and {Metallicity}” (2010, {ApJ}, 724, 189),} The
  Astrophysical Journal, 844, 89, \dodoi{10.3847/1538-4357/aa79f7}

% type= article
\bibitem[{K.~L. Cahoy {et~al.}(2010)Cahoy, Marley, \&
  Fortney}]{cahoyExoplanetAlbedoSpectra2010}
Cahoy, K.~L., Marley, M.~S., \& Fortney, J.~J. 2010, \bibinfo{title}{Exoplanet
  {Albedo} {Spectra} and {Colors} as a {Function} of {Planet} {Phase},
  {Separation}, and {Metallicity},} The Astrophysical Journal, 724, 189,
  \dodoi{10.1088/0004-637X/724/1/189}

% type= article
\bibitem[{J.~L. {Christiansen} {et~al.}(2025){Christiansen}, {McElroy},
  {Harbut}, {Ciardi}, {Crane}, {Good}, {Hardegree-Ullman}, {Kesseli}, {Lund},
  {Lynn}, {Muthiar}, {Nilsson}, {Oluyide}, {Papin}, {Rivera}, {Swain},
  {Susemiehl}, {Tam}, {van Eyken}, \&
  {Beichman}}]{Christiansen2025ExoplanetArchive}
{Christiansen}, J.~L., {McElroy}, D.~L., {Harbut}, M., {et~al.} 2025,
  \bibinfo{title}{{The NASA Exoplanet Archive and Exoplanet Follow-up Observing
  Program: Data, Tools, and Usage},} \psj, 6, 186, \dodoi{10.3847/PSJ/ade3c2}

% type= article
\bibitem[{A.~R. {Costa Silva} {et~al.}(2020){Costa Silva}, {Delgado Mena}, \&
  {Tsantaki}}]{CostaSilvia2020ChemicalAbuncances}
{Costa Silva}, A.~R., {Delgado Mena}, E., \& {Tsantaki}, M. 2020,
  \bibinfo{title}{{Chemical abundances of 1111 FGK stars from the HARPS-GTO
  planet search sample. III. Sulfur},} Astron. \& Astrophys, 634, A136,
  \dodoi{10.1051/0004-6361/201936523}

% type= inproceedings
\bibitem[{J.~D. {Crane} {et~al.}(2010){Crane}, {Shectman}, {Butler},
  {Thompson}, {Birk}, {Jones}, \& {Burley}}]{Crane2010PFS}
{Crane}, J.~D., {Shectman}, S.~A., {Butler}, R.~P., {et~al.} 2010,
  \bibinfo{title}{{The Carnegie Planet Finder Spectrograph: integration and
  commissioning},} in Society of Photo-Optical Instrumentation Engineers (SPIE)
  Conference Series, Vol. 7735, Ground-based and Airborne Instrumentation for
  Astronomy III, ed. I.~S. {McLean}, S.~K. {Ramsay}, \& H.~{Takami}, 773553,
  \dodoi{10.1117/12.857792}

% type= article
\bibitem[{T. {Currie} {et~al.}(2013){Currie}, {Burrows}, {Madhusudhan},
  {Fukagawa}, {Girard}, {Dawson}, {Murray-Clay}, {Kenyon}, {Kuchner},
  {Matsumura}, {Jayawardhana}, {Chambers}, \&
  {Bromley}}]{Currie2013betaPicbAtm}
{Currie}, T., {Burrows}, A., {Madhusudhan}, N., {et~al.} 2013,
  \bibinfo{title}{{A Combined Very Large Telescope and Gemini Study of the
  Atmosphere of the Directly Imaged Planet, {\ensuremath{\beta}} Pictoris b},}
  \apj, 776, 15, \dodoi{10.1088/0004-637X/776/1/15}

% type= article
\bibitem[{T. {Currie} {et~al.}(2021){Currie}, {Brandt}, {Kuzuhara}, {Chilcote},
  {Cashman}, {Liu}, {Lawson}, {Tobin}, {Brandt}, {Guyon}, {Lozi}, {Deo},
  {Vievard}, {Ahn}, \& {Skaf}}]{Currie2021SurveyofAcceleratingStars}
{Currie}, T., {Brandt}, T., {Kuzuhara}, M., {et~al.} 2021, \bibinfo{title}{{A
  New Type of Exoplanet Direct Imaging Search: The SCExAO/CHARIS Survey of
  Accelerating Stars},} arXiv e-prints, arXiv:2109.09745.
\newblock \doarXiv{2109.09745}

% type= article
\bibitem[{T. {Currie} {et~al.}(2023){Currie}, {Brandt}, {Brandt}, {Lacy},
  {Burrows}, {Guyon}, {Tamura}, {Liu}, {Sagynbayeva}, {Tobin}, {Chilcote},
  {Groff}, {Marois}, {Thompson}, {Murphy}, {Kuzuhara}, {Lawson}, {Lozi}, {Deo},
  {Vievard}, {Skaf}, {Uyama}, {Jovanovic}, {Martinache}, {Kasdin}, {Kudo},
  {McElwain}, {Janson}, {Wisniewski}, {Hodapp}, {Nishikawa}, {He{\l}miniak},
  {Kwon}, \& {Hayashi}}]{Currie2023-HIP99770b}
{Currie}, T., {Brandt}, G.~M., {Brandt}, T.~D., {et~al.} 2023,
  \bibinfo{title}{{Direct imaging and astrometric detection of a gas giant
  planet orbiting an accelerating star},} Science, 380, 198,
  \dodoi{10.1126/science.abo6192}

% type= article
\bibitem[{J. {Dallant} {et~al.}(2023){Dallant}, {Langlois}, {Flasseur}, \&
  {Thi{\'e}baut}}]{Dallant2023PACOME}
{Dallant}, J., {Langlois}, M., {Flasseur}, O., \& {Thi{\'e}baut}, {\'E}. 2023,
  \bibinfo{title}{{PACOME: Optimal multi-epoch combination of direct imaging
  observations for joint exoplanet detection and orbit estimation},} \aap, 679,
  A38, \dodoi{10.1051/0004-6361/202346427}

% type= article
\bibitem[{R. {de la Reza} {et~al.}(2023){de la Reza}, {Chavero},
  {Roca-F{\`a}brega}, {Llorente de Andr{\'e}s}, {Cruz}, \&
  {Cifuentes}}]{delaReza2023Searching}
{de la Reza}, R., {Chavero}, C., {Roca-F{\`a}brega}, S., {et~al.} 2023,
  \bibinfo{title}{{Searching for the nature of stars with debris disks and
  planets},} \aap, 671, A136, \dodoi{10.1051/0004-6361/202245222}

% type= inproceedings
\bibitem[{M. {Endl} {et~al.}(2011){Endl}, {Cochran}, {Hatzes}, \&
  {Wittenmyer}}]{Endl2011NewsFromGammaCep}
{Endl}, M., {Cochran}, W.~D., {Hatzes}, A.~P., \& {Wittenmyer}, R.~A. 2011,
  \bibinfo{title}{{News from the {\ensuremath{\gamma}} Cephei Planetary
  System},} in American Institute of Physics Conference Series, Vol. 1331,
  Planetary Systems Beyond the Main Sequence, ed. S.~{Schuh}, H.~{Drechsel}, \&
  U.~{Heber} (AIP), 88--94, \dodoi{10.1063/1.3556187}

% type= article
\bibitem[{M. {Endl} {et~al.}(2016){Endl}, {Brugamyer}, {Cochran}, {MacQueen},
  {Robertson}, {Meschiari}, {Ramirez}, {Shetrone}, {Gullikson}, {Johnson},
  {Wittenmyer}, {Horner}, {Ciardi}, {Horch}, {Simon}, {Howell}, {Everett},
  {Caldwell}, \& {Castanheira}}]{Endl2016TwoNewLongPeriod}
{Endl}, M., {Brugamyer}, E.~J., {Cochran}, W.~D., {et~al.} 2016,
  \bibinfo{title}{{Two New Long-period Giant Planets from the McDonald
  Observatory Planet Search and Two Stars with Long-period Radial Velocity
  Signals Related to Stellar Activity Cycles},} \apj, 818, 34,
  \dodoi{10.3847/0004-637X/818/1/34}

% type= article
\bibitem[{F. {Feng} {et~al.}(2022){Feng}, {Butler}, {Vogt}, {Clement},
  {Tinney}, {Cui}, {Aizawa}, {Jones}, {Bailey}, {Burt}, {Carter}, {Crane},
  {Flammini Dotti}, {Holden}, {Ma}, {Ogihara}, {Oppenheimer}, {O'Toole},
  {Shectman}, {Wittenmyer}, {Wang}, {Wright}, \& {Xuan}}]{Feng2022-3dSelection}
{Feng}, F., {Butler}, R.~P., {Vogt}, S.~S., {et~al.} 2022, \bibinfo{title}{{3D
  Selection of 167 Substellar Companions to Nearby Stars},} \apjs, 262, 21,
  \dodoi{10.3847/1538-4365/ac7e57}

% type= article
\bibitem[{R. {Ferrer-Chavez} {et~al.}(2026){Ferrer-Chavez}, {Wang}, {Wagner},
  {Lawson}, {Carter}, {Biller}, {Bendahan-West}, {McCreery}, {Luppe}, {Ertel},
  {James}, {Bogat}, {Kane}, {Sutlieff}, {Strampelli}, {Balmer}, {Bowens-Rubin},
  {Bruinsma}, {Skemer}, {Girard}, {Booth}, {Subbotina Stephenson}, {Crotts},
  {Marino}, {Sanghi}, {Liu}, {Skaf}, {Rickman}, {Perrin}, {Pueyo}, {Bowler},
  {Mang}, {Rebollido}, {Meyer}, {Pearce}, {Bonse}, {Fontanive}, {Carlson},
  {Rees}, {Rupert}, {Weible}, {Isbell}, {Guerra}, {Chen}, {Belle Parkinson},
  {Squicciarini}, {Chauvin}, {Matthews}, {Franson}, \&
  {Millar-Blanchaer}}]{Ferrer-Chavez2026Sub-JupitersSurvey}
{Ferrer-Chavez}, R., {Wang}, J.~J., {Wagner}, K., {et~al.} 2026,
  \bibinfo{title}{{The JWST Sub-Jupiters Survey: Direct Imaging Discovery of a
  Giant Planet and a Debris Disk Around the Young M-dwarf RX J0534.0-0221},}
  arXiv e-prints, arXiv:2609.20748, \dodoi{10.48550/arXiv.2609.20748}

% type= article
\bibitem[{K. {Franson} {et~al.}(2022){Franson}, {Bowler}, {Brandt}, {Dupuy},
  {Tran}, {Brandt}, {Li}, \& {Kraus}}]{Franson2022DynamicalMass}
{Franson}, K., {Bowler}, B.~P., {Brandt}, T.~D., {et~al.} 2022,
  \bibinfo{title}{{Dynamical Mass of the Young Substellar Companion HD 984 B},}
  AJ, 163, 50, \dodoi{10.3847/1538-3881/ac35e8}

% type= article
\bibitem[{K. {Franson} {et~al.}(2023){Franson}, {Bowler}, {Bonavita}, {Brandt},
  {Chen}, {Samland}, {Zhang}, {Lueber}, {Heng}, {Kitzmann}, {Wolf}, {Jones},
  {Tran}, {Bardalez Gagliuffi}, {Biller}, {Chilcote}, {Crepp}, {Dupuy},
  {Faherty}, {Fontanive}, {Groff}, {Gratton}, {Guyon}, {Jensen-Clem},
  {Jovanovic}, {Kasdin}, {Lozi}, {Magnier}, {Mu{\v{z}}i{\'c}}, {Sanghi}, \&
  {Theissen}}]{Franson2023-AFLepb}
{Franson}, K., {Bowler}, B.~P., {Bonavita}, M., {et~al.} 2023,
  \bibinfo{title}{{Astrometric Accelerations as Dynamical Beacons: Discovery
  and Characterization of HIP 21152 B, the First T-dwarf Companion in the
  Hyades},} \aj, 165, 39, \dodoi{10.3847/1538-3881/aca408}

% type= article
\bibitem[{G. {Fu} {et~al.}(2026){Fu}, {Lothringer}, {Mukherjee}, {Sing},
  {Manjavacas}, {Stevenson}, \& {Balmer}}]{Fu2026JWSTColorSequence}
{Fu}, G., {Lothringer}, J.~D., {Mukherjee}, S., {et~al.} 2026,
  \bibinfo{title}{{JWST color sequence of transiting exoplanets and directly
  imaged substellar objects},} arXiv e-prints, arXiv:2608.25993,
  \dodoi{10.48550/arXiv.2608.25993}

% type= article
\bibitem[{E. {Gaidos} {et~al.}(2023){Gaidos}, {Claytor}, {Dungee}, {Ali}, \&
  {Feiden}}]{Gaido2023TimeTable}
{Gaidos}, E., {Claytor}, Z., {Dungee}, R., {Ali}, A., \& {Feiden}, G.~A. 2023,
  \bibinfo{title}{{The TIME Table: rotation and ages of cool exoplanet host
  stars},} \mnras, 520, 5283, \dodoi{10.1093/mnras/stad343}

% type= article
\bibitem[{M.~R. {Giovinazzi} {et~al.}(2025){Giovinazzi}, {Blake}, {Robertson},
  {Lin}, {Gupta}, {Mahadevan}, {Wright}, {Bardalez Gagliuffi}, {Dong},
  {Fernandes}, {Fitzmaurice}, {Halverson}, {Kanodia}, {Logsdon}, {Luhn},
  {McElwain}, {Monson}, {Ninan}, {Rajagopal}, {Roy}, {Schwab},
  {Stef{\'a}nsson}, {Terrien}, {Eastman}, {Horner}, {Plavchan}, {Wang},
  {Wilson}, \& {Wittenmyer}}]{Giovinazzi2025NEIDEarthTwinII}
{Giovinazzi}, M.~R., {Blake}, C.~H., {Robertson}, P., {et~al.} 2025,
  \bibinfo{title}{{The NEID Earth Twin Survey. II. Dynamical Masses in Seven
  High-acceleration Star Systems},} \aj, 170, 52,
  \dodoi{10.3847/1538-3881/add922}

% type= article
\bibitem[{M.~R. {Giovinazzi} {et~al.}(2026){Giovinazzi}, {Fitzmaurice},
  {Gupta}, {Robertson}, {Mahadevan}, {Ford}, {Alvarado-Montes}, {Bender},
  {Blake}, {Dong}, {Fernandes}, {Halverson}, {Han}, {Kanodia}, {Krolikowski},
  {Logsdon}, {Ninan}, {Roy}, {Schwab}, {Stefansson}, {Terrien}, \&
  {Wright}}]{Giovinazzi2026NEIDEarthTwinIV}
{Giovinazzi}, M.~R., {Fitzmaurice}, E., {Gupta}, A.~F., {et~al.} 2026,
  \bibinfo{title}{{The NEID Earth Twin Survey. IV. Confirming an 89 days,
  msini=10M{\ensuremath{\oplus}} Planet Orbiting a Nearby Sun-like Star},} \aj,
  171, 286, \dodoi{10.3847/1538-3881/ae48f1}

% type= article
\bibitem[{E.~J. {Greenfield} {\&} J.~E. {Owen}(2026){Greenfield} \&
  {Owen}}]{Greenfield2026GaiaOrbitMultiplicityBias}
{Greenfield}, E.~J., \& {Owen}, J.~E. 2026, \bibinfo{title}{{Detecting multiple
  planets in Gaia data using Bayesian model selection},} arXiv e-prints,
  arXiv:2610.07149.
\newblock \doarXiv{2610.07149}

% type= article
\bibitem[{A.~F. {Gupta} {et~al.}(2021){Gupta}, {Wright}, {Robertson},
  {Halverson}, {Luhn}, {Roy}, {Mahadevan}, {Ford}, {Bender}, {Blake}, {Hearty},
  {Kanodia}, {Logsdon}, {McElwain}, {Monson}, {Ninan}, {Schwab},
  {Stef{\'a}nsson}, \& {Terrien}}]{Gupta2021TargetPrioritizationNEID}
{Gupta}, A.~F., {Wright}, J.~T., {Robertson}, P., {et~al.} 2021,
  \bibinfo{title}{{Target Prioritization and Observing Strategies for the NEID
  Earth Twin Survey},} \aj, 161, 130, \dodoi{10.3847/1538-3881/abd79e}

% type= article
\bibitem[{C.~K. {Harada} {et~al.}(2024){Harada}, {Dressing}, {Kane}, \&
  {Ardestani}}]{Harada2024SettingTheStage}
{Harada}, C.~K., {Dressing}, C.~D., {Kane}, S.~R., \& {Ardestani}, B.~A. 2024,
  \bibinfo{title}{{Setting the Stage for the Search for Life with the Habitable
  Worlds Observatory: Properties of 164 Promising Planet-survey Targets},}
  \apjs, 272, 30, \dodoi{10.3847/1538-4365/ad3e81}

% type= article
\bibitem[{C.~K. {Harada} {et~al.}(2025){Harada}, {Dressing}, {Turtelboom},
  {Kane}, {Blunt}, {Dietrich}, {Hinkel}, {Li}, {Mamajek}, {Rice}, {Tuchow},
  {Wittenmyer}, {Chin}, {Fernandez}, {Kulkarni}, {Lin}, {Liu}, {Liu}, {Nathan},
  \& {Zbriger}}]{Harada2025SPORES-HWOII}
{Harada}, C.~K., {Dressing}, C.~D., {Turtelboom}, E.~V., {et~al.} 2025,
  \bibinfo{title}{{SPORES-HWO. II. Companion Mass Limits and Updated Planet
  Properties for 120 Future Exoplanet Imaging Targets from 35 yr of Precise
  Doppler Monitoring},} \aj, 170, 343, \dodoi{10.3847/1538-3881/ae0b62}

% type= article
\bibitem[{K.~K. Hardegree-Ullman {et~al.}(2023)Hardegree-Ullman, Apai,
  Bergsten, Pascucci, \& López-Morales}]{hardegree-ullman_bioverse_2023}
Hardegree-Ullman, K.~K., Apai, D., Bergsten, G.~J., Pascucci, I., \&
  López-Morales, M. 2023, \bibinfo{title}{Bioverse: {A} {Comprehensive}
  {Assessment} of the {Capabilities} of {Extremely} {Large} {Telescopes} to
  {Probe} {Earth}-like {O2} {Levels} in {Nearby} {Transiting} {Habitable}-zone
  {Exoplanets},} The Astronomical Journal, 165, 267,
  \dodoi{10.3847/1538-3881/acd1ec}

% type= article
\bibitem[{K.~K. {Hardegree-Ullman} {et~al.}(2025){Hardegree-Ullman}, {Apai},
  {Haffert}, {Schlecker}, {Kasper}, {Kammerer}, \&
  {Wagner}}]{Hardegree-Ullman2025BioverseO2-ELTs}
{Hardegree-Ullman}, K.~K., {Apai}, D., {Haffert}, S.~Y., {et~al.} 2025,
  \bibinfo{title}{{Bioverse: Giant Magellan Telescope and Extremely Large
  Telescope Direct Imaging and High-resolution Spectroscopy
  Assessment{\textemdash}Surveying Exo-Earth O$_{2}$ and Testing the Habitable
  Zone Oxygen Hypothesis},} \aj, 169, 171, \dodoi{10.3847/1538-3881/adb02f}

% type= article
\bibitem[{C.~R. Harris {et~al.}(2020)Harris, Millman, van~der Walt, Gommers,
  Virtanen, Cournapeau, Wieser, Taylor, Berg, Smith, Kern, Picus, Hoyer, van
  Kerkwijk, Brett, Haldane, del Río, Wiebe, Peterson, Gérard-Marchant,
  Sheppard, Reddy, Weckesser, Abbasi, Gohlke, \& Oliphant}]{harris_array_2020}
Harris, C.~R., Millman, K.~J., van~der Walt, S.~J., {et~al.} 2020,
  \bibinfo{title}{Array programming with {NumPy},} Nature, 585, 357,
  \dodoi{10.1038/s41586-020-2649-2}

% type= article
\bibitem[{S.~N. {Hasler} {et~al.}(2026){Hasler}, {Greenbaum}, {Bryden},
  {Bailey}, {Llop-Sayson}, {Lane}, {Limbach}, {Pearce}, {Ingalls}, \&
  {Lowrance}}]{Hasler2026RingDirectImaging}
{Hasler}, S.~N., {Greenbaum}, A.~Z., {Bryden}, G., {et~al.} 2026,
  \bibinfo{title}{{Entering the Era of Circumplanetary Ring Detection with
  Direct Imaging},} arXiv e-prints, arXiv:2610.23850.
\newblock \doarXiv{2610.23850}

% type= article
\bibitem[{C. {Hemmann} {et~al.}(2026){Hemmann}, {M{\"u}ller}, \&
  {Helled}}]{Hemmann2026GiantPlanetsNotFullyMixed}
{Hemmann}, C., {M{\"u}ller}, S., \& {Helled}, R. 2026, \bibinfo{title}{{Giant
  exoplanets are not fully mixed},} arXiv e-prints, arXiv:2608.15717,
  \dodoi{10.48550/arXiv.2608.15717}

% type= article
\bibitem[{J.~D. Hunter(2007)Hunter}]{hunter_matplotlib_2007}
Hunter, J.~D. 2007, \bibinfo{title}{Matplotlib: {A} 2d graphics environment,}
  Computing in Science {\textbackslash}\& Engineering, 9, 90,
  \dodoi{10.1109/MCSE.2007.55}

% type= inproceedings
\bibitem[{M. {Johns}(2006){Johns}}]{Johns2006GMT}
{Johns}, M. 2006, \bibinfo{title}{{The Giant Magellan Telescope (GMT)},} in
  Society of Photo-Optical Instrumentation Engineers (SPIE) Conference Series,
  Vol. 6267, Ground-based and Airborne Telescopes, ed. L.~M. {Stepp}, 626729,
  \dodoi{10.1117/12.670839}

% type= article
\bibitem[{N.~A. {Kaib} {\&} J.~E. {Chambers}(2016){Kaib} \&
  {Chambers}}]{KaibChambers2016Fragility}
{Kaib}, N.~A., \& {Chambers}, J.~E. 2016, \bibinfo{title}{{The fragility of the
  terrestrial planets during a giant-planet instability},} \mnras, 455, 3561,
  \dodoi{10.1093/mnras/stv2554}

% type= article
\bibitem[{J. {Kammerer} {et~al.}(2024){Kammerer}, {Lawson}, {Perrin},
  {Rebollido}, {Stark}, {Stolker}, {Girard}, {Pueyo}, {Balmer}, {Worthen},
  {Chen}, {van der Marel}, {Lewis}, {Ward-Duong}, {Valenti}, {Clampin}, \&
  {Mountain}}]{Kammerer2024JWST/NIRCam-Beta-Pic-b}
{Kammerer}, J., {Lawson}, K., {Perrin}, M.~D., {et~al.} 2024,
  \bibinfo{title}{{JWST-TST High Contrast: JWST/NIRCam Observations of the
  Young Giant Planet {\ensuremath{\beta}} Pic b},} \aj, 168, 51,
  \dodoi{10.3847/1538-3881/ad4ffe}

% type= article
\bibitem[{J. {Kammerer} {et~al.}(2025){Kammerer}, {Winterhalder}, {Lacour},
  {Stolker}, {Marleau}, {Balmer}, {Moore}, {Piscarreta}, {Toci}, {M{\'e}rand},
  {Nowak}, {Rickman}, {Pueyo}, {Pourr{\'e}}, {Nasedkin}, {Wang}, {Bourdarot},
  {Eisenhauer}, {Henning}, {Garcia Lopez}, {van Dishoeck}, {Forveille},
  {Monnier}, {Abuter}, {Amorim}, {Benisty}, {Berger}, {Beust}, {Blunt},
  {Boccaletti}, {Bonnefoy}, {Bonnet}, {Bordoni}, {Brandner}, {Cantalloube},
  {Caselli}, {Ceva}, {Charnay}, {Chauvin}, {Chavez}, {Chomez}, {Choquet},
  {Christiaens}, {Cl{\'e}net}, {Coud{\'e} du Foresto}, {Cridland}, {Davies},
  {Dembet}, {Dexter}, {Drescher}, {Duvert}, {Eckart}, {Fontanive}, {F{\"o}rster
  Schreiber}, {Garcia}, {Gendron}, {Genzel}, {Gillessen}, {Girard}, {Grant},
  {Hagelberg}, {Haubois}, {Hei{\ss}el}, {Hinkley}, {Hippler}, {Houll{\'e}},
  {Hubert}, {Jocou}, {Keppler}, {Kervella}, {Kreidberg}, {Kurtovic},
  {Lagrange}, {Lapeyr{\`e}re}, {Le Bouquin}, {Lutz}, {Maire}, {Mang},
  {Matthews}, {Molli{\`e}re}, {Mordasini}, {Mouillet}, {Ott}, {Otten},
  {Paladini}, {Paumard}, {Perraut}, {Perrin}, {Pfuhl}, {Ribeiro},
  {Rustamkulov}, {S{\'e}gransan}, {Shangguan}, {Shimizu}, {Samland}, {Sing},
  {Stadler}, {Straub}, {Straubmeier}, {Sturm}, {Tacconi}, {Udry}, {Vigan},
  {Vincent}, {von Fellenberg}, {Widmann}, {Woillez}, \&
  {Yazici}}]{Kammerer2025ExoGRAVITYSurvey}
{Kammerer}, J., {Winterhalder}, T.~O., {Lacour}, S., {et~al.} 2025,
  \bibinfo{title}{{The ExoGRAVITY survey: A K-band spectral library of giant
  exoplanet and brown dwarf companions},} \aap, 704, A318,
  \dodoi{10.1051/0004-6361/202556860}

% type= article
\bibitem[{E. {Karkoschka}(1994){Karkoschka}}]{Karkoschka1994JovianPlanets}
{Karkoschka}, E. 1994, \bibinfo{title}{{Spectrophotometry of the Jovian Planets
  and Titan at 300- to 1000-nm Wavelength: The Methane Spectrum},} \icarus,
  111, 174, \dodoi{10.1006/icar.1994.1139}

% type= article
\bibitem[{M. {Kasper} {et~al.}(2021){Kasper}, {Cerpa Urra}, {Pathak}, {Bonse},
  {Nousiainen}, {Engler}, {Heritier}, {Kammerer}, {Leveratto}, {Rajani},
  {Bristow}, {Le Louarn}, {Madec}, {Str{\"o}bele}, {Verinaud}, {Glauser},
  {Quanz}, {Helin}, {Keller}, {Snik}, {Boccaletti}, {Chauvin}, {Mouillet},
  {Kulcs{\'a}r}, \& {Raynaud}}]{Kasper2021-ELTPCS}
{Kasper}, M., {Cerpa Urra}, N., {Pathak}, P., {et~al.} 2021,
  \bibinfo{title}{{PCS {\textemdash} A Roadmap for Exoearth Imaging with the
  ELT},} The Messenger, 182, 38, \dodoi{10.18727/0722-6691/5221}

% type= article
\bibitem[{D. {Keithly} {et~al.}(2021){Keithly}, {Savransky}, \&
  {Spohn}}]{Keithly2021IntegrationTimeAdjustedCompleteness}
{Keithly}, D., {Savransky}, D., \& {Spohn}, C. 2021,
  \bibinfo{title}{{Integration time adjusted completeness},} Journal of
  Astronomical Telescopes, Instruments, and Systems, 7, 037002,
  \dodoi{10.1117/1.JATIS.7.3.037002}

% type= article
\bibitem[{R.~K. {Kopparapu} {et~al.}(2014){Kopparapu}, {Ramirez},
  {SchottelKotte}, {Kasting}, {Domagal-Goldman}, \& {Eymet}}]{Kopparapu2014-HZ}
{Kopparapu}, R.~K., {Ramirez}, R.~M., {SchottelKotte}, J., {et~al.} 2014,
  \bibinfo{title}{{Habitable Zones around Main-sequence Stars: Dependence on
  Planetary Mass},} \apjl, 787, L29, \dodoi{10.1088/2041-8205/787/2/L29}

% type= article
\bibitem[{C. {Lammers} {\&} J.~N. {Winn}(2026){Lammers} \&
  {Winn}}]{LammersWinn2026GaiaDR4}
{Lammers}, C., \& {Winn}, J.~N. 2026, \bibinfo{title}{{On the Exoplanet Yield
  of Gaia Astrometry},} \aj, 171, 18, \dodoi{10.3847/1538-3881/ae21de}

% type= article
\bibitem[{E. {Lane} {et~al.}(2026){Lane}, {Limbach}, {Pearce}, {Hasler}, \&
  {Greenbaum}}]{Lane2026RomanRings}
{Lane}, E., {Limbach}, M.~A., {Pearce}, L., {Hasler}, S., \& {Greenbaum}, A.~Z.
  2026, \bibinfo{title}{{Detecting Exoplanetary Rings with Roman CGI Reflected
  Light Phase Curve Observations},} arXiv e-prints, arXiv:2610.23835.
\newblock \doarXiv{2610.23835}

% type= inproceedings
\bibitem[{H. {Le Coroller} {et~al.}(2015){Le Coroller}, {Nowak}, {Arnold},
  {Dohlen}, {Fusco}, {Sauvage}, \& {Vigan}}]{LeCoroller2015KStacker}
{Le Coroller}, H., {Nowak}, M., {Arnold}, L., {et~al.} 2015,
  \bibinfo{title}{{K-Stacker, a new way of detecting and characterizing
  exoplanets with high contrast imaging instruments},} in Twenty Years of Giant
  Exoplanets, ed. I.~{Boisse}, O.~{Demangeon}, F.~{Bouchy}, \& L.~{Arnold},
  59--65, \dodoi{10.48550/arXiv.1510.06331}

% type= article
\bibitem[{H. {Le Coroller} {et~al.}(2020){Le Coroller}, {Nowak}, {Delorme},
  {Chauvin}, {Gratton}, {Devinat}, {Bec-Canet}, {Schneeberger}, {Estevez},
  {Arnold}, {Beust}, {Bonnefoy}, {Boccaletti}, {Desgrange}, {Desidera},
  {Galicher}, {Lagrange}, {Langlois}, {Maire}, {Menard}, {Vernazza}, {Vigan},
  {Zurlo}, {Fenouillet}, {Lambert}, {Bonavita}, {Cheetham}, {D'orazi}, {Feldt},
  {Janson}, {Ligi}, {Mesa}, {Meyer}, {Samland}, {Sissa}, {Beuzit}, {Dohlen},
  {Fusco}, {Le Mignant}, {Mouillet}, {Ramos}, {Rochat}, \&
  {Sauvage}}]{LeCoroller2020KStacker}
{Le Coroller}, H., {Nowak}, M., {Delorme}, P., {et~al.} 2020,
  \bibinfo{title}{{K-Stacker: an algorithm to hack the orbital parameters of
  planets hidden in high-contrast imaging. First applications to VLT/SPHERE
  multi-epoch observations},} \aap, 639, A113,
  \dodoi{10.1051/0004-6361/202037605}

% type= article
\bibitem[{C. {Lehmann} {et~al.}(2025){Lehmann}, {Murphy}, {Liu}, \&
  {Flynn}}]{Lehmann2025ChemodynamicEvolution}
{Lehmann}, C., {Murphy}, M.~T., {Liu}, F., \& {Flynn}, C. 2025,
  \bibinfo{title}{{Chemodynamic evolution of Sun-like stars in nearby moving
  groups},} \mnras, 536, 498, \dodoi{10.1093/mnras/stae2656}

% type= article
\bibitem[{M. {Lietzow} {\&} S. {Wolf}(2023){Lietzow} \&
  {Wolf}}]{LietzowWol2023RingPolarization}
{Lietzow}, M., \& {Wolf}, S. 2023, \bibinfo{title}{{Scattered polarized
  radiation of extrasolar circumplanetary rings},} \aap, 671, A113,
  \dodoi{10.1051/0004-6361/202245474}

% type= article
\bibitem[{M.~A. {Limbach} {et~al.}(2024){Limbach}, {Lustig-Yaeger},
  {Vanderburg}, {Vos}, {Heller}, \& {Robinson}}]{Limbach2024ExomoonsRingHWO-I}
{Limbach}, M.~A., {Lustig-Yaeger}, J., {Vanderburg}, A., {et~al.} 2024,
  \bibinfo{title}{{Exomoons and Exorings with the Habitable Worlds Observatory.
  I. On the Detection of Earth─Moon Analog Shadows and Eclipses},} \aj, 168,
  57, \dodoi{10.3847/1538-3881/ad4a75}

% type= article
\bibitem[{J. {Llop-Sayson} {et~al.}(2026){Llop-Sayson}, {Fogal}, {Mang},
  {Thompson}, {Do 'O}, {Sanghi}, {Huckabee}, {Pueyo}, {Beichman}, {Bryden},
  {Ygouf}, {G'asp'ar}, {Wolff}, {Leisenring}, {Mawet}, {Meshkat}, {Morley}, \&
  {Ruffio}}]{Llop-Sayson2026EpsEri}
{Llop-Sayson}, J., {Fogal}, A., {Mang}, J., {et~al.} 2026, \bibinfo{title}{{An
  Updated Model for Epsilon Eridani b and Prospects for Imaging with the Roman
  Coronagraph},} arXiv e-prints, arXiv:2609.19131.
\newblock \doarXiv{2609.19131}

% type= article
\bibitem[{M. {Lucas} {et~al.}(2024){Lucas}, {Norris}, {Guyon}, {Bottom}, {Deo},
  {Vievard}, {Lozi}, {Ahn}, {Ashcraft}, {Currie}, {Doelman}, {Kudo},
  {Leboulleux}, {Lilley}, {Millar-Blanchaer}, {Safonov}, {Tuthill}, {Uyama},
  {Walk}, \& {Zhang}}]{Lucas2024SCExAO-Vampires}
{Lucas}, M., {Norris}, B., {Guyon}, O., {et~al.} 2024,
  \bibinfo{title}{{Visible-light High-contrast Imaging and Polarimetry with
  SCExAO/VAMPIRES},} \pasp, 136, 114504, \dodoi{10.1088/1538-3873/ad89af}

% type= article
\bibitem[{R. {Luque} {et~al.}(2019){Luque}, {Trifonov}, {Reffert},
  {Quirrenbach}, {Lee}, {Albrecht}, {Fredslund Andersen}, {Antoci}, {Grundahl},
  {Schwab}, \& {Wolthoff}}]{Luque2019PreciseRadialVelocities}
{Luque}, R., {Trifonov}, T., {Reffert}, S., {et~al.} 2019,
  \bibinfo{title}{{Precise radial velocities of giant stars. XIII. A second
  Jupiter orbiting in 4:3 resonance in the 7 CMa system},} \aap, 631, A136,
  \dodoi{10.1051/0004-6361/201936464}

% type= article
\bibitem[{R.~J. MacDonald {et~al.}(2018)MacDonald, Marley, Fortney, \&
  Lewis}]{macdonaldExploringH2OProminence2018}
MacDonald, R.~J., Marley, M.~S., Fortney, J.~J., \& Lewis, N.~K. 2018,
  \bibinfo{title}{Exploring {H2O} {Prominence} in {Reflection} {Spectra} of
  {Cool} {Giant} {Planets},} ApJ, 858, 69, \dodoi{10.3847/1538-4357/aabb05}

% type= article
\bibitem[{N. {Madhusudhan} {\&} A. {Burrows}(2012){Madhusudhan} \&
  {Burrows}}]{Madhusudhan2012AnalyticModels}
{Madhusudhan}, N., \& {Burrows}, A. 2012, \bibinfo{title}{{Analytic Models for
  Albedos, Phase Curves, and Polarization of Reflected Light from Exoplanets},}
  \apj, 747, 25, \dodoi{10.1088/0004-637X/747/1/25}

% type= article
\bibitem[{V.~V. {Makarov}(2025){Makarov}}]{Makarov2025GaiaInclBias}
{Makarov}, V.~V. 2025, \bibinfo{title}{{Orbital Inclination of Astrometric
  Binaries and the Dearth of Face-on Orbits in Gaia DR3 Solutions},} \apjl,
  981, L23, \dodoi{10.3847/2041-8213/adb840}

% type= article
\bibitem[{J. {Maldonado} {et~al.}(2025){Maldonado}, {Mirouh},
  {Mendigut{\'\i}a}, {Montesinos}, {Gragera-M{\'a}s}, \&
  {Villaver}}]{Maldonado2025IntermeidateMassStars}
{Maldonado}, J., {Mirouh}, G.~M., {Mendigut{\'\i}a}, I., {et~al.} 2025,
  \bibinfo{title}{{Intermediate-mass stars and the origin of the gas-giant
  planet-metallicity correlation},} \aap, 695, A27,
  \dodoi{10.1051/0004-6361/202453328}

% type= inproceedings
\bibitem[{J.~R. {Males} {et~al.}(2015){Males}, {Belikov}, \&
  {Bendek}}]{Males2015ODI}
{Males}, J.~R., {Belikov}, R., \& {Bendek}, E. 2015, \bibinfo{title}{{Orbital
  Differential Imaging: a new high-contrast post-processing technique for
  direct imaging of exoplanets},} in Society of Photo-Optical Instrumentation
  Engineers (SPIE) Conference Series, Vol. 9605, Techniques and Instrumentation
  for Detection of Exoplanets VII, ed. S.~{Shaklan}, 960518,
  \dodoi{10.1117/12.2188766}

% type= article
\bibitem[{J.~R. Males {et~al.}(2013)Males, Skemer, \&
  Close}]{malesDirectImagingHabitable2013}
Males, J.~R., Skemer, A.~J., \& Close, L.~M. 2013, \bibinfo{title}{Direct
  {Imaging} in the {Habitable} {Zone} and the {Problem} of {Orbital} {Motion},}
  The Astrophysical Journal, 771, 10, \dodoi{10.1088/0004-637X/771/1/10}

% type= article
\bibitem[{J.~R. {Males} {et~al.}(2024{\natexlab{a}}){Males}, {Close},
  {Haffert}, {Kautz}, {Kelly}, {Fletcher}, {Salanski}, {Durney}, {Noenickx},
  {Ford}, {Gasho}, {Pearce}, {Kueny}, {Guyon}, {Weinberger}, {Bowler}, {Kraus},
  \& {Batalha}}]{Males2024GMagAOX}
{Males}, J.~R., {Close}, L.~M., {Haffert}, S.~Y., {et~al.} 2024{\natexlab{a}},
  \bibinfo{title}{{High-Contrast Imaging at First-Light of the GMT: The
  Preliminary Design of GMagAO-X},} arXiv e-prints, arXiv:2407.13014,
  \dodoi{10.48550/arXiv.2407.13014}

% type= article
\bibitem[{J.~R. {Males} {et~al.}(2024{\natexlab{b}}){Males}, {Close},
  {Haffert}, {Kautz}, {Kueny}, {Long}, {McEwen}, {Swimmer}, {Bailey}, {Foster},
  {Mazin}, {Pearce}, {Liberman}, {Twitchell}, {Weinberger}, {Guyon}, {Hedglen},
  {McLeod}, {Roberts}, {Van Gorkom}, {Li}, {Doty}, \&
  {Gasho}}]{Males2024MagAOXPhaseIICommisioningResults}
{Males}, J.~R., {Close}, L.~M., {Haffert}, S.~Y., {et~al.} 2024{\natexlab{b}},
  \bibinfo{title}{{MagAO-X: Commissioning Results and Status of Ongoing
  Upgrades},} arXiv e-prints, arXiv:2407.13007,
  \dodoi{10.48550/arXiv.2407.13007}

% type= article
\bibitem[{J. {Mang} {et~al.}(2026{\natexlab{a}}){Mang}, {Chachan}, {Morley},
  {Batalha}, {Wogan}, {Mukherjee}, {Fortney}, {Marley}, {Visscher}, \&
  {Gharib-Nezhad}}]{Mang2026FlameSkimmer}
{Mang}, J., {Chachan}, Y., {Morley}, C.~V., {et~al.} 2026{\natexlab{a}},
  \bibinfo{title}{{The Sonora Substellar Atmosphere Models. VII. Flame Skimmer:
  Cloud-free Atmospheric and Evolutionary Models for the Coldest Substellar
  Objects},} arXiv e-prints, arXiv:2608.06454,
  \dodoi{10.48550/arXiv.2608.06454}

% type= article
\bibitem[{J. {Mang} {et~al.}(2026{\natexlab{b}}){Mang}, {Batalha}, {Morley},
  {Wogan}, {Mukherjee}, {Visscher}, {Marley}, {Fortney}, {Chubb}, {Gao}, \&
  {Malsky}}]{Mang2026PICASO40}
{Mang}, J., {Batalha}, N.~E., {Morley}, C.~V., {et~al.} 2026{\natexlab{b}},
  \bibinfo{title}{{PICASO 4.0: Clouds and Photochemistry in Climate Models of
  Brown Dwarfs and Exoplanets},} \apj, 1000, 98,
  \dodoi{10.3847/1538-4357/ae47ff}

% type= article
\bibitem[{E.~C. {Matthews} {et~al.}(2024){Matthews}, {Carter}, {Pathak},
  {Morley}, {Phillips}, {P.~M.}, {Feng}, {Bonse}, {Boogaard}, {Burt},
  {Crossfield}, {Douglas}, {Henning}, {Hom}, {Ko}, {Kasper}, {Lagrange}, {Petit
  dit de la Roche}, \& {Philipot}}]{Matthews2024epsIndiAb}
{Matthews}, E.~C., {Carter}, A.~L., {Pathak}, P., {et~al.} 2024,
  \bibinfo{title}{{A temperate super-Jupiter imaged with JWST in the
  mid-infrared},} \nat, 633, 789, \dodoi{10.1038/s41586-024-07837-8}

% type= article
\bibitem[{M. {Mayor} {et~al.}(2004){Mayor}, {Udry}, {Naef}, {Pepe}, {Queloz},
  {Santos}, \& {Burnet}}]{Mayor2004}
{Mayor}, M., {Udry}, S., {Naef}, D., {et~al.} 2004, \bibinfo{title}{{The
  CORALIE survey for southern extra-solar planets. XII. Orbital solutions for
  16 extra-solar planets discovered with CORALIE},} \aap, 415, 391,
  \dodoi{10.1051/0004-6361:20034250}

% type= article
\bibitem[{B.~E. {McArthur} {et~al.}(2010){McArthur}, {Benedict}, {Barnes},
  {Martioli}, {Korzennik}, {Nelan}, \& {Butler}}]{McArthur2010UpsAndHSTFGS}
{McArthur}, B.~E., {Benedict}, G.~F., {Barnes}, R., {et~al.} 2010,
  \bibinfo{title}{{New Observational Constraints on the {\ensuremath{\upsilon}}
  Andromedae System with Data from the Hubble Space Telescope and Hobby-Eberly
  Telescope},} \apj, 715, 1203, \dodoi{10.1088/0004-637X/715/2/1203}

% type= article
\bibitem[{B.~E. {McArthur} {et~al.}(2014){McArthur}, {Benedict}, {Henry},
  {Hatzes}, {Cochran}, {Harrison}, {Johns-Krull}, \& {Nelan}}]{McArthur2014}
{McArthur}, B.~E., {Benedict}, G.~F., {Henry}, G.~W., {et~al.} 2014,
  \bibinfo{title}{{Astrometry, Radial Velocity, and Photometry: The HD 128311
  System Remixed with Data from HST, HET, and APT},} \apj, 795, 41,
  \dodoi{10.1088/0004-637X/795/1/41}

% type= inproceedings
\bibitem[{K.~L. Miller {et~al.}(2026)Miller, Douglas, Gorkom, Franks, Stefan,
  West, Durney, Ford, Koehler, Males, Stone, Agarwal, Anche, Ard,
  Brynjegard-Bialik, Choi, Derby, Eiklenborg, Gasho, Haughwout, Huie, Ingraham,
  Kaye, Kim, Lambert, Meyer, Milani, Morrison, Nicolas, Olivas, Rinaldi,
  Rosenbluth, Saedi-Marghmaleki, Schilperoort, Subramanian, Yoo, Wong,
  Fucetola, \& Merrill}]{Miller2026Lazuli-ESC}
Miller, K.~L., Douglas, E., Gorkom, K. J.~V., {et~al.} 2026,
  \bibinfo{title}{{The Lazuli Space Observatory’s ExtraSolar Coronagraph},}
  in Space Telescopes and Instrumentation 2026: Optical, Infrared, and
  Millimeter Wave, ed. L.~E. Coyle, S.~Matsuura, \& M.~D. Perrin, Vol. 14145,
  International Society for Optics and Photonics (SPIE), 1414563,
  \dodoi{10.1117/12.3114857}

% type= article
\bibitem[{S.~E. {Moran} {et~al.}(2025){Moran}, {Lodge}, {Batalha}, {Ohno},
  {Vahidinia}, {Marley}, {Wakeford}, \& {Leinhardt}}]{Moran2025Virga2}
{Moran}, S.~E., {Lodge}, M.~G., {Batalha}, N.~E., {et~al.} 2025,
  \bibinfo{title}{{Fractal Aggregate Aerosols in the Virga Cloud Code. I. Model
  Description and Application to a Benchmark Cloudy Exoplanet},} \apj, 994,
  116, \dodoi{10.3847/1538-4357/ae0583}

% type= misc
\bibitem[{C.~V. Morley {et~al.}(2024)Morley, Mukherjee, Marley, Fortney,
  Visscher, Lupu, Gharib-Nezhad, Thorngren, Freedman, \&
  7}]{morleySonoraSubstellarAtmosphere2024}
Morley, C.~V., Mukherjee, S., Marley, M.~S., {et~al.} 2024, The {Sonora}
  {Substellar} {Atmosphere} {Models}. {III}. {Diamondback}: {Atmospheric}
  {Properties}, {Spectra}, and {Evolution} for {Warm} {Cloudy} {Substellar}
  {Objects}, arXiv.
\newblock \url{http://arxiv.org/abs/2402.00758}

% type= misc
\bibitem[{{}{NASA Exoplanet Archive}(2026){NASA Exoplanet
  Archive}}]{PSCompPars}
{NASA Exoplanet Archive}. 2026, Planetary Systems Composite Parameters,
  Version: 2026-10-05 NExScI-Caltech/IPAC, \dodoi{10.26133/NEA13}

% type= book
\bibitem[{{}{National Academies of Sciences, Engineering, and
  Medicine}(2021){National Academies of Sciences, Engineering, and
  Medicine}}]{Astro2020Decadal}
{National Academies of Sciences, Engineering, and Medicine}. 2021, {Pathways to
  Discovery in Astronomy and Astrophysics for the 2020s},
  \dodoi{10.17226/26141}

% type= article
\bibitem[{M. Nayak {et~al.}(2017)Nayak, Lupu, Marley, Fortney, Robinson, \&
  Lewis}]{nayak_atmospheric_2017}
Nayak, M., Lupu, R., Marley, M.~S., {et~al.} 2017, \bibinfo{title}{Atmospheric
  {Retrieval} for {Direct} {Imaging} {Spectroscopy} of {Gas} {Giants} in
  {Reflected} {Light}. {II}. {Orbital} {Phase} and {Planetary} {Radius},}
  Publications of the Astronomical Society of the Pacific, 129, 034401,
  \dodoi{10.1088/1538-3873/129/973/034401}

% type= article
\bibitem[{B.~E. {Nelson} {et~al.}(2014){Nelson}, {Ford}, {Wright}, {Fischer},
  {von Braun}, {Howard}, {Payne}, \& {Dindar}}]{Nelson201455Cancri}
{Nelson}, B.~E., {Ford}, E.~B., {Wright}, J.~T., {et~al.} 2014,
  \bibinfo{title}{{The 55 Cancri planetary system: fully self-consistent N-body
  constraints and a dynamical analysis},} \mnras, 441, 442,
  \dodoi{10.1093/mnras/stu450}

% type= article
\bibitem[{M. {Nowak} {et~al.}(2018){Nowak}, {Le Coroller}, {Arnold}, {Dohlen},
  {Estevez}, {Fusco}, {Sauvage}, \& {Vigan}}]{Nowak2018KStacker}
{Nowak}, M., {Le Coroller}, H., {Arnold}, L., {et~al.} 2018,
  \bibinfo{title}{{K-Stacker: Keplerian image recombination for the direct
  detection of exoplanets},} \aap, 615, A144,
  \dodoi{10.1051/0004-6361/201629531}

% type= article
\bibitem[{G. {Orban de Xivry} {et~al.}(2026){Orban de Xivry}, {Absil},
  {Hammond}, {Bertram}, {van Boekel}, {Bon{\'e}}, {Chauvin}, {Christiaens},
  {Dolkens}, {Otten}, {Pathak}, \& {Pereira}}]{deXivry2026METIS-simultations}
{Orban de Xivry}, G., {Absil}, O., {Hammond}, I., {et~al.} 2026,
  \bibinfo{title}{{METIS high-contrast imaging simulations: From instrument
  modelling to science readiness},} arXiv e-prints, arXiv:2608.30988,
  \dodoi{10.48550/arXiv.2608.30988}

% type= article
\bibitem[{P. {Padovani} {\&} M. {Cirasuolo}(2023){Padovani} \&
  {Cirasuolo}}]{Paolo2023ELT}
{Padovani}, P., \& {Cirasuolo}, M. 2023, \bibinfo{title}{{The Extremely Large
  Telescope},} Contemporary Physics, 64, 47,
  \dodoi{10.1080/00107514.2023.2266921}

% type= article
\bibitem[{L.~A. {Pearce} {et~al.}(2025{\natexlab{a}}){Pearce}, {Males}, \&
  {Limbach}}]{Pearce2025projecc}
{Pearce}, L.~A., {Males}, J.~R., \& {Limbach}, M.~A. 2025{\natexlab{a}},
  \bibinfo{title}{{Direct Detection of Known Exoplanets in Reflected Light:
  Predicting Sky Position with Literature Orbit Solutions},} \pasp, 137,
  094401, \dodoi{10.1088/1538-3873/ae015e}

% type= article
\bibitem[{L.~A. {Pearce} {et~al.}(2025{\natexlab{b}}){Pearce}, {Males},
  {Haffert}, {Close}, {Long}, {McEwen}, {Liberman}, {Kautz}, {Kueny},
  {Weinberger}, {Li}, {Tonucci}, {Twitchell}, {McLeod}, {Foster}, {Guyon},
  {Hedglen}, {Van Gorkom}, {Lumbres}, {Schatz}, {Gasho}, {Morzinski}, \&
  {Hinz}}]{Pearce2025SLS}
{Pearce}, L.~A., {Males}, J.~R., {Haffert}, S.~Y., {et~al.} 2025{\natexlab{b}},
  \bibinfo{title}{{Five New Sirius-like White Dwarf─Main-sequence Star
  Systems with MagAO-X},} \aj, 170, 58, \dodoi{10.3847/1538-3881/addb50}

% type= article
\bibitem[{F. {Philipot} {et~al.}(2023{\natexlab{a}}){Philipot}, {Lagrange},
  {Kiefer}, {Rubini}, {Delorme}, \& {Chomez}}]{Philipot2023Multitechnique}
{Philipot}, F., {Lagrange}, A.-M., {Kiefer}, F., {et~al.} 2023{\natexlab{a}},
  \bibinfo{title}{{A multi-technique approach to identifying and/or
  constraining radial-velocity substellar companions},} \aap, 678, A107,
  \dodoi{10.1051/0004-6361/202346612}

% type= article
\bibitem[{F. {Philipot} {et~al.}(2023{\natexlab{b}}){Philipot}, {Lagrange},
  {Rubini}, {Kiefer}, \& {Chomez}}]{Philipot2023UpdatedCharacterization}
{Philipot}, F., {Lagrange}, A.-M., {Rubini}, P., {Kiefer}, F., \& {Chomez}, A.
  2023{\natexlab{b}}, \bibinfo{title}{{Updated characterization of long-period
  single companion by combining radial velocity, relative astrometry, and
  absolute astrometry},} \aap, 670, A65, \dodoi{10.1051/0004-6361/202245396}

% type= article
\bibitem[{G. {Piccinini} {et~al.}(2026){Piccinini}, {Petralia}, {Sozzetti},
  {Benatti}, {Gandolfi}, \& {Micela}}]{Piccinini2026piMen}
{Piccinini}, G., {Petralia}, A., {Sozzetti}, A., {et~al.} 2026,
  \bibinfo{title}{{True masses using radial velocity data with HIPPARCOS and
  Gaia astrometry},} \aap, 707, A104, \dodoi{10.1051/0004-6361/202556735}

% type= article
\bibitem[{M. {Pinamonti} {et~al.}(2026){Pinamonti}, {Sozzetti}, {Barbato},
  {Desidera}, {Biazzo}, {Bonomo}, {Lanza}, {Naponiello}, {Affer}, {Anche},
  {Andreuzzi}, {Basilicata}, {Brinjikji}, {Brogi}, {Cabona}, {Carolo},
  {Colombo}, {Damasso}, {D'Arpa}, {Di Filippo}, {Harutyunyan}, {Hom},
  {Mancini}, {Mantovan}, {Nardiello}, {Santhakumari}, \&
  {Zingales}}]{Pinamonti2026HD128717}
{Pinamonti}, M., {Sozzetti}, A., {Barbato}, D., {et~al.} 2026,
  \bibinfo{title}{{The GAPS programme at TNG: LXX. HD128717 B/Gaia-6 B: A
  long-period eccentric low-mass brown dwarf from astrometry and radial
  velocities},} \aap, 707, A67, \dodoi{10.1051/0004-6361/202557414}

% type= article
\bibitem[{D. {Pourbaix}(2001){Pourbaix}}]{Pourbaix2001Hip-Planet-Masses}
{Pourbaix}, D. 2001, \bibinfo{title}{{The Hipparcos observations and the mass
  of sub-stellar objects},} \aap, 369, L22, \dodoi{10.1051/0004-6361:20010252}

% type= article
\bibitem[{L.~J. {Rosenthal} {et~al.}(2021){Rosenthal}, {Fulton}, {Hirsch},
  {Isaacson}, {Howard}, {Dedrick}, {Sherstyuk}, {Blunt}, {Petigura}, {Knutson},
  {Behmard}, {Chontos}, {Crepp}, {Crossfield}, {Dalba}, {Fischer}, {Henry},
  {Kane}, {Kosiarek}, {Marcy}, {Rubenzahl}, {Weiss}, \&
  {Wright}}]{Rosenthal2021CalLegSurvey}
{Rosenthal}, L.~J., {Fulton}, B.~J., {Hirsch}, L.~A., {et~al.} 2021,
  \bibinfo{title}{{The California Legacy Survey. I. A Catalog of 178 Planets
  from Precision Radial Velocity Monitoring of 719 Nearby Stars over Three
  Decades},} ApJs, 255, 8, \dodoi{10.3847/1538-4365/abe23c}

% type= article
\bibitem[{A. {Roy} {et~al.}(2026){Roy}, {Feldman}, {Klupar}, {DiPalma},
  {Perlmutter}, {Douglas}, {Aldering}, {Furesz}, {Ingraham}, {Stefansson},
  {Kelly}, {Yang}, {Wevers}, {Arulanantham}, {Lasker}, {Rigault}, {Schlawin},
  {Zandbergen}, {Worden}, {Anche}, {Choi}, {Crossfield}, {Derby}, {Edelstein},
  {Eiklenborg}, {Gezari}, {Giuliano}, {Hom}, {Hoyt}, {Kang}, {Kim},
  {Kunnumkai}, {Lacroix}, {Males}, {Maccarone}, {Milani}, {Miller}, {Miller},
  {Nicolas}, {Palmese}, {Pero}, {Pueyo}, {Rinaldi}, {Sand}, {Schneider},
  {Sabhlok}, {Smith}, {Stefan}, {Kalyani Subramanian}, {Van Gorkom}, {Wong},
  {Yoo}, {Abdullah Al Zaman}, \& {the Lazuli Science
  Team}}]{Roy2026LazuliArchitechture}
{Roy}, A., {Feldman}, S., {Klupar}, P., {et~al.} 2026, \bibinfo{title}{{The
  Lazuli Space Observatory: Architecture \& Capabilities},} arXiv e-prints,
  arXiv:2601.02556, \dodoi{10.48550/arXiv.2601.02556}

% type= article
\bibitem[{A. {Sanghi} {et~al.}(2026){Sanghi}, {Mang}, {Llop-Sayson}, {Mamajek},
  {Thompson}, {Sur}, {Beichman}, {Bryden}, {G{\'a}sp{\'a}r}, {Leisenring},
  {Mawet}, {Morley}, {Ruffio}, {Wolff}, \& {Ygouf}}]{Sanghi2026epsEri}
{Sanghi}, A., {Mang}, J., {Llop-Sayson}, J., {et~al.} 2026,
  \bibinfo{title}{{Worlds Next Door. III. Indirect Evidence for Enhanced
  Atmospheric Metallicity and/or the Presence of Water Clouds in the Nearest
  Jupiter-analog ϵ Eri b},} \aj, 171, 225, \dodoi{10.3847/1538-3881/ae4909}

% type=
\bibitem[{D. Savransky {\&} H. Grigoryan(2025)Savransky \&
  Grigoryan}]{corgietc}
Savransky, D., \& Grigoryan, H. 2025, {corgietc}, 1.3.0 Github.
\newblock \url{https://github.com/roman-corgi/corgietc}

% type= article
\bibitem[{S. {Schlagenhauf} {et~al.}(2024){Schlagenhauf}, {Mugrauer}, {Ginski},
  {Buder}, {Fern{\'a}ndez}, \& {Bischoff}}]{Schlegenhauf2024SearchForStellar}
{Schlagenhauf}, S., {Mugrauer}, M., {Ginski}, C., {et~al.} 2024,
  \bibinfo{title}{{Search for Stellar Companions of Exoplanet Host Stars with
  AstraLux/CAHA 2.2 m},} \mnras, 529, 4768, \dodoi{10.1093/mnras/stae520}

% type= inproceedings
\bibitem[{C. {Schwab} {et~al.}(2016){Schwab}, {Rakich}, {Gong}, {Mahadevan},
  {Halverson}, {Roy}, {Terrien}, {Robertson}, {Hearty}, {Levi}, {Monson},
  {Wright}, {McElwain}, {Bender}, {Blake}, {St{\"u}rmer}, {Gurevich},
  {Chakraborty}, \& {Ramsey}}]{Schwab2016NEID}
{Schwab}, C., {Rakich}, A., {Gong}, Q., {et~al.} 2016, \bibinfo{title}{{Design
  of NEID, an extreme precision Doppler spectrograph for WIYN},} in Society of
  Photo-Optical Instrumentation Engineers (SPIE) Conference Series, Vol. 9908,
  Ground-based and Airborne Instrumentation for Astronomy VI, ed. C.~J.
  {Evans}, L.~{Simard}, \& H.~{Takami}, 99087H, \dodoi{10.1117/12.2234411}

% type= article
\bibitem[{D. {Spergel} {et~al.}(2013){Spergel}, {Gehrels}, {Breckinridge},
  {Donahue}, {Dressler}, {Gaudi}, {Greene}, {Guyon}, {Hirata}, {Kalirai},
  {Kasdin}, {Moos}, {Perlmutter}, {Postman}, {Rauscher}, {Rhodes}, {Wang},
  {Weinberg}, {Centrella}, {Traub}, {Baltay}, {Colbert}, {Bennett},
  {Kiessling}, {Macintosh}, {Merten}, {Mortonson}, {Penny}, {Rozo},
  {Savransky}, {Stapelfeldt}, {Zu}, {Baker}, {Cheng}, {Content}, {Dooley},
  {Foote}, {Goullioud}, {Grady}, {Jackson}, {Kruk}, {Levine}, {Melton},
  {Peddie}, {Ruffa}, \& {Shaklan}}]{Spergel2013WFIRST}
{Spergel}, D., {Gehrels}, N., {Breckinridge}, J., {et~al.} 2013,
  \bibinfo{title}{{Wide-Field InfraRed Survey Telescope-Astrophysics Focused
  Telescope Assets WFIRST-AFTA Final Report},} arXiv e-prints, arXiv:1305.5422,
  \dodoi{10.48550/arXiv.1305.5422}

% type= article
\bibitem[{M.~R. {Standing} {et~al.}(2026){Standing}, {Barnes}, {Haswell},
  {Stevenson}, {Faria}, {Quintin}, {Ross}, {Fossati}, {Jenkins}, {Alves}, \&
  {Staab}}]{Standing2026DMPP}
{Standing}, M.~R., {Barnes}, J.~R., {Haswell}, C.~A., {et~al.} 2026,
  \bibinfo{title}{{The dispersed matter planet project sample ─ detection
  limits, occurrence rates, and new planets},} \mnras, 547, stag370,
  \dodoi{10.1093/mnras/stag370}

% type= article
\bibitem[{{}{STScI Development Team}(2013){STScI Development
  Team}}]{pysynphot-citation}
{STScI Development Team}. 2013, \bibinfo{title}{{pysynphot: Synthetic
  photometry software package},}

% type= article
\bibitem[{B.~J. {Sutlieff} {et~al.}(2026){Sutlieff}, {Bonse}, {Christiaens},
  {Fontanive}, {Matthews}, {Parker}, {Pearce}, {Birkby}, {Biller}, {Dupuy},
  {Garvin}, {Iskandarli}, {Kammerer}, {Zhou}, {De Rosa}, {Carter}, {Hinkley},
  {Kenworthy}, {Balmer}, {Hammond}, {Mang}, {Morley}, {Neeser}, {Absil},
  {Boccaletti}, {Bonavita}, {Bowler}, {Chen}, {Dannert}, {Girard}, {Kasper},
  {Lagrange}, {Liu}, {Orban de Xivry}, {Poon}, {Quanz}, {Serra}, {Vos},
  {Wagner}, {Wang}, {Sch{\"o}lkopf}, {Agapito}, {Agudo Berbel}, {Apai},
  {Baruffolo}, {Black}, {Bonaglia}, {Briguglio}, {Cao}, {Carbonaro}, {Chapman},
  {Cresci}, {Dallilar}, {Davies}, {Deysenroth}, {Di Antonio}, {Di Cianno}, {Di
  Rico}, {Doelman}, {Dolci}, {Eisenhauer}, {Esposito}, {Ferruzzi},
  {Feuchtgruber}, {F{\"o}rster-Schreiber}, {Franson}, {Genzel}, {Gillessen},
  {Gonzales}, {Hartl}, {Hayoz}, {Huber}, {Keller}, {Kravchenko}, {Leisenring},
  {Lightfoot}, {Lunney}, {Lutz}, {Macintosh}, {Mannucci}, {Metchev}, {Ott},
  {Pearson}, {Puglisi}, {Rabien}, {Rau}, {Riccardi}, {Salasnich}, {Shimizu},
  {Snik}, {Sturm}, {Su{\'a}rez}, {Tacconi}, {Tan}, {Taylor}, {Waring}, \&
  {Xompero}}]{Sutlieff2026betaPicd}
{Sutlieff}, B.~J., {Bonse}, M.~J., {Christiaens}, V., {et~al.} 2026,
  \bibinfo{title}{{Direct Imaging Discovery of Giant Exoplanet
  {\ensuremath{\beta}} Pictoris d: A Decade-long Game of Hide-and-seek},}
  \apjl, 1006, L10, \dodoi{10.3847/2041-8213/ae80a0}

% type= article
\bibitem[{W. Thompson {et~al.}(2023)Thompson, Lawrence, Blakely, Marois, Wang,
  Giordano, Brandt, Johnstone, Ruffio, Ammons, Crotts, \'o, Gonzales, \&
  Rice}]{thompsonOctofitterFastFlexible2023}
Thompson, W., Lawrence, J., Blakely, D., {et~al.} 2023,
  \bibinfo{title}{Octofitter: {Fast}, {Flexible}, and {Accurate} {Orbit}
  {Modelling} to {Detect} {Exoplanets},} AJ, 166, 164,
  \dodoi{10.3847/1538-3881/acf5cc}

% type= article
\bibitem[{A.~K. {Veenstra} {et~al.}(2025){Veenstra}, {Zuluaga},
  {Alvarado-Montes}, {Sucerquia}, \&
  {Stam}}]{Veenstra2025RingPolarizationModel}
{Veenstra}, A.~K., {Zuluaga}, J.~I., {Alvarado-Montes}, J.~A., {Sucerquia}, M.,
  \& {Stam}, D.~M. 2025, \bibinfo{title}{{A general polarimetric model for
  transiting and nontransiting ringed exoplanets},} \aap, 693, A310,
  \dodoi{10.1051/0004-6361/202347194}

% type= article
\bibitem[{A.~J. {Winter} {et~al.}(2020){Winter}, {Kruijssen}, {Longmore}, \&
  {Chevance}}]{Winter2020StarClustering}
{Winter}, A.~J., {Kruijssen}, J.~M.~D., {Longmore}, S.~N., \& {Chevance}, M.
  2020, \bibinfo{title}{{Stellar clustering shapes the architecture of
  planetary systems},} \nat, 586, 528, \dodoi{10.1038/s41586-020-2800-0}

% type= article
\bibitem[{S. {Witte} {et~al.}(2011){Witte}, {Helling}, {Barman}, {Heidrich}, \&
  {Hauschildt}}]{Witte2011DustInBDs}
{Witte}, S., {Helling}, C., {Barman}, T., {Heidrich}, N., \& {Hauschildt},
  P.~H. 2011, \bibinfo{title}{{Dust in brown dwarfs and extra-solar planets.
  III. Testing synthetic spectra on observations},} \aap, 529, A44,
  \dodoi{10.1051/0004-6361/201014105}

% type= article
\bibitem[{S.~G. {Wolff} {et~al.}(2026){Wolff}, {Bailey}, {Hom}, {Biller},
  {Savransky}, {Girard}, {Bogat}, {Carri{\'o}n-Gonz{\'a}lez}, {Do {\'O}},
  {Kuzuhara}, {Xie}, {Millar-Blanchaer}, {Redmond}, {Mazoyer}, {Vega-Pallauta},
  {Chauvin}, {Ravet}, {Greenbaum}, {Cady}, {Anche}, {Ingalls}, \& {the Roman
  Coronagraph Community Participation Program Team}}]{Wolff2026RomanCGI-CPP}
{Wolff}, S.~G., {Bailey}, V.~P., {Hom}, J., {et~al.} 2026, \bibinfo{title}{{The
  Roman Coronagraph Community Participation Program: trials and triumphs of
  designing an observing program for a technology demonstration instrument},}
  arXiv e-prints, arXiv:2608.17077, \dodoi{10.48550/arXiv.2608.17077}

% type= article
\bibitem[{Y. {Wu} {et~al.}(2026{\natexlab{a}}){Wu}, {Xiao}, {Butler}, {Feng},
  {Shectman}, {Crane}, {Teske}, {Wang}, {Beletsky}, {Burt}, {Daylan}, {Diaz},
  {Dragomir}, {Flowers}, {Jenkins}, {Kanodia}, {Nandakumar}, {Rice}, {Shporer},
  {Yee}, {Zhou}, \& {Lin}}]{Wu2026DetectionOfFourColdJupiters}
{Wu}, Y., {Xiao}, G.-Y., {Butler}, R.~P., {et~al.} 2026{\natexlab{a}},
  \bibinfo{title}{{Detection of Four Cold Jupiters through Combined Analyses of
  Radial Velocity and Astrometry Data},} \aj, 171, 189,
  \dodoi{10.3847/1538-3881/ae40f5}

% type= article
\bibitem[{G.-Y. {Xiao} {et~al.}(2023){Xiao}, {Liu}, {Teng}, {Wang}, {Brandt},
  {Zhao}, {Zhao}, {Zhai}, \& {Gao}}]{Xioa2023MassesOfASample}
{Xiao}, G.-Y., {Liu}, Y.-J., {Teng}, H.-Y., {et~al.} 2023, \bibinfo{title}{{The
  Masses of a Sample of Radial-velocity Exoplanets with Astrometric
  Measurements},} Research in Astronomy and Astrophysics, 23, 055022,
  \dodoi{10.1088/1674-4527/accb7e}

% type= article
\bibitem[{J.~W. {Xuan} {et~al.}(2024){Xuan}, {Perrin}, {Mawet}, {Knutson},
  {Mukherjee}, {Zhang}, {Hoch}, {Wang}, {Inglis}, {Wallack}, \&
  {Ruffio}}]{Xuan2024Gl229}
{Xuan}, J.~W., {Perrin}, M.~D., {Mawet}, D., {et~al.} 2024,
  \bibinfo{title}{{Atmospheric Abundances and Bulk Properties of the Binary
  Brown Dwarf Gliese 229Bab from JWST/MIRI Spectroscopy},} \apjl, 977, L32,
  \dodoi{10.3847/2041-8213/ad92f9}

% type= article
\bibitem[{J.~W. {Xuan} {et~al.}(2026{\natexlab{a}}){Xuan}, {Balmer}, {Chachan},
  {Ruffio}, {Ohno}, {De Rosa}, {Baburaj}, {Nielsen}, {Murray-Clay}, {Fortney},
  {Perrin}, {Wang}, {Bowler}, {Madurowicz}, {Macintosh}, {Zhang}, {Benneke},
  {Bidot}, {Blake}, {Franson}, {He}, {Inglis}, {Knutson}, {Mawet}, {Pueyo},
  {Rickman}, {Sanghi}, \& {Wallack}}]{Xuan2026JWST-AFLepb}
{Xuan}, J.~W., {Balmer}, W.~O., {Chachan}, Y., {et~al.} 2026{\natexlab{a}},
  \bibinfo{title}{{A Sulfur-Rich Atmosphere for the Young Jupiter Analog AF Lep
  b Reveals Significant Solid Accretion},} arXiv e-prints, arXiv:2608.10068,
  \dodoi{10.48550/arXiv.2608.10068}

% type= article
\bibitem[{J.~W. {Xuan} {et~al.}(2026{\natexlab{b}}){Xuan}, {Ruffio}, {Chachan},
  {Ohno}, {Kesseli}, {Murray-Clay}, {Lee}, {Moses}, {Balmer}, {Baburaj},
  {Blake}, {Johnstone}, {Zhang}, {Knutson}, {Mawet}, {Beichman}, {Hodapp},
  {Perrin}, {Konopacky}, {Meyer}, {Bryden}, {Greene}, {Leisenring}, {Ygouf},
  {Benneke}, {Inglis}, \& {Wallack}}]{Xuan2026CompositionsOfHR8799}
{Xuan}, J.~W., {Ruffio}, J.-B., {Chachan}, Y., {et~al.} 2026{\natexlab{b}},
  \bibinfo{title}{{The Compositions of the HR 8799 Planets Reflect Accretion of
  Both Solids and Metal-enriched Gas},} \apj, 1000, 27,
  \dodoi{10.3847/1538-4357/ae448f}

\end{thebibliography}
\bibliographystyle{aasjournalv7.1}



\end{document}